# The Chemistry and Kinematics of Two Molecular Clouds near Sagittarius A*

John A. P. Lopez[1,2★], Maria R. Cunningham[1], Paul A. Jones[1], Jonathan P. Marshall[1,3], Leonardo Bronfman[4], Nadia Lo[4], Andrew J. Walsh[5]
[1]*School of Physics, UNSW Australia, Sydney NSW 2052, Australia*
[2]*CSIRO Astronomy & Space Science, Australia Telescope National Facility, PO Box 76, Epping, NSW 1710, Australia*
[3]*Australian Centre for Astrobiology, UNSW Australia, Sydney NSW 2052, Australia*
[4]*Departamento de Astronomía, Universidad de Chile, Santiago, Chile*
[5]*International Centre for Radio Astronomy Research, Curtin University, GPO Box U1987, Perth, WA 6845, Australia*

**ABSTRACT**
We have analysed the chemical and kinematic properties of the 20 and 50 km s$^{-1}$ molecular clouds in the Central Molecular Zone of the Milky Way Galaxy, as well as those of the molecular ridge bridging these two clouds. Our work has utilized 37 molecular transitions in the 0.65, 3 and 7-mm wavebands, from the Mopra and NANTEN2 telescopes. The 0.65-mm NANTEN2 data highlights a dense condensation of emission within the western part of the 20 km s$^{-1}$ cloud, visible in only four other transitions, which are 3-mm $H^{13}CN$ (1–0), $H^{13}CO^+$ (1–0), HNC (1–0) and $N_2H^+$ (1–0), suggesting that the condensation is moderately optically thick and cold. We find that while the relative chemical abundances between both clouds are alike in many transitions, suggesting little variation in the chemistry between both clouds; the 20 km s$^{-1}$, cold cloud is brighter than the 50 km s$^{-1}$ cloud in shock and high density tracers. The spatial distribution of enhanced emission is widespread in the 20 km s$^{-1}$ cloud, as shown via line ratio maps. The position velocity diagrams across both clouds indicate that the gas is well mixed. We show that the molecular ridge is most likely part of the 20 km s$^{-1}$ cloud and that both of them may possibly extend to include the 50 km s$^{-1}$ cloud, as part of one larger cloud. Furthermore, we expect that the 20 km s$^{-1}$ cloud is being tidally sheared as a result of the gravitational potential from Sgr A*.

**Key words:** ISM: abundances – ISM: clouds – ISM: kinematics and dynamics – ISM: molecules – Galaxy: centre – radio lines: ISM

## 1 INTRODUCTION

Molecular clouds located in the centre of our Galaxy, the Milky Way, have unique properties and are exposed to an extreme environment dissimilar to that we are familiar with in the spiral arms. They have a rotational (and comparable kinetic) temperature ranging from 60–120 K in $NH_3$ (Guesten et al. 1985), a mean density of $n(H_2) \geqslant 10^4$ cm$^{-3}$, as derived from CS emission (Bally et al. 1987), and an internal velocity dispersion of 15–50 km s$^{-1}$ (Bally et al. 1988). These values are significantly greater than those of clouds in the disk of our Galaxy, which have average values of 10 K (kinetic temperature) and $n(H_2) = 300$ cm$^{-3}$ (Solomon, Sanders, & Scoville 1979), as well as a velocity dispersion of $\leqslant 8$ km s$^{-1}$ (Solomon et al. 1987); which were all obtained from CO emission. The Galactic Centre is also the location of a supermassive black hole known as Sagittarius A* (hereafter Sgr A*; Schödel et al. 2002). Within this environment, high-mass stars are observed in close proximity to Sgr A* (Paumard et al. 2006), and nearby regions are in the process of forming new high-mass stars, like Sgr B2 (Zhao & Wright 2011) and Sgr C (Kendrew et al. 2013). Critical to understanding the star formation process in this dense and dynamic environment are measurements of the physical and chemical properties of the clouds themselves.

In this paper, we have analysed molecular clouds in a region close to Sgr A*, focusing on molecular chemistry (temperature, density and abundances) and kinematics. These clouds are found in the Sagittarius A (Sgr A) region, which is located within the centre of our Galaxy, in a neighbourhood known as the Central Molecular Zone (CMZ), which contains large scale complex kinematical properties (Henshaw et al. 2016), whilst also exhibiting high levels of turbulence and a low star formation rate, as described in Kruijssen et al. 2014. This region harbours a vast reservoir of molecular gas, both rich in chemistry and diverse in nature (Morris & Serabyn 1996; Requena-Torres et al. 2006, 2008; Riquelme et al. 2010; Martín et al. 2012).

★ E-mail: j.lopez@unsw.edu.au

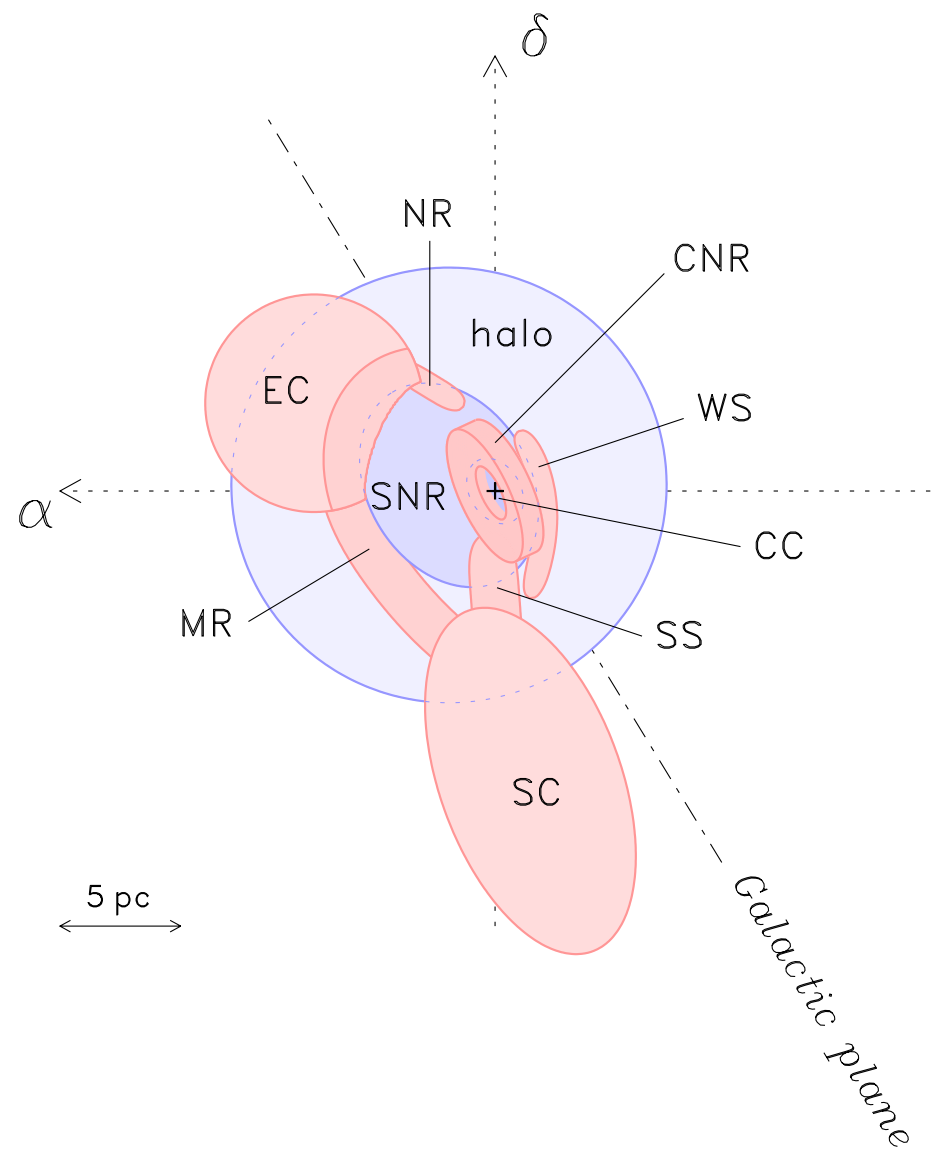

**Figure 1.** Illustration of gas within 10 pc of Sgr A*. The dotted vertical line $\delta$ denotes the declination and $\alpha$ is the right ascension. The abbreviations are as follows: EC (Eastern Cloud; 50 km $s^{-1}$ cloud), SC (Southern Cloud; 20 km $s^{-1}$ cloud), MR (Molecular Ridge), CC (Central Cavity), CNR (Circumnuclear Ring), NR (Northern Ridge), Radio Halo (Halo), WS (Western Streamer), SNR (Sagittarius A East Supernova Remnant), SS (Southern Streamer) and '+' is the location of Sgr A*. Note that the red and blue colours correspond to primarily molecular and ionised gas respectively. We are focusing on the EC, SC and MR features. This diagram has been reproduced from fig. 4c in Ferrière 2012.

The molecular clouds within Sgr A have been observed in many molecules at a range of wavelengths (Minh, Irvine, & Friberg 1992; Martin et al. 2004; Jones et al. 2012, 2013; Mills & Morris 2013; Armijos-Abendaño et al. 2015; Hsieh, Ho, & Hwang 2015), including para-formaldehyde (Ao et al. 2013; Ginsburg et al. 2016), continuum emission (Liu et al. 2013), Paschen-alpha images (Mills et al. 2011) and the H92α radio recombination line (Wong et al. 2016). In Fig. 1, we have reproduced a model of different structures in the region found in Ferrière (2012), made from many past observational studies of the Galactic Centre. It shows the relative positions and interplay between molecular and ionised components within 10 pc of Sgr A*, such as the supernova remnant, molecular clouds and the circumnuclear molecular ring, to illustrate our current understanding of them. Similar models can be found in Lee et al. (2008); Amo-Baladrón, Martín-Pintado, & Martín (2011).

In this work, we use molecular line millimetre data to primarily focus on two molecular clouds near Sgr A*, known as GCM–0.02–0.07 and GCM–0.13–0.08 (Güesten, Walmsley, & Pauls 1981), which are part of the Sgr A molecular cloud complex. These two clouds are now commonly referred to as the 50 and 20 km $s^{-1}$ clouds respectively. In Fig. 1, this corresponds to the parts labelled EC (Eastern Cloud; 50 km $s^{-1}$ cloud) and SC (Southern Cloud; 20 km $s^{-1}$ cloud). The molecular ridge connecting both of the clouds, denoted as MR in Fig. 1, is also discussed. Emission from molecular transitions, allow us to investigate the properties of these clouds, such as temperature and density (Bally 1986). By combining a total of 18 molecules and some of their isotopologues, covering 37 transitions, across three different wavebands, we have an unparalleled view of these two molecular clouds.

In § 2, we discuss the multi-wavelength radio datasets that we have used, and the reduction process. This comprises archival 3 and 7-mm molecular transitions, described in Jones et al. (2012) and Jones et al. (2013), as well as new 3-mm molecular transitions from the Mopra 22-metre radio telescope and CO data from the NANTEN2 telescope. In § 2.1 we present the CO (4–3) data from NANTEN2 and in § 2.2, new 3-mm CO (1–0) data from Mopra. A comparison between the NANTEN2 and Mopra data is found in § 2.3. In § 2.4, we explain the calculation of the abundance ratios for molecular transitions in the 20 and 50 km $s^{-1}$ clouds.

Within § 3, we analyse and discuss our key findings. In § 3.1 we derive column densities and $H_2$ conversion factors for a selection of molecules, while in § 3.2 our integrated emission findings are compared with previous studies. The results of the line ratio plots between the 20 and 50 km $s^{-1}$ clouds and regions within those clouds are examined in § 3.3. A comparison of peak and moment 0 images from a selection of enhanced molecules is discussed in § 3.4; and in § 3.5, images of line ratio maps are analyzed to determine the extent of enhanced emission in the 20 and 50 km $s^{-1}$ cloud. In § 3.6, we interpret the kinematics of the two aforementioned clouds as well as their relationship to the molecular ridge. In § 3.7, we investigate the possibility of expecting tidal effects from Sgr A* on the 20 km $s^{-1}$ cloud.

**Table 1.** Summary of radio molecular line datasets. $T_{sys}$ is the system temperature.

| Frequency range (GHz) | Beam size (arcsec) | Velocity resolution (km $s^{-1}$) | mean $T_{sys}$ (K) | RMS noise (mK) |
|---|---|---|---|---|
| 42–49[a] | 75 | ≈ 1.86 | 84–126 | 24–76 |
| 85–94[b] | 39 | ≈ 1.86 | 210–225 | ≈ 50 |
| 108–116[c] | 36 | ≈ 1.86 | 325–586 | 82–322 |
| 460[c] | 38 | ≈ 0.68 | ≈ 900 | 480 |

[a] Jones et al. (2013). [b] Jones et al. (2012). [c] This work. The $T_{sys}$ value at 460 GHz is an approximation based upon a sample spectrum from the data, and previous observations with the same telescope (e.g. Mizuno et al. 2010).

## 2 OBSERVATIONAL DATA AND REDUCTION

We have utilized four molecular line emission datasets, as shown in Table 1. The molecular lines from 42–116 GHz were observed with the Mopra 22-metre radio telescope, which is located near Coonabarabran in northern New South Wales, Australia. It is operated by CSIRO Astronomy and Space Science, which forms part of the Australia Telescope National Facility. The observation and reduction methods of the 42–49 and 85–94 GHz molecular transitions were thoroughly discussed in Jones et al. (2013) and Jones et al. (2012) respectively. The 108–116 GHz transtions were observed in June 2009. A continuous noise diode was used in the calibration of the system temperature for the 3-mm (Ladd et al. 2005) and 7-mm observations (Urquhart et al. 2010). In the 3-mm observations, an ambient temperature load (paddle) was used (Jones et al. 2012); however, in the 7-mm observations, the paddle was not utilised, so a few extra corrections were required to account for atmospheric opacity, as described in (Jones et al. 2013).

The data were processed by converting on-the-fly maps to FITS

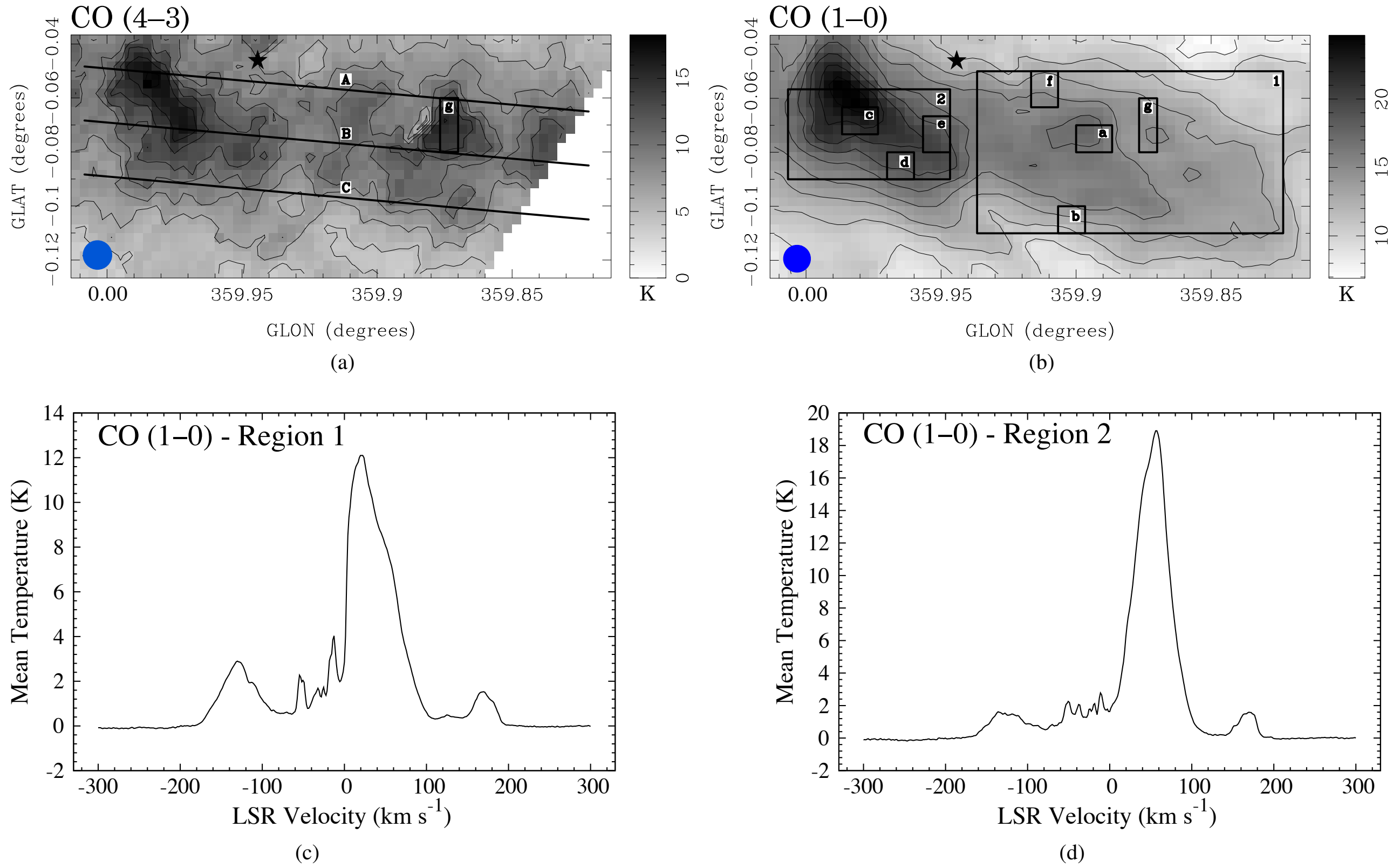

**Figure 2.** NANTEN2 and Mopra CO temperature maps. (a) CO (4–3) peak brightness temperature emission map, in the $T_A^*$ (K) scale, which is the uncorrected antenna temperature, as surveyed with the NANTEN2 telescope. Contour levels are: 0.9, 2.7, 4.6, 6.4, 8.2, 10.1, 11.9, 13.7, 15.5, 17.4 (K). The star symbol refers to the position of Sgr A*, whereas region 'g' indicates the location of the 'western peak', centred at coordinates of 359.873°, -0.070°; and the beam full width at half maximum (hereafter FWHM) is represented by the circle on the bottom left corner. The three solid lines labelled A, B and C, as shown in (a), illustrate the 'slices' used to make the PV diagrams, as shown in § 3.6. (b) 3-mm CO (1–0) peak brightness image, also in the $T_A^*$ (K) scale. The contour levels are: 7.8, 9.6, 11.3, 13.1, 14.9, 16.7, 18.4, 20.2, 22.0, 23.8 (K). The labels in (b) are as follows: region '1': this rectangular area is the region we have defined as the '20 km s$^{-1}$ molecular cloud'. Regions 'a' and 'b' are the respective peak and off peak positions within the 20 km s$^{-1}$ cloud. Region '2': this area covers the region we have designated as the '50 km s$^{-1}$ molecular cloud'. Regions 'c' and 'd' are the respective peak and off peak positions within the 50 km s$^{-1}$ cloud. Regions 'e' and 'f' are the locations of previous observations of the 50 and 20 km s$^{-1}$ clouds respectively, as discussed in Amo-Baladrón, Martín-Pintado, & Martín (2011). The star symbol, region 'g' and the circle are the same as described in (a). Further information on the properties of the regions, can be found in Table 3. Fig. (c) and (d) are the mean brightness temperature spectral lines of the Mopra CO (1–0) data, for regions 1 and 2 respectively, in $T_A^*$ (K). The Doppler reference frame is the Local Standard of Rest (LSR). The integrated emission was performed as discussed in § 2.4. An absorption feature is present at 0 km s$^{-1}$ in (c).

data cubes with GRIDZILLA and LIVEDATA[1]. A bandpass correction was applied to the raw spectra, which were also calibrated using an appropriate reference spectra that was at an emission free off-source position, with Galactic coordinates of $l$ = 1.093° and $b$ = -0.735°, for both the 3 and 7-mm data (Jones et al. 2012, 2013). A good second order polynomial was then fitted to and consequently subtracted from the baseline of the spectra; these steps were carried out utilising LIVEDATA. After doing this, the data cubes were created by gridding the spectra and applying a median filter (to avoid outliers from bad data) with GRIDZILLA.

The corresponding data cube FITS files from GRIDZILLA were also subsequently processed through MIRIAD[2]. In particular, the data cubes were hanning smoothed and then Nyquist sampled to the resolutions shown in Table 1; such that it is smaller than the broad lines which we expect to see in the CMZ (> 10 km s$^{-1}$), therefore making it appropriate for use in this analysis. We note that for molecules which have characteristically narrow properties, such as methanol maser emission, the cubes that were not smoothed were used.

The 460 GHz dataset was obtained with the NANTEN2[3] 4-metre sub-millimetre telescope. It is located in the Atacama desert, on Pampa La Bola in Chile. The observations were made in November 2008. Data were collected using the KOSMA SMART receiver. The data were calibrated using KOSMA-SMART software and gridded with CLASS, which is part of the GILDAS[4] software package.

This section provides an outline of our procedure in analysing the radio data. This methodology has been applied to all the molecular transitions given in Table 2 and the extensive set of results will be discussed in the appropriate sections throughout this paper.

The CMZ maps of Jones et al. (2012) cover a Galactic longitude range ($l$) from $-0.7^\circ$ to $1.8^\circ$ and latitude ($b$) of $-0.3^\circ$ to $0.2^\circ$. We chose a sub region to analyse, which includes the location of molecular clouds close to Sgr A*, with $l$ from $\approx 359.815^\circ$ to $0.012^\circ$ and $b$ between $\approx -0.125^\circ$ and $-0.038^\circ$. The dimensions of our box in $l \times b$ are $\approx 0.197^\circ \times 0.087^\circ$ respectively; as can be seen in Fig. 2(b).

[1] http://www.atnf.csiro.au/computing/software/livedata/
[2] http://www.atnf.csiro.au/computing/software/miriad/
[3] http://www.astro.uni-koeln.de/nanten2/
[4] http://www.iram.fr/IRAMFR/GILDAS

**Table 2.** 7, 3 and 0.65-mm molecular lines and transitions discussed in this paper. This table was made by reproducing and editing table 1 from both Jones et al. 2012 and Jones et al. 2013, in addition to including our new datasets, which utilized the *Splatalogue* molecular line database (Remijan, Markwick-Kemper, & ALMA Working Group on Spectral Line Frequencies 2007). The 'gp' which has been appended to the aforementioned transitions, simply indicate a 'group' of transitions. Furthermore, we also shortened table 1 from Jones et al. 2012, by adding 'gp' to their molecules which had groups of transitions. The asterisks in each wavelength group, signify lines which have hyperfine transitions.

| Wavelength | Molecule | Transition | Rest Frequency (GHz) |
|---|---|---|---|
| 7-mm | $NH_2CHO$ | 2(0,2)–1(0,1) gp | 42.386070* |
| | SiO | 1–0 v = 0 | 43.423864 |
| | HNCO | 2(0,2)–1(0,1) gp | 43.962998* |
| | $CH_3OH$ | 7(0,7)–6(1,6) A++ | 44.069476 |
| | $H^{13}CCCN$ | 5–4 | 44.084172 |
| | CCS | 4,3–3,2 | 45.379029 |
| | $HC_3N$ | 5–4 gp | 45.490316* |
| | $^{13}CS$ | 1–0 | 46.247580 |
| | $C^{34}S$ | 1–0 | 48.206946 |
| | $CH_3OH$ | 1(0,1)–0(0,0) A++ | 48.372467* |
| | $CH_3OH$ | 1(0,1)–0(0,0) E | 48.376889 |
| | OCS | 4–3 | 48.651604 |
| | CS | 1–0 | 48.990957 |
| 3-mm | $c\text{-}C_3H_2$ | 2(1,2)–1(0,1) | 85.338906 |
| | $CH_3CCH$ | 5–4 gp | 85.457299* |
| | $H^{13}CN$ | 1–0 gp | 86.340167* |
| | $H^{13}CO^+$ | 1–0 | 86.754330 |
| | SiO | 2–1 v = 0 | 86.847010 |
| | $C_2H$ | 1–0 3/2–1/2 gp | 87.316925* |
| | HNCO | 4(0,4)–3(0,3) | 87.925238 |
| | HCN | 1–0 gp | 88.631847* |
| | $HCO^+$ | 1–0 | 89.188526 |
| | HNC | 1–0 gp | 90.663572* |
| | $HC_3N$ | 10–9 | 90.978989 |
| | $CH_3CN$ | 5–4 gp | 91.987089* |
| | $^{13}CS$ | 2–1 | 92.494303 |
| | $N_2H^+$ | 1–0 gp | 93.173480* |
| | $CH_3OH$ $v_t$ = 0 | 0(0,0)–1(-1,1) | 108.893929 |
| | $HC_3N$ v = 0 | 12–11 gp | 109.173638* |
| | $C^{18}O$ | 1–0 | 109.782173 |
| | HNCO | 5(0,5)–4(0,4) gp | 109.905753* |
| | $^{13}CO$ v = 0 | 1–0 | 110.201353 |
| | $CH_3CN$ v = 0 | 6(1)–5(1) gp | 110.381404* |
| | CN v = 0 | 1–0 1/2–1/2, F = 3/2–1/2 | 113.170528 |
| | CN v = 0 | 1–0 3/2–1/2, F = 5/2–3/2 | 113.490982 |
| | CO v = 0 | 1–0 | 115.271202 |
| 0.65-mm | CO v = 0 | 4–3 | 461.040808 |

The data cubes are regridded to the respective resolutions listed in Table 1 with MIRIAD version 1.5 (Sault, Teuben, & Wright 1995). The data were further corrected for the beam efficiency.

The 3-mm data were corrected based on the efficiency calculated in Ladd et al. (2005). Their results determined the extended beam efficiency of the Mopra telescope, in the 86–115 GHz range to be between 0.65–0.55 respectively. We applied a linear interpolation to find the beam efficiency factor for each of our respective molecular line frequencies. The extended beam efficiency is a quantification of the antenna radiation pattern and brightness distribution (both of which are modeled), for an extended object (in comparison to the size of the main beam of the telescope). In particular, it accounts for an error beam (side lobes) in addition to the main beam (main lobe) (Ulich & Haas 1976; Wilson, Rohlfs & Hüttemeister 2013).

To apply an extended beam efficiency correction to the 7-mm data, we extrapolated from Urquhart et al. (2010). Their paper provides a fourth order polynomial fitting equation for the main-beam efficiency of the Mopra telescope and an estimate is given for the extended beam efficiency. In order to replicate their results, we increased the main-beam efficiency corrections by a factor of $\approx 30$ per cent to get the extended beam efficiency and have confirmed this to be correct (Urquhart, J.S., private communication).

## 2.1 NANTEN2 Results

In Fig. 2(a), we present a CO (4–3) peak brightness temperature map, in $T_A^*$ (K), which was converted from raw NANTEN2 data to the $T_A^*$ scale, using continuum observations from Jupiter, as described in Kramer et al. 2008 and references therein. We have not corrected this data for the beam efficiency of the antenna, as it is not used in any quantitative analysis. Rather, we have only used it for comparing the distribution of molecular line emission between images, for all our datasets, as will be described in § 2.3 and § 3.6. This image is annotated with the position of Sgr A* (star symbol). The coordinates for this position was taken from the SIMBAD database[5] (Wenger et al. 2000). Region 'g' is also marked in this figure and it denotes the position of the 'western peak', which is discussed in § 2.3. The beam FWHM is the circle located on the corner of the bottom left hand size of the image. There are three parallel diagonal lines of equal length annotated in Fig. 2(a), which are marked as A, B and C. These three lines are used to make the position velocity (hereafter, PV) diagrams in Fig. 11 and 12; where three PV diagrams are generated for six bright molecular transitions, tracing the path of A, B and C. The three lines were chosen such that together, they 'slice' through the majority of emission from the 20 and 50 km s$^{-1}$ cloud. The positions of the two aforementioned clouds are labelled as '1' and '2' in Fig. 2(b). The spatial and velocity resolution of the CO (4–3) data are as found in Table 1.

## 2.2 Mopra Results

Fig. 2(b) is one example from the new 3-mm data (108.89–115.27 GHz) showing CO (1–0). Similar to Fig. 2(a), it is also a peak brightness emission image, in $T_A^*$ (K), which is the antenna temperature scale that has not been corrected for the efficiency of the antenna (e.g. Ladd et al. 2005); and the star symbol and beam FWHM are at the same positions. As seen in Table 1, the 3 and 7-mm datasets (42–116 GHz) have the same velocity spacing, but differing spatial resolutions. Additionally, we have also added the positions of the 20 and 50 km s$^{-1}$ clouds, as well as regions of interest within the clouds themselves. The molecular emission detected from these clouds (and the whole region) is extended and goes beyond the positions we have labelled above. For consistency, we have selected 6 regions for analysis which encapsulate the molecular line emission from both the 20 and 50 km s$^{-1}$ clouds, across all molecular transitions in our datasets. These 6 regions are: 1 (20 km s$^{-1}$ cloud), 2 (50 km s$^{-1}$ cloud), 'a' and 'b' (the respective peak and off peak regions from the 20 km s$^{-1}$ cloud); as well as 'c' and 'd' (the corresponding peak and off peak regions from the 50 km s$^{-1}$ cloud. We have also briefly discussed 3 other positions within the 2 clouds; 'e' and 'f' are regions which have been previously analysed in the 50 and 20 km s$^{-1}$ clouds respectively (Amo-Baladrón, Martín-Pintado, & Martín 2011); and 'g' is the location of a possible dense core, which we have named the 'western peak'. The positions, dimensions and brief descriptions of each region can be found in Table 3.

[5] http://simbad.u-strasbg.fr/simbad/

**Table 3.** Summary of regions that are discussed in this work, in the 20 and 50 km s$^{-1}$ clouds.

| Region | $\Delta l$ | $\Delta b$ | Central coordinates |
|---|---|---|---|
| 1$^{i}$ | 0.113 ° | 0.060 ° | 359.880 ° , -0.080 ° |
| 2$^{ii}$ | 0.060 ° | 0.033 ° | 359.977 ° , -0.073 ° |
| a$^{iii}$ | 0.013 ° | 0.010 ° | 359.893 ° , -0.075 ° |
| b$^{iv}$ | 0.010 ° | 0.010 ° | 359.902 ° , -0.105 ° |
| c$^{v}$ | 0.013 ° | 0.010 ° | 359.980 ° , -0.068 ° |
| d$^{vi}$ | 0.010 ° | 0.010 ° | 359.965 ° , -0.085 ° |
| e$^{vii}$ | 0.010 ° | 0.013 ° | 359.952 ° , -0.073 ° |
| f$^{viii}$ | 0.010 ° | 0.013 ° | 359.912 ° , -0.057 ° |
| g$^{ix}$ | 0.007 ° | 0.020 ° | 359.873 ° , -0.070 ° |

[i] The 20 km s$^{-1}$ cloud. [ii] The 50 km s$^{-1}$ cloud. [iii] The peak position in the 20 km s$^{-1}$ cloud. This area covers the $H_2$ peak in this cloud (Tsuboi et al. 2011) and G-0.11-0.08 (Martín et al. 2008; Requena-Torres et al. 2008). [iv] An off peak position from the 20 km s$^{-1}$ cloud. [v] The peak position in the 50 km s$^{-1}$ cloud. This area covers the $H_2$ peak in this cloud (Tsuboi et al. 2011) and G–0.02–0.07 (Martín et al. 2008; Requena-Torres et al. 2008). [vi] An off peak position from the 50 km s$^{-1}$ cloud. [vii] This area covers SiO emission from velocity integrated maps of the '50 GMC' from Amo-Baladrón, Martín-Pintado, & Martín 2011, which corresponds to emission from the 50 km s$^{-1}$ cloud i.e. within region 2. [viii] This area covers SiO emission from velocity integrated maps of the '20 GMC' from Amo-Baladrón, Martín-Pintado, & Martín 2011 which corresponds to emission from the 20 km s$^{-1}$ cloud i.e. within region 1. [ix] This area covers the 'western peak' in the 20 km s$^{-1}$ cloud.

Peak brightness emission images were used instead of 0th-moment images to avoid issues with baseline ripples. The errors associated with this in the 42–49, 85–94 and 108–116 GHz datasets are: ≈14 mK, ≈ 12 mK and ≈ 39 mK respectively. All peak brightness emission images shown in this paper and in the online supplementary material are in the $T_A^*$ (K) scale, although the data in Tables 4–16 and for all calculations in this paper, are corrected for the extended beam efficiency (Ladd et al. 2005; Urquhart et al. 2010), as described early on in § 2 and this will be briefly explained in § 2.4.

The peak brightness emission images, for each of the remaining molecular transitions (as well as including CO 1–0 and CO 4–3), can be found in the online supplementary material.

## 2.3 Comparison of NANTEN2 CO (4–3) to 3-mm Mopra CO (1–0)

Both Fig. 2(a) and 2(b) are similar in tracing the distribution of molecular gas in our chosen region. However, region 'g', the location covering the 'western peak' is visible in the NANTEN2 CO (4–3) image but not present in the CO (1–0) image.

Martin et al. (2004) observed the Galactic Centre in CO (4–3), CO (7–6) and [C I] $^3P_1$–$^3P_0$ with beam sizes of 103–109 arcsec using the Antarctic Submillimeter Telescope and Remote Observatory (AST/RO) (Stark et al. 2001). The NANTEN2 'western peak' is discernible in their integrated emission map of CO (4–3), less so in CO (7–6) and blended with surrounding gas in [C I] $^3P_1$–$^3P_0$. The 'western peak' is brightest between ≈ 2 to 8 km s$^{-1}$.

A search through the images for the transitions presented in Table 2, revealed that bright emission at the position of the 'western peak' is also clearly visible in 3-mm $H^{13}CN$ (1–0), $H^{13}CO^+$ (1–0), HNC (1–0) and $N_2H^+$ (1–0). We have calculated the optical depth of 3-mm $H^{13}CN$ (1–0) and $H^{13}CO^+$ (1–0) from region 'g', which covers an area surrounding the 'western peak', and found $\tau$ to be 0.59 and ≈ 0.32 respectively. The spectral lines of 3-mm $H^{13}CO^+$ (1–0) and $H^{13}CN$ (1–0) do not show any evidence of self absorption, but the 3-mm $HCO^+$ (1–0) and HCN (1–0) spectra do. In particular, while self absorption is present at 0 km s$^{-1}$ (which likely originates from gas outside the Galactic Centre, as will be discussed in § 3.6), 'minor' self absorption is also present at a velocity range between 10–15 km s$^{-1}$. As this 'western peak' is not prominent in $HCO^+$ (1–0), $^{13}CO$ (1–0), HCN (1–0), $HCO^+$ (1–0) and $C^{18}O$ (1–0), we can state that within region 'g', the gas is not very optically thick; and therefore we infer that the 'western peak' is likely to be deeply embedded cold dense gas. The strong detection of $N_2H^+$ (1–0), known to trace dense gas where CO is depleted, and HNC, which is preferentially formed over HCN in gas < 20 K is evidence for this core being cold (Lo et al. 2009).

## 2.4 Calculation of abundance ratios

The python package LMFIT[6] was utilised to determine the Gaussian fitting parameters of each spectral line, which included a calculation for the integrated emission and a corresponding ≈ 1 sigma uncertainty. This was done for regions 1 and 2 (the 20 and 50 km s$^{-1}$ clouds), 'a' and 'b' (the peak and off peak position in the 20 km s$^{-1}$ cloud); as well as 'c' and 'd' (the peak and off peak positions in the 50 km s$^{-1}$ cloud), for all molecules in Table 2, except for CO (4–3).

The ratios are presented in pairs for each molecule, in both the 20 and 50 km s$^{-1}$ clouds and only plotted when the signal to noise (hereafter S/N) is ⩾ 3. We selected $^{13}CS$ as the molecule to create our line ratios, making the assumption that it is optically thin and relatively insensitive to local conditions, and is therefore an appropriate molecule to determine the relative abundances of molecular gas in our selected regions. In addition, this molecule also emits in both the 3 and 7-mm bands. As a further check to corroborate the consistency of this assumption, we have also plotted the 3-mm line ratios with respect to another optically thin molecule, $C^{18}O$ (1–0). This comparison is not possible at 7-mm, as this CO isotopologue does not have any transitions in the 7-mm band. The 3-mm line ratio plots are found in Fig. 3(a), 3(b) and the 7-mm ratio is found in Fig. 4. Furthermore, we have also presented $^{13}CS$ ratios for region 'a' vs. region 1 (peak position in 20 km s$^{-1}$ cloud vs entirety of that cloud) in Fig. 5(a) and 5(b). In addition, we have also shown $^{13}$CS ratios for region 'c' vs. region 2 (peak position in 50 km s$^{-1}$ cloud vs. the entirety of that cloud), in Fig. 6(a) and 6(b). Lastly, we have also compared the peak vs. off peak position for both clouds, but we have not calculated the $^{13}$CS line ratio in these cases; instead, opting to simply compare the $\log_{10}$ integration emission results between the peak and off peak position for each molecule. This is demonstrated for the 20 km s$^{-1}$ cloud (region 'a' vs. 'b'), as shown in Fig. 7(a) and 7(b) and for the 50 km s$^{-1}$ cloud (region 'c' vs. 'd'), which can be found in Fig. 8(a) and 8(b). These are discussed further in § 3.3.

[6] http://dx.doi.org/10.5281/zenodo.11813

Each plot has been sorted in ascending frequency. The line ratios for both regions 1 and 2 follow:

$$\log_{10} \frac{\int T_{\mathrm{A}}\, \mathrm{dV}_{\mathrm{molecule}}}{\int T_{\mathrm{A}}\, \mathrm{dV}_{\mathrm{reference\ molecule}}} \tag{1}$$

where $\int T_{\mathrm{A}}\, \mathrm{dV}_{\mathrm{molecule}}$ and $\int T_{\mathrm{A}}\, \mathrm{dV}_{\mathrm{reference\ molecule}}$ are the respective integrated emissions of an individual molecule; whereas the reference molecule is the optically thin isotopologue e.g. $^{13}$CS; both of which include the extended beam efficiency calculation. We have defined $T_{\mathrm{A}}$ as $T_{\mathrm{A}}^{*}/\eta$, where $\eta$ is the extended beam efficiency (which was implemented following the discussion earlier in § 2.

The error ($\sigma$) in the line ratios are combined in quadrature, taking into account the fractional uncertainties of $\int T_{\mathrm{A}}\, \mathrm{dV}_{\mathrm{molecule}}$ and $\int T_{\mathrm{A}}\, \mathrm{dV}_{\mathrm{reference\ molecule}}$. This is in the form of:

$$\sigma = 0.4343 \times \sqrt{\left(\frac{\Delta \int T_{\mathrm{A}}\, \mathrm{dV}_{\mathrm{reference\ molecule}}}{\int T_{\mathrm{A}}\, \mathrm{dV}_{\mathrm{reference\ molecule}}}\right)^2 + \left(\frac{\Delta \int T_{\mathrm{A}}\, \mathrm{dV}_{\mathrm{molecule}}}{\int T_{\mathrm{A}}\, \mathrm{dV}_{\mathrm{molecule}}}\right)^2} \tag{2}$$

The factor of 0.4343, arises from the definition of the uncertainty in $\log_{10}$X, which is defined as $\frac{1}{\ln 10} \times \frac{\Delta X}{X}$, where $\frac{1}{\ln 10}$ is $\approx$ 0.4343 and $\Delta X$ is the uncertainty in the integrated emission of the molecule divided by the value of the integrated emission in that same molecule (X). Similarly, in the plots where we compared the $\log_{10}$ value of the integrated emission from the molecular lines, the uncertainty is defined as $0.4343 \times \frac{\Delta X}{X}$, where $\frac{\Delta X}{X}$ is the uncertainty in the integrated emission of that molecule, divided by the value of the integrated emission for that molecule.

Lastly, the integrated emission and molecular line ratios for regions 1 and 2, 'a' and 'b'; as well as 'c' and 'd', in the 3 and 7-mm data are presented in Tables 6-7, 9-10 and 12-13 respectively. Note that all errors discussed in this section have been included in the aforementioned tables. The NANTEN2 data are not included as part of the molecular line ratio section, but it is used as part of the kinematic analysis in § 3.6.

## 3 ANALYSIS AND DISCUSSION

### 3.1 Density, optical depth and $H_2$ conversion factors

We have evaluated the column densities ($N_{\mathrm{u}}$, $N$), optical depth ($\tau$) and the $H_2$ conversion factor ($X_{\mathrm{R}}$) for molecules in Table 5. We followed the same method as described in Jones et al. (2012), in which we have calculated the number of molecules in the upper quantum level, denoted by the term $N_{\mathrm{u}}$:

$$N_{\mathrm{u}} = \frac{8\pi k \nu^2}{A h c^3} \int T_{\mathrm{A}} \left(\frac{\tau}{1 - \exp(-\tau)}\right) dV \tag{3}$$

$N_{\mathrm{u}}$ is the average upper quantum level column density for each molecule in the boxed regions. The $T_{\mathrm{A}}$ value, in our case is $T_{\mathrm{xb}}$, which is the temperature after we have applied the extended beam efficiency correction. The units for the variables are as follows: $k$ (Boltzmann constant, erg K$^{-1}$), $h$ (Planck constant, erg s), $c$ (speed of light, cm s$^{-1}$), $\nu$ (frequency, s$^{-1}$), $\int T_{\mathrm{A}}\, dV$ (integrated emission of individual molecular transition, K cm s$^{-1}$) and $A$ (Einstein coefficient, s$^{-1}$). In determining the value for optical depth ($\tau$), we have used a similar method to equation 1 from Wong et al. (2008). This involved numerically solving for $\tau$, using a known main-beam temperature ($T_{\mathrm{mb}}$) ratio e.g. $^{12}$CO/C$^{18}$O. As a result, the optical depth of both molecules used in the ratio can be easily established. Our method only differs in the fact that we used the $T_{\mathrm{xb}}$ ratio, instead of $T_{\mathrm{mb}}$. A comparison of both ratios has been completed and shows small variations between them. As we have used the extended beam efficiency for $N_{\mathrm{u}}$, we decided to use the $T_{\mathrm{xb}}$ ratio to ensure consistency. The results of this can be found in Table 4. When we assume that molecules in our dataset are optically thin, $\tau/(1 - \exp(-\tau))$ will be equal to unity, i.e. 1. The final value for $N_{\mathrm{u}}$ is in cm$^{-2}$.

Next, we calculated the total column density ($N$), using:

$$N = \left(\frac{N_{\mathrm{u}} Q(T)}{g_{\mathrm{u}}}\right) \exp \frac{E_{\mathrm{u}}}{k T_{\mathrm{ex}}} \tag{4}$$

$Q(T)$ is defined as:

$$Q(T) = \frac{2kT_{\mathrm{ex}}}{h\nu} \tag{5}$$

In equation (4), we have $N_{\mathrm{u}}$ from earlier and have assumed that $T = T_{\mathrm{ex}}$ (the excitation temperature, K). $g_{\mathrm{u}} = 2J + 1$ (for linear molecules), where $J$ is the upper level quantum number. The units for $k$ are cm$^{-1}$ K$^{-1}$ (which is still the Boltzmann constant, but in units that differ to those used when calcuating $N_{\mathrm{u}}$) and for $E_{\mathrm{u}}$ (upper energy, cm$^{-1}$). This results in $N$ (cm$^{-2}$). In equation (5), the units for $k$ are found in equation (3). We retrieved our values for $A$ and $E_{\mathrm{u}}$, by matching the transitions and frequencies listed in Table 2, as closely as possible, to those from *Splatalogue*[7], specifically utilising the CDMS (Müller et al. 2005) catalogue. In the instances where a transition has multiple hyperfine components e.g. $HC_3N$ (5–4), $N_2H^+$ (1–0), these components are blended, due to the large velocity widths that are present in the Galactic Centre.

A comprehensive search of the literature yielded possible multiple excitation temperatures for the molecules in our datasets. We made a comparison of these values, allowing us to estimate a general value of $T_{\mathrm{ex}}$ for our molecules. A $T_{\mathrm{ex}}$ of 20 K was calculated for H$^{13}$CN (1–0) in the Galactic Centre region (Lee & Lee 2003). Martín et al. (2008) have used an excitation temperature of 10 K for the 20 km s$^{-1}$ cloud (note that their transition was J = 3–2). Tsuboi et al. (2011) have used Large Velocity Gradient (LVG) calculations and determined that H$^{13}$CO$^{+}$ (1–0) has $T_{\mathrm{ex}}$ of $\approx$ 5–8 K and SiO (2–1, v = 0) is $\approx$ 4–20 K, in the 20 and 50 km s$^{-1}$ clouds. The HNCO 5(0,5)–4(0,4) excitation temperature for the 20 km s$^{-1}$ cloud was found to be varying between 9–16 K (Amo-Baladrón, Martín-Pintado, & Martín 2011). Jones et al. (2013) have found that while some molecules such as SiO and $^{13}$CS are sub-thermal, strong detections in molecules such as HNCO and $HC_3N$ have a higher $T_{\mathrm{ex}}$ value of $\approx$ 10 K. We note that Jones et al. (2013), have also found that when comparing 3 and 7-mm molecular lines, sub-thermal excitation ($T_{\mathrm{ex}}$ < $T_{\mathrm{kin}}$) was determined for the molecules we considered here; and a low excitation temperature was also confirmed with the non-LTE model RADEX[8]. Other molecules in higher energy states, are likely to probe different properties of the gas; however, this is not an inconsistency in choosing $T_{ex}$. Changing the excitation temperature between values of 5–25 K, will increase the column densities by factors that range from ≈ 1.7 to 3.6, for nearly all the molecules in Table 5, where the only exceptions are $HC_3N$ (10–9) and (12-11). Using the same excitation temperature range for $HC_3N$ (10–9) and (12-11), will decrease the column densities by factors of ≈ 9.3 and 46.5 respectively, due to their high $J$ and $E_u$ values. However, we note that Jones et al. 2013 calculated the excitation temperature of $HC_3N$ in Sgr A to be $9.4^{+1.4}_{-1.1}$ K, meaning that the excitation temperature range of 5–25 K is not valid for this molecule. Taking these results into consideration and noting the range of values for $T_{ex}$ in the literature, we have assumed that an average value of $T_{ex}$ = 10 K is suitable for our calculations, when we account for the fact that our integrated emission values from the 20 km $s^{-1}$ cloud, for all molecules, are also averaged over a region. This was subsequently applied in equations (4) and (5).

[7] http://www.cv.nrao.edu/php/splat/

**Table 4.** Optical depth of selected molecules in region 1 (20 km $s^{-1}$ cloud). The following Galactocentric ratios were used to obtain our value of $\tau$, as explained in § 3.1: $^{12}C/^{13}C$ = 24 (Langer & Penzias 1990) and $^{12}CO/C^{18}O$ = 250 (Penzias 1981). Our low values of $\tau$ for $C^{18}O$, $^{13}CO$ and $^{13}CS$, arise from the fact that we have averaged over the entirety of region 1; so, these reflect values we expect to find in the outer parts of the cloud, which have a greater projected surface area.

| Molecule | Ratio Molecules | Average $T_{xb}$ ratio (this work) | $\tau$ (this work) |
|---|---|---|---|
| $H^{13}CCCN$ (5–4) | $HC_3N$ (5–4) / $H^{13}CCCN$ (5–4) | 18.5 | 0.02 |
| $HC_3N$ (5–4) | $HC_3N$ (5–4) / $H^{13}CCCN$ (5–4) | 18.5 | 0.57 |
| $^{13}CS$ (1–0) | CS (1–0) / $^{13}CS$ (1–0) | 13.1 | 0.06 |
| CS (1–0) | CS (1–0) / $^{13}CS$ (1–0) | 13.1 | 1.44 |
| $H^{13}CN$ (1–0) | HCN (1–0) / $H^{13}CN$ (1–0) | 4.3 | 0.26 |
| $H^{13}CO^+$ (1–0) | $HCO^+$ (1–0) / $H^{13}CO^+$ (1–0) | 7.1 | 0.15 |
| HCN (1–0) | HCN (1–0) / $H^{13}CN$ (1–0) | 4.3 | 6.30 |
| $HCO^+$ (1–0) | $HCO^+$ (1–0) / $H^{13}CO^+$ (1–0) | 7.1 | 3.53 |
| $C^{18}O$ (1–0) | $^{12}CO$ (1–0) / $C^{18}O$ (1–0) | 45.1 | 0.02 |
| $^{13}CO$ (1–0) | $^{12}CO$ (1–0) / $^{13}CO$ (1–0) | 4.8 | 0.23 |
| $^{12}CO$ (1–0) | $^{12}CO$ (1–0) / $^{13}CO$ (1–0) | 4.8 | 5.56 |

Consequently, we also derived $H_2$ conversion factors, which are commonly defined as:

$$X_R = \frac{N(\text{molecule})}{N(H_2)} \quad (6)$$

The $H_2$ column density for the 20 km $s^{-1}$ cloud was established using $^{13}CO$ (1–0), as it has a high S/N and is optically thin. Using $N(^{13}CO)$ from Table 5, in combination with the Galactocentric $^{12}C/^{13}C$ ratio of 24 (Langer & Penzias 1990) and $CO/H_2 = 10^{-4}$ (Frerking, Langer, & Wilson 1982), we found $N(H_2) \approx 5.4 \times 10^{22}$ $cm^{-2}$. In contrast, pointed observations with the Mopra telescope, towards the 20 km $s^{-1}$ cloud found $N(H_2)$ using $C^{18}O$ to be 2.4 (0.2) $\times 10^{22}$ $cm^{-2}$ (Armijos-Abendaño et al. 2015). They also calculated a lower $N$ for $C^{18}O$ compared to our work. Note that our values are for the entirety of region 1, depicted in Fig. 2(a), compared to their pointed observations. The difference in $N(H_2)$ between our results, can be explained, when you consider that we used a different set of assumptions to that found in Armijos-Abendaño et al. 2015, which includes the galactocentric ratio conversion factors, optical depth calculations and errors from telescope calibration.

[8] http://var.sron.nl/radex/radex.php

**Table 5.** Physical properties of the 20 km $s^{-1}$ cloud. The derivation of values for $\tau$, $N_u$, $N$ and $X_R$ were explained in Table 4 and § 3.1 respectively. HNCO (Martín et al. 2008) and SiO (Amo-Baladrón, Martín-Pintado, & Martín 2011) were shown to be optically thin and we have assumed $\tau \ll 1$ for $C^{34}S$ (1–0), $HC_3N$ (10–9), $^{13}CS$ (2–1), $N_2H^+$ (1–0) and $HC_3N$ (12–11). $X_R$ has been determined by $N$(molecule) / $N(H_2)$, where $N(H_2)$ has been calculated from $^{13}CO$ (1–0). The spectra used for these calculations have a S/N ⩾ 3. The largest value of uncertainty arises from the calibration uncertainty in the Mopra telescope (≈ 10 %), the baseline ripples and the excitation temperature approximation.

| Molecule | $\tau$ | $N_u$ ($cm^{-2}$) $\times 10^{14}$ | $N$ ($cm^{-2}$) $\times 10^{14}$ | $X_R$ $\times 10^{-9}$ |
|---|---|---|---|---|
| SiO (1–0) | ≪ 1 | 0.32 | 1.27 | 2.36 |
| $H^{13}CCCN$ (5–4) | 0.02 | 0.01 | 0.02 | 0.04 |
| $HC_3N$ (5–4) | 0.57 | 0.32 | 0.51 | 0.95 |
| $^{13}CS$ (1–0) | 0.06 | 0.24 | 0.92 | 1.70 |
| $C^{34}S$ (1–0) | ≪ 1 | 0.45 | 1.81 | 3.37 |
| CS (1–0) | 1.4 | 6.48 | 23.26 | 43.12 |
| $H^{13}CN$ (1–0) | 0.26 | 0.21 | 0.51 | 0.95 |
| $H^{13}CO^+$ (1–0) | 0.15 | 0.03 | 0.08 | 0.15 |
| SiO (2–1) | ≪ 1 | 0.10 | 0.18 | 0.34 |
| HCN (1–0) | 6.3 | 6.91 | 16.58 | 30.74 |
| $HCO^+$ (1–0) | 3.5 | 1.30 | 3.11 | 5.78 |
| $HC_3N$ (10–9) | ≪ 1 | 0.07 | 0.17 | 0.32 |
| $^{13}CS$ (2–1) | ≪ 1 | 0.10 | 0.17 | 0.31 |
| $N_2H^+$ (1–0) | ≪ 1 | 0.25 | 0.59 | 1.10 |
| $HC_3N$ (12–11) | ≪ 1 | 0.05 | 0.22 | 0.40 |
| $C^{18}O$ (1–0) | 0.02 | 76 | 163 | 302 |
| $^{13}CO$ (1–0) | 0.23 | 1051 | 2247 | 4167 |
| $^{12}CO$ (1–0) | 5.6 | 28473 | 59662 | 110614 |

### 3.2 Integrated emission comparison

Amo-Baladrón, Martín-Pintado, & Martín (2011) mapped the central 12 parsecs of the Galactic Centre and determined the integrated emission for their molecules from positions of local maxima in SiO (2–1) emission. A comparison of the integrated emission between their '20 GMC' and '50 GMC' positions (which correspond to the 20 and 50 km $s^{-1}$ clouds respectively), show that SiO (2–1), $H^{13}CO^+$ (1–0), HNCO 5(0,5)–4(0,4) are brighter in the 20 km $s^{-1}$ cloud whereas CS (1–0) and $C^{18}O$ (1–0) is brighter in the 50 km

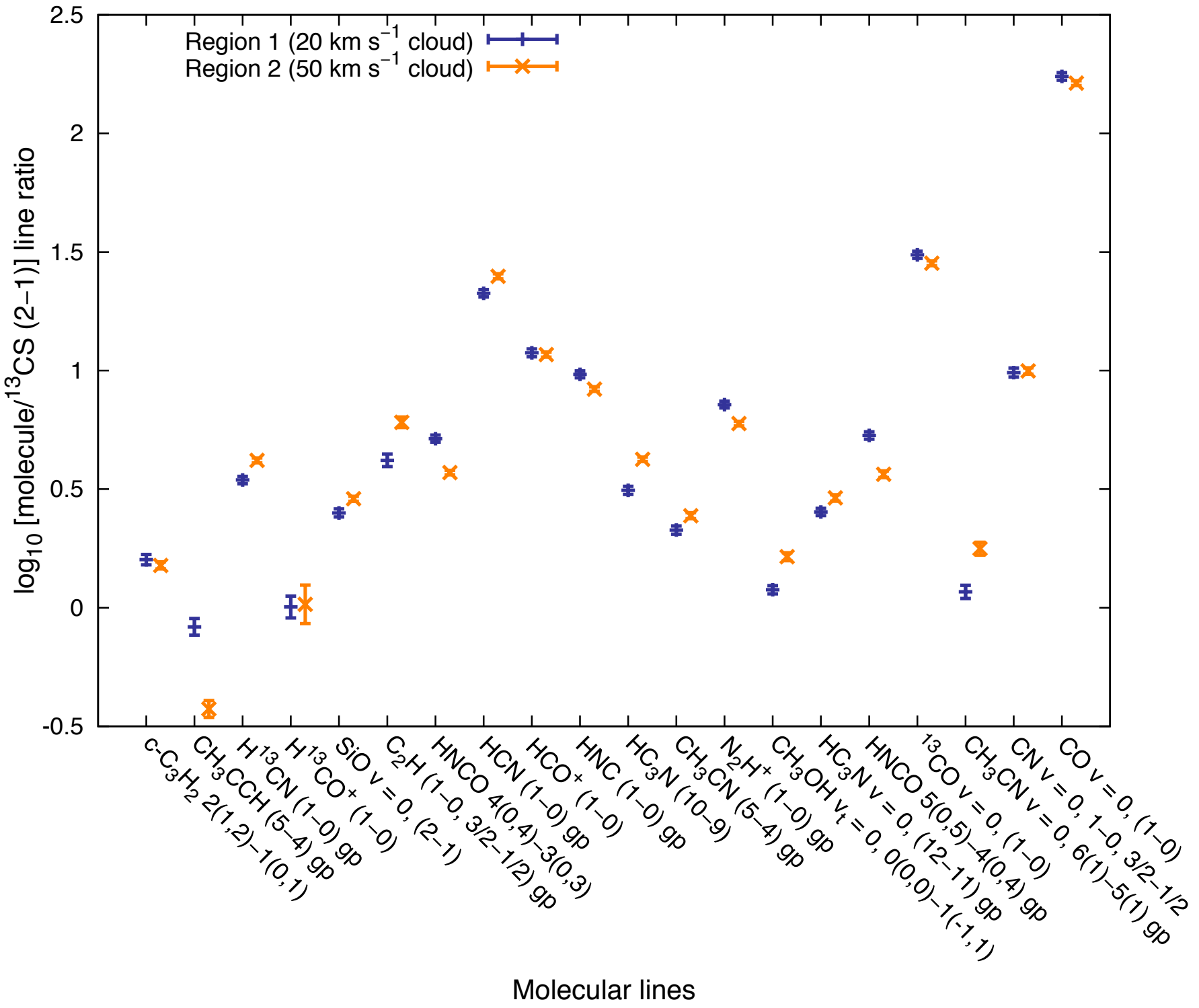

(a)

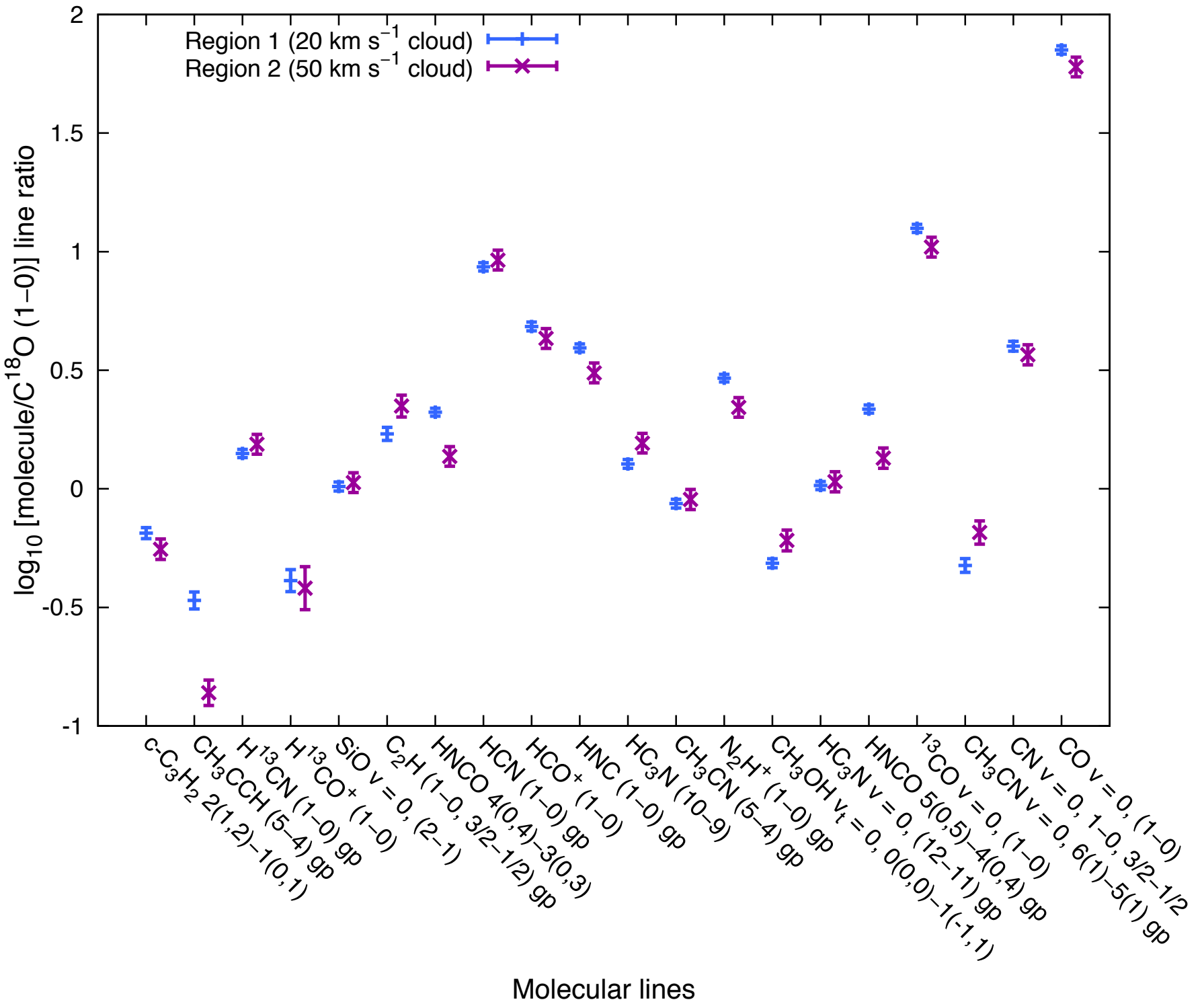

(b)

**Figure 3.** 3(a) 3-mm $\log_{10}$ $^{13}$CS (2–1) line ratio plot and 3(b) 3-mm $\log_{10}$ C$^{18}$O (1–0) line ratio plot. The pairs of points in the plots, for each molecular transition, have been taken from Table 6. In regions 1 and 2, we averaged over 612 and 180 pixels respectively. In addition, the molecules have been sorted from left to right, in ascending frequency. All the molecules here have a S/N $\geqslant$ 3.

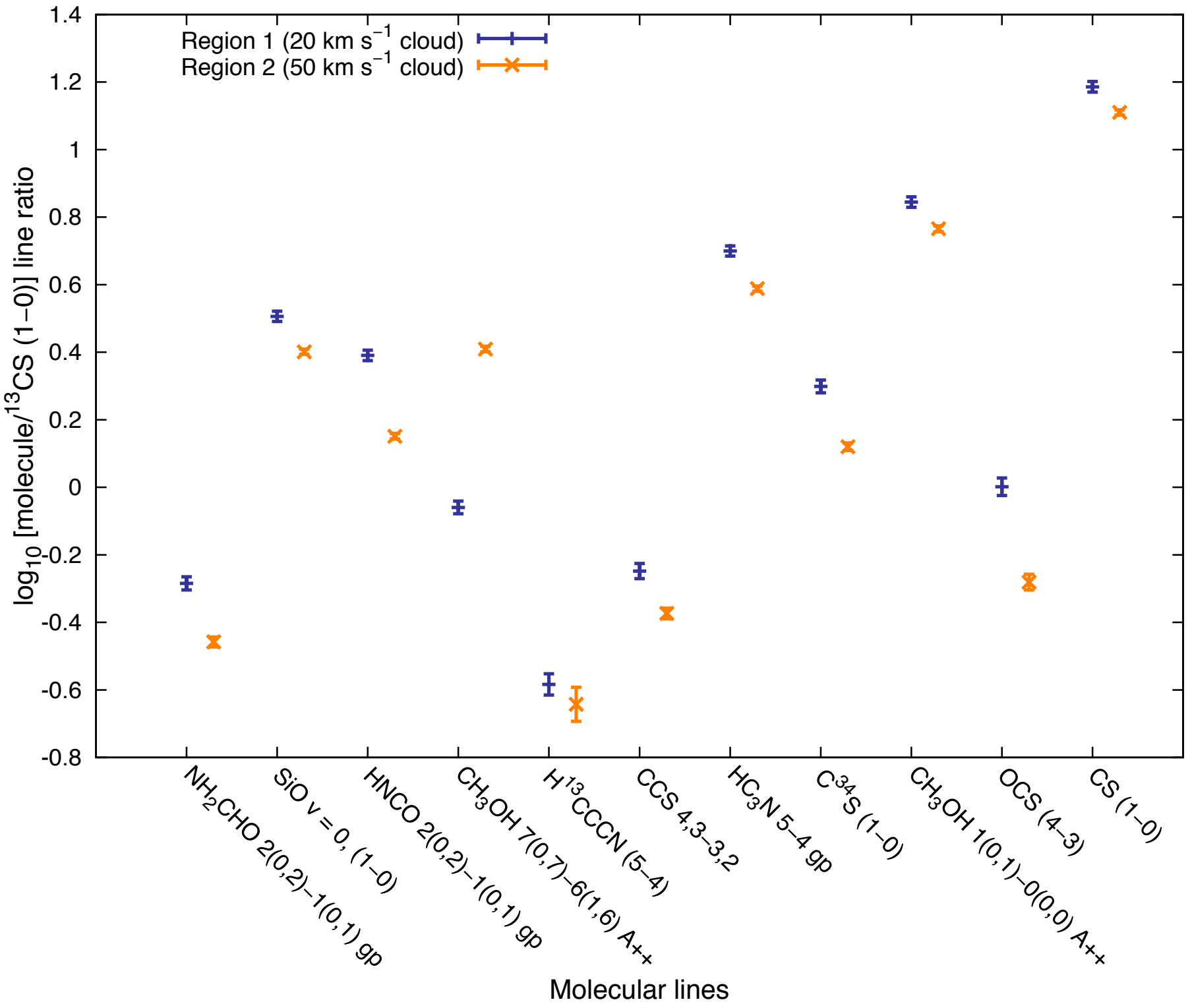

**Figure 4.** 7-mm log$_{10}$ $^{13}$CS (1–0) line ratio plot. Similarly to Fig. 3(a) and 3(b), we have sorted the data points in the plots, from left to right, in ascending frequency; and have averaged over the same number of pixels from regions 1 and 2 (612 and 180 pixels respectively). The data for these plots was sourced from Table 7. All the molecules here have a S/N $\geqslant$ 3.

**Table 6.** 3-mm line statistics of regions 1 and 2, the 20 and 50 km s$^{-1}$ clouds respectively: Integrated emissions, $^{13}$CS (2–1) and C$^{18}$O (1–0) molecular line ratios; see equation (1).

| Molecule | Region 1 $\int T_A$ dV (K km s$^{-1}$) | Region 2 $\int T_A$ dV (K km s$^{-1}$) | Region 1 log$_{10}$ $^{13}$CS line ratio | Region 2 log$_{10}$ $^{13}$CS line ratio | Region 1 log$_{10}$ C$^{18}$O line ratio | Region 2 log$_{10}$ C$^{18}$O line ratio |
|---|---|---|---|---|---|---|
| c-C$_3$H$_2$ 2(1,2)–1(0,1) | 13.1 ± 0.5 | 16.2 ± 0.5 | 0.203 ± 0.022 | 0.178 ± 0.015 | -0.187 ± 0.023 | -0.255 ± 0.043 |
| CH$_3$CCH (5–4) gp | 6.8 ± 0.5 | 4.0 ± 0.3 | -0.081 ± 0.035 | -0.427 ± 0.035 | -0.471 ± 0.036 | -0.860 ± 0.054 |
| H$^{13}$CN (1–0) gp | 28.3 ± 0.5 | 45.0 ± 0.5 | 0.539 ± 0.016 | 0.621 ± 0.010 | 0.149 ± 0.017 | 0.188 ± 0.042 |
| H$^{13}$CO$^+$ (1–0) | 8.2 ± 0.8 | 11.1 ± 2.1 | 0.003 ± 0.046 | 0.014 ± 0.081 | -0.387 ± 0.046 | -0.419 ± 0.091 |
| SiO v = 0, (2–1) | 20.5 ± 0.5 | 31.0 ± 0.4 | 0.400 ± 0.017 | 0.459 ± 0.010 | 0.010 ± 0.019 | 0.026 ± 0.042 |
| C$_2$H (1–0, 3/2–1/2) gp | 34.2 ± 1.8 | 65.2 ± 3.1 | 0.622 ± 0.026 | 0.782 ± 0.022 | 0.232 ± 0.027 | 0.349 ± 0.046 |
| HNCO 4(0,4)–3(0,3) | 42.2 ± 0.5 | 40.0 ± 0.3 | 0.713 ± 0.015 | 0.570 ± 0.009 | 0.323 ± 0.017 | 0.137 ± 0.041 |
| HCN (1–0) gp | 173.2 ± 3.0 | 269.2 ± 3.9 | 1.326 ± 0.016 | 1.398 ± 0.011 | 0.936 ± 0.017 | 0.964 ± 0.042 |
| HCO$^+$ (1–0) | 97.1 ± 2.0 | 125.8 ± 1.8 | 1.075 ± 0.016 | 1.067 ± 0.011 | 0.685 ± 0.018 | 0.634 ± 0.042 |
| HNC (1–0) gp | 78.8 ± 0.8 | 90.0 ± 0.9 | 0.984 ± 0.015 | 0.922 ± 0.010 | 0.594 ± 0.016 | 0.489 ± 0.042 |
| HC$_3$N (10–9) | 25.6 ± 0.6 | 45.5 ± 0.3 | 0.495 ± 0.017 | 0.626 ± 0.009 | 0.105 ± 0.019 | 0.192 ± 0.041 |
| CH$_3$CN (5–4) gp | 17.4 ± 0.4 | 26.3 ± 0.5 | 0.328 ± 0.017 | 0.388 ± 0.013 | -0.062 ± 0.019 | -0.045 ± 0.042 |
| $^{13}$CS (2–1) | 8.2 ± 0.3 | 10.8 ± 0.2 | ... | ... | ... | ... |
| N$_2$H$^+$ (1–0) gp | 58.7 ± 0.4 | 64.4 ± 0.4 | 0.856 ± 0.014 | 0.777 ± 0.009 | 0.466 ± 0.016 | 0.343 ± 0.041 |
| CH$_3$OH v$_t$ = 0, 0(0,0)–1(-1,1) | 9.7 ± 0.2 | 17.7 ± 0.6 | 0.076 ± 0.017 | 0.215 ± 0.017 | -0.314 ± 0.019 | -0.218 ± 0.044 |
| HC$_3$N v = 0, (12–11) gp | 20.7 ± 0.3 | 31.3 ± 0.7 | 0.404 ± 0.015 | 0.463 ± 0.013 | 0.014 ± 0.017 | 0.030 ± 0.043 |
| C$^{18}$O (1–0) | 20.1 ± 0.7 | 29.2 ± 2.8 | ... | ... | ... | ... |
| HNCO 5(0,5)–4(0,4) gp | 43.5 ± 0.7 | 39.4 ± 1.0 | 0.726 ± 0.016 | 0.563 ± 0.014 | 0.336 ± 0.017 | 0.130 ± 0.043 |
| $^{13}$CO v = 0, (1–0) | 251.6 ± 3.8 | 305.3 ± 3.5 | 1.488 ± 0.015 | 1.452 ± 0.010 | 1.098 ± 0.017 | 1.019 ± 0.042 |
| CH$_3$CN v = 0, 6(1)–5(1) gp | 9.5 ± 0.5 | 19.1 ± 1.1 | 0.067 ± 0.028 | 0.249 ± 0.027 | -0.323 ± 0.029 | -0.184 ± 0.049 |
| CN v = 0, 1–0, 1/2–1/2, F=3/2–1/2 | ... | ... | ... | ... | ... | ... |
| CN v = 0, 1–0, 3/2–1/2, F=5/2–3/2 | 80.1 ± 2.5 | 107.3 ± 2.5 | 0.991 ± 0.019 | 0.998 ± 0.013 | 0.601 ± 0.021 | 0.565 ± 0.043 |
| CO v = 0, (1–0) | 1422.5 ± 26.1 | 1755.2 ± 18.2 | 2.241 ± 0.016 | 2.212 ± 0.010 | 1.851 ± 0.018 | 1.779 ± 0.042 |

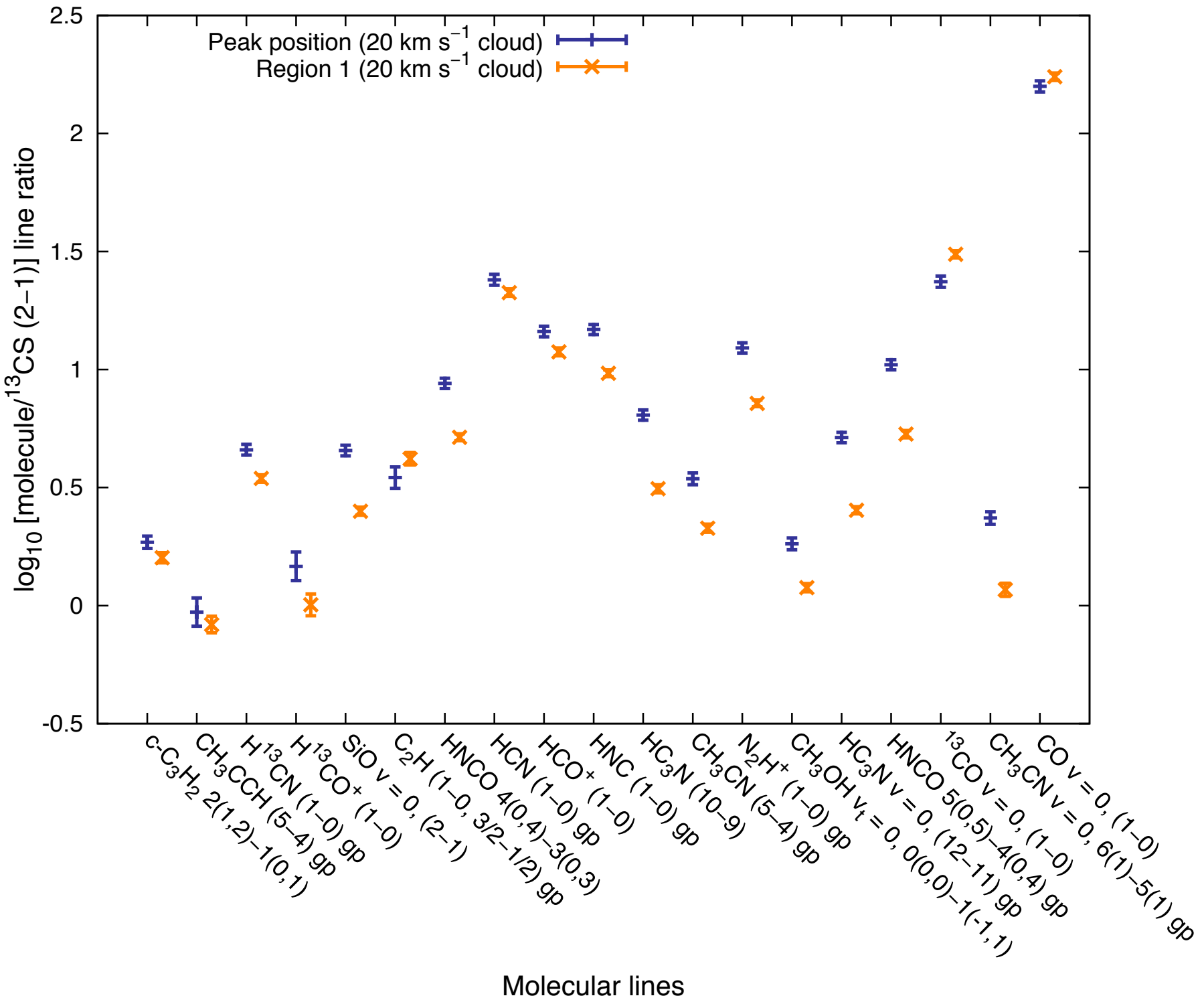

(a)

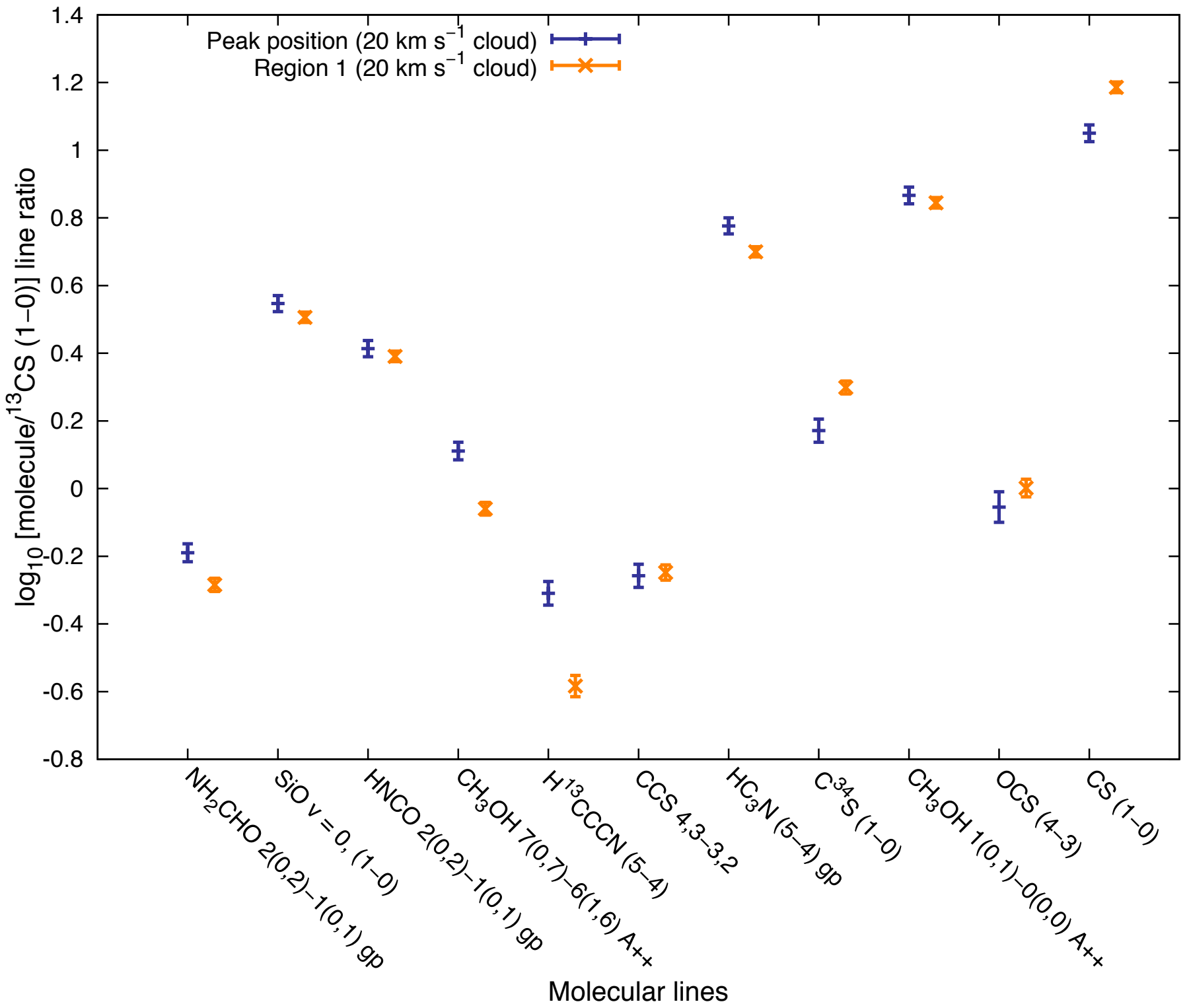

(b)

**Figure 5.** 5(a) 3-mm log$_{10}$ $^{13}$CS (2–1) line ratio plot and 5(b) 7-mm log$_{10}$ $^{13}$CS (1–0) line ratio plot. Both plots compare the peak position of the 20 km s$^{-1}$ cloud (region 'a') with respect to region 1 i.e. the entire 20 km s$^{-1}$ cloud. The pairs of points in the plots, for each molecular transition, have been taken from Tables 6, 7, 9 and 10. In regions 'a' and '1', we have averaged over 12 and 612 pixels respectively. In addition, the molecules have been sorted from left to right, in ascending frequency.

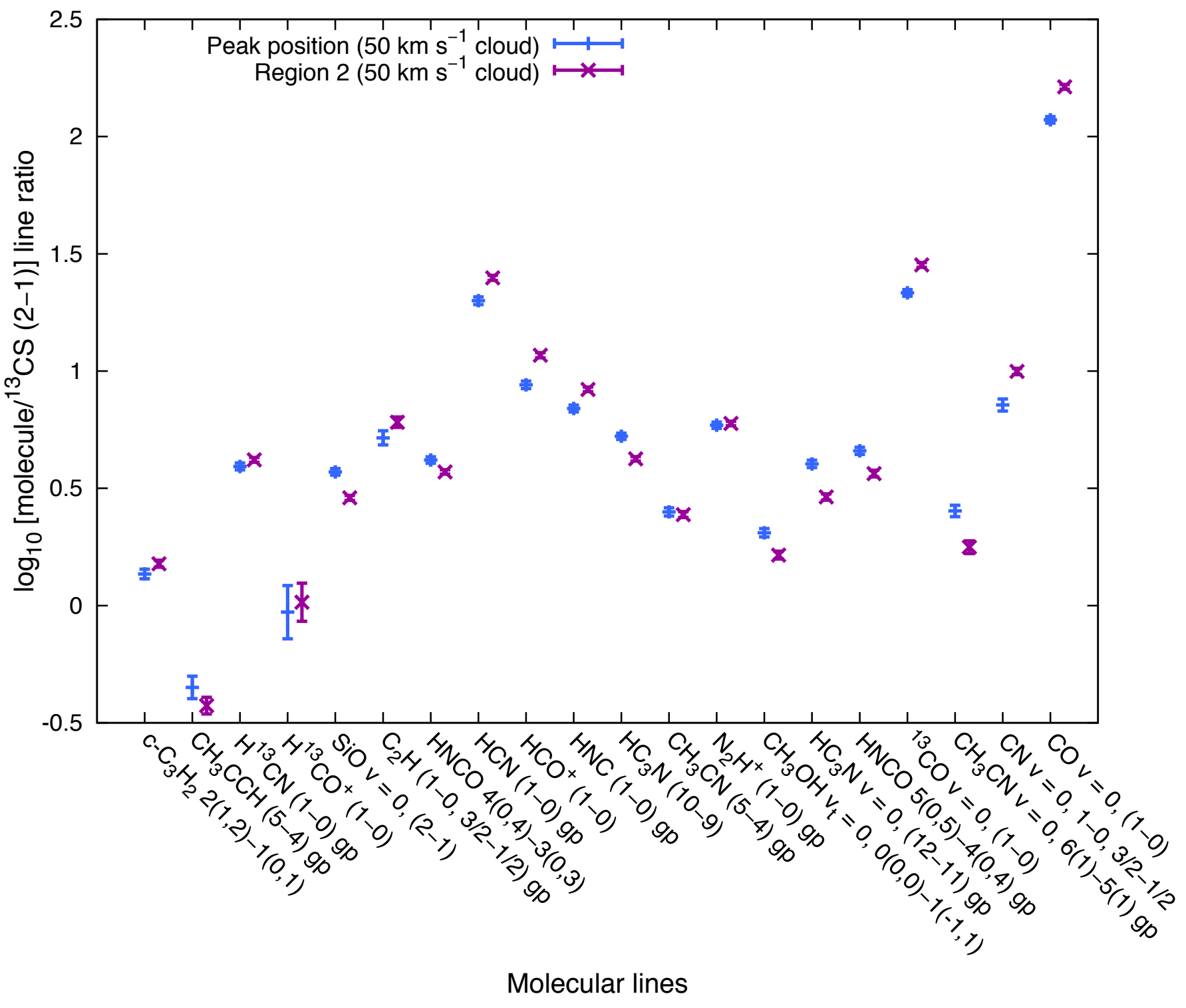

(a)

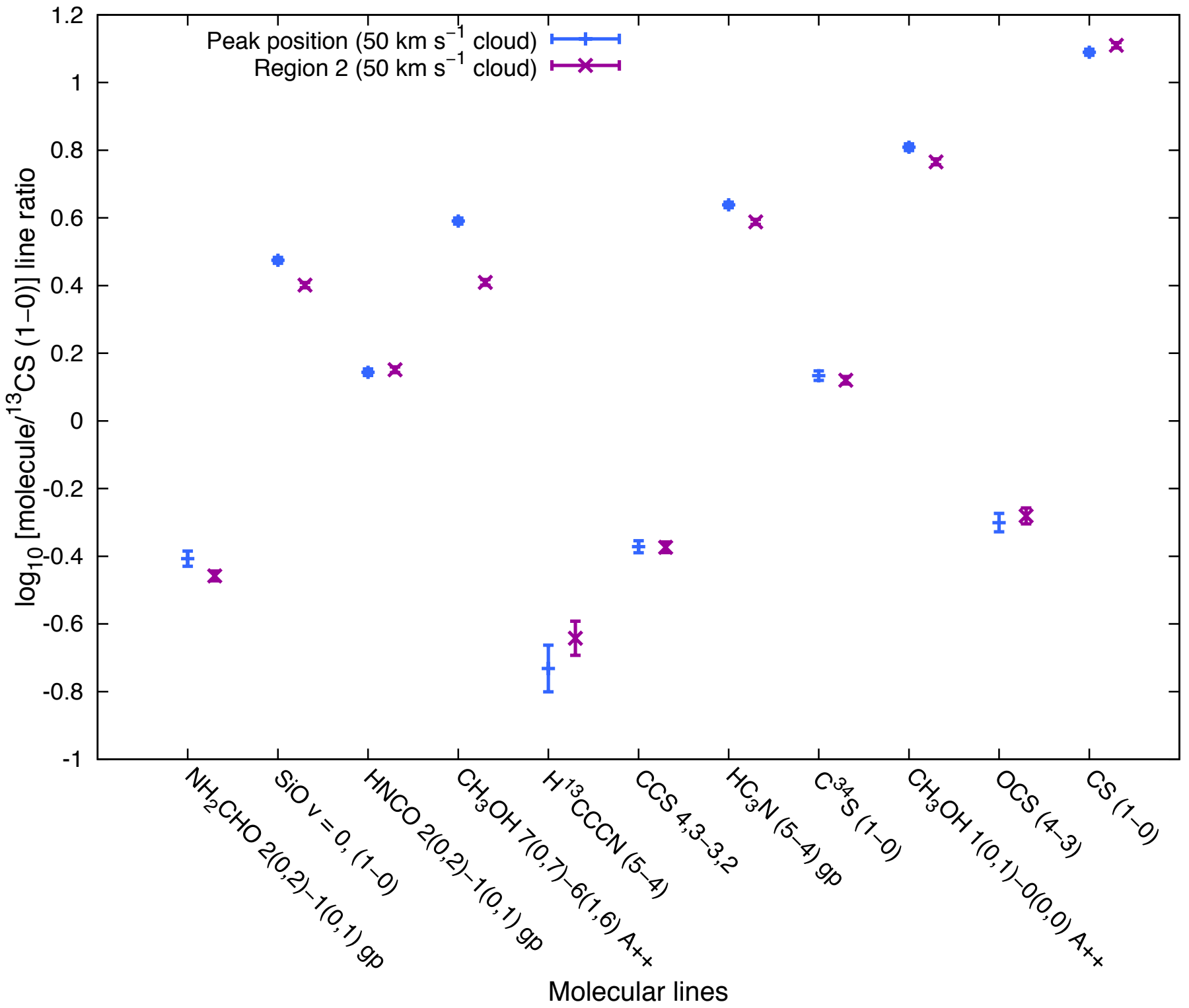

(b)

**Figure 6.** 6(a) 3-mm log$_{10}$ $^{13}$CS (2–1) line ratio plot and 6(b) 7-mm log$_{10}$ $^{13}$CS (1–0) line ratio plot. Both plots compare the peak position of the 50 km s$^{-1}$ cloud (region 'c') with respect to region 2 i.e. the entire 50 km s$^{-1}$ cloud. The pairs of points in the plots, for each molecular transition, have been taken from Tables 6, 7, 12 and 13. In regions 'c' and '2', we have averaged over 12 and 180 pixels respectively. In addition, the molecules have been sorted from left to right, in ascending frequency.

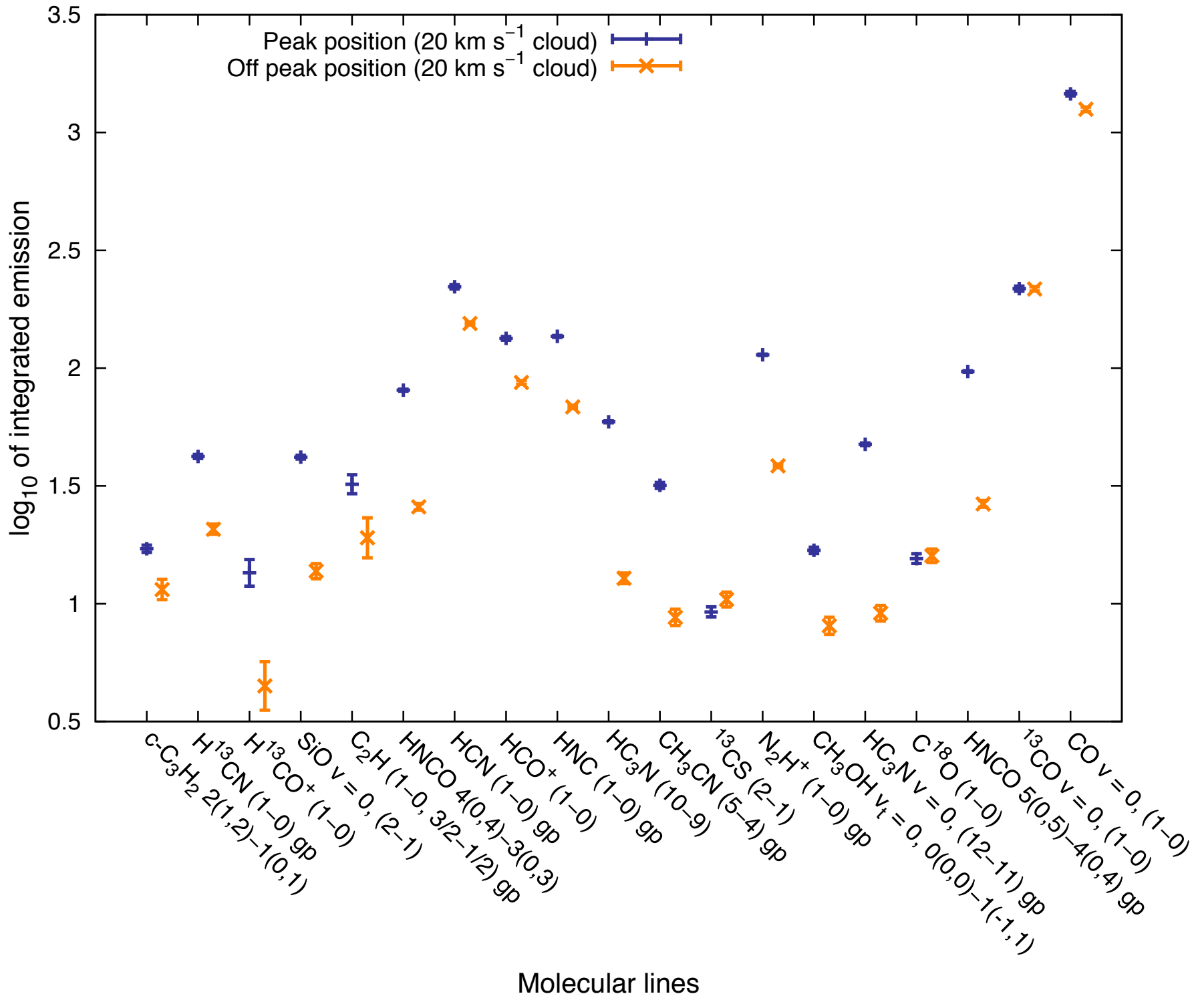

(a)

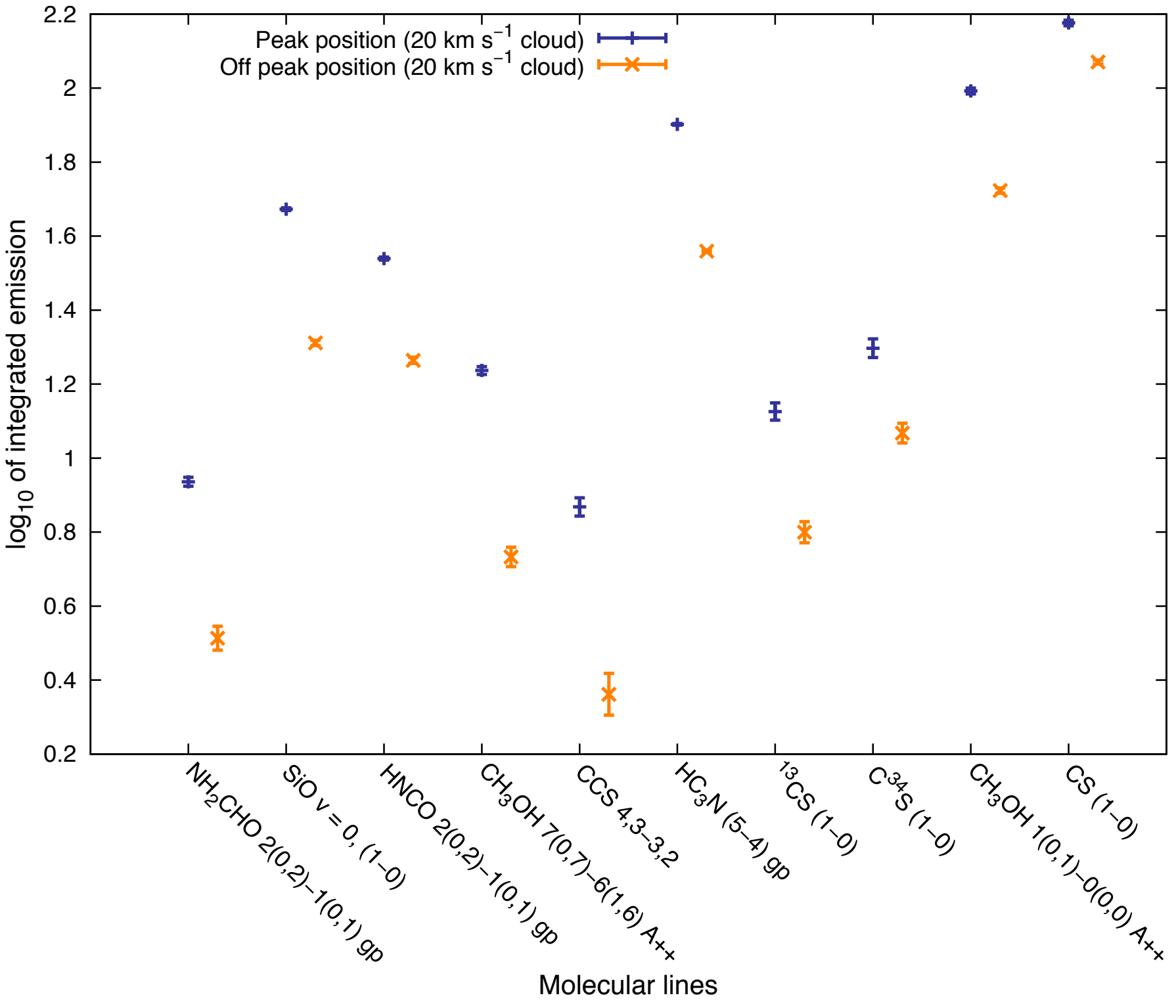

(b)

**Figure 7.** 7(a) 3-mm log$_{10}$ integrated emission of molecular lines plot and 7(b) 7-mm log$_{10}$ integrated emission of molecular lines plot. Both plots compare the peak (region 'a') and off peak (region 'b') positions within the 20 km s$^{-1}$ cloud. The pairs of points in the plots, for each molecular transition, have been taken from Tables 9 and 10. In regions 'a' and 'b', we have averaged over 12 and 9 pixels respectively. In addition, the molecules have been sorted from left to right, in ascending frequency.

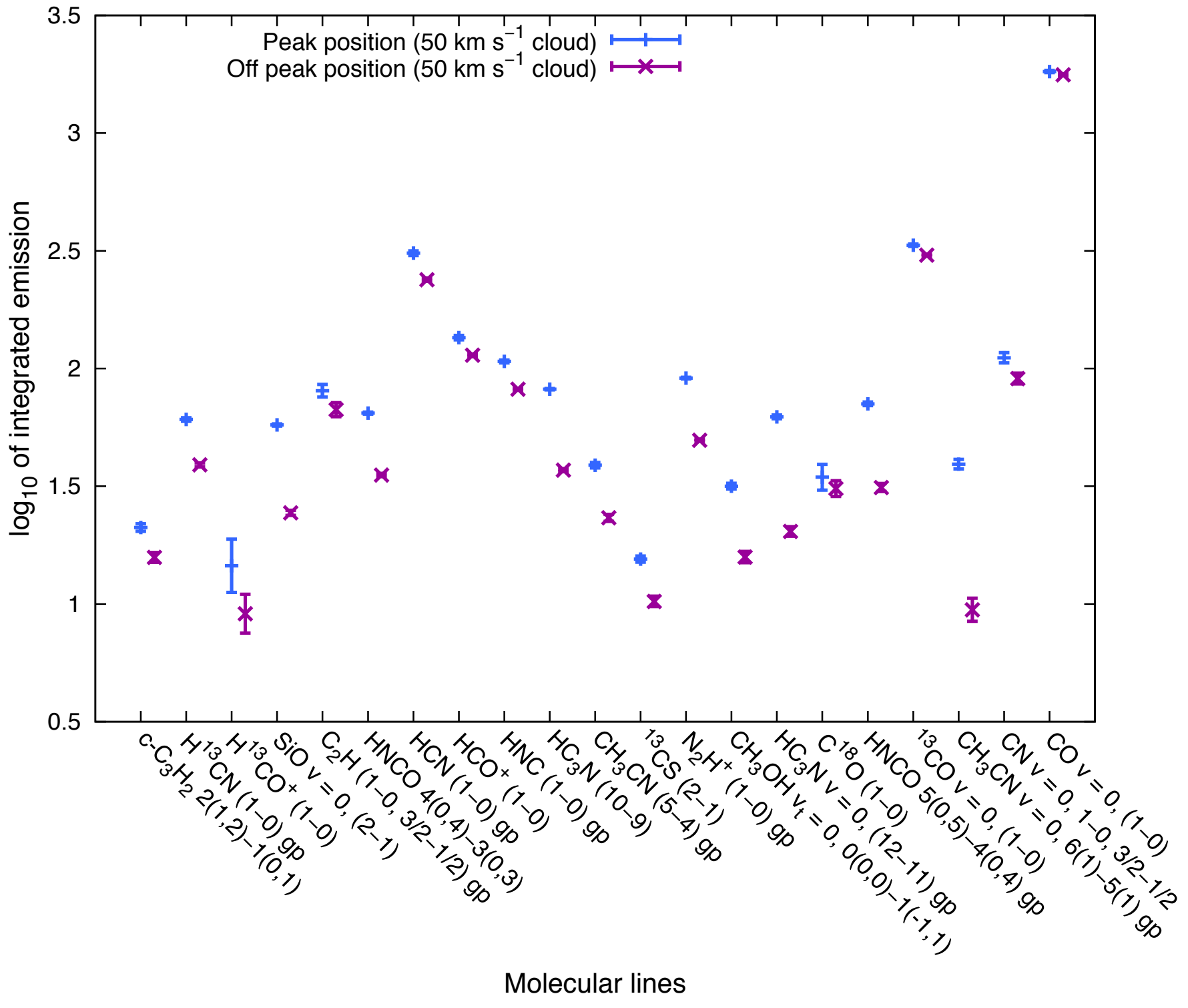

(a)

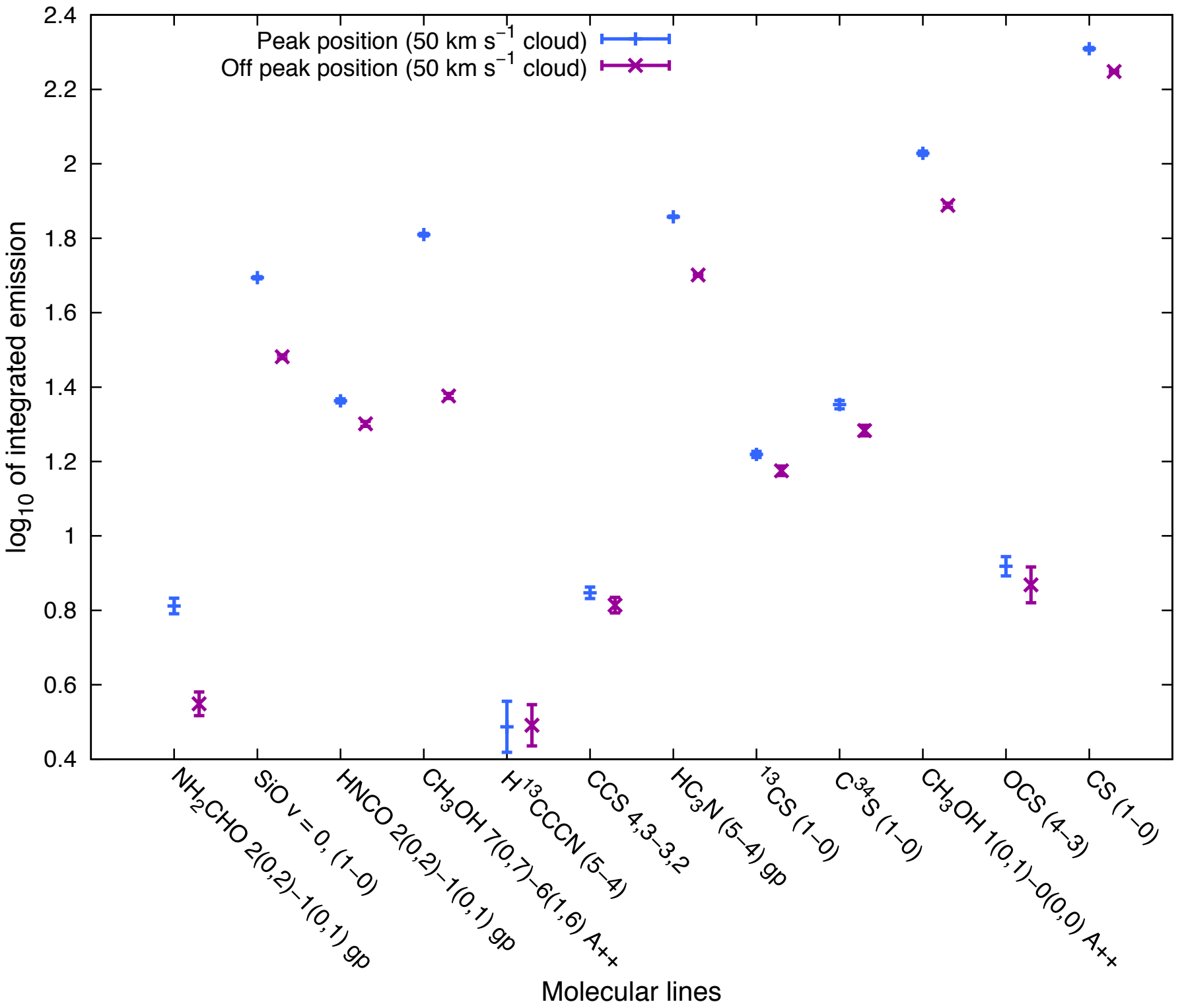

(b)

**Figure 8.** 8(a) 3-mm log$_{10}$ integrated emission of molecular lines plot and 8(b) 7-mm log$_{10}$ integrated emission of molecular lines plot. Both plots compare the peak (region 'c') and off peak (region 'd') positions within the 50 km s$^{-1}$ cloud. The pairs of points in the plots, for each molecular transition, have been taken from Tables 12 and 13. In regions 'c' and 'd', we have averaged over 12 and 9 pixels respectively. In addition, the molecules have been sorted from left to right, in ascending frequency.

**Table 7.** 7-mm line statistics of regions 1 and 2, the 20 and 50 km s$^{-1}$ clouds respectively: Integrated emissions, $^{13}$CS (1–0) molecular line ratios; see equation (1).

| Molecule | Region 1 $\int T_A$ dV (K km s$^{-1}$) | Region 2 $\int T_A$ dV (K km s$^{-1}$) | Region 1 $\log_{10}$ $^{13}$CS line ratio | Region 2 $\log_{10}$ $^{13}$CS line ratio |
|---|---|---|---|---|
| $NH_2CHO$ 2(0,2)–1(0,1) gp | 4.4 ± 0.1 | 4.7 ± 0.1 | -0.285 ± 0.019 | -0.458 ± 0.015 |
| SiO v = 0, (1–0) | 26.9 ± 0.2 | 33.8 ± 0.2 | 0.506 ± 0.015 | 0.401 ± 0.008 |
| HNCO 2(0,2)–1(0,1) gp | 20.6 ± 0.2 | 19.0 ± 0.2 | 0.391 ± 0.015 | 0.151 ± 0.008 |
| $CH_3OH$ 7(0,7)–6(1,6) A++ | 7.3 ± 0.2 | 34.5 ± 0.3 | -0.060 ± 0.019 | 0.409 ± 0.008 |
| H$^{13}$CCCN (5–4) | 2.2 ± 0.1 | 3.1 ± 0.4 | -0.584 ± 0.031 | -0.642 ± 0.050 |
| CCS 4,3–3,2 | 4.7 ± 0.2 | 5.7 ± 0.2 | -0.248 ± 0.022 | -0.374 ± 0.016 |
| $HC_3N$ 5–4 gp | 42.0 ± 0.3 | 52.0 ± 0.3 | 0.700 ± 0.015 | 0.588 ± 0.008 |
| $^{13}$CS(1–0) | 8.4 ± 0.3 | 13.4 ± 0.2 | ... | ... |
| C$^{34}$S (1–0) | 16.7 ± 0.5 | 17.7 ± 0.3 | 0.299 ± 0.019 | 0.120 ± 0.011 |
| $CH_3OH$ 1(0,1)–0(0,0) A++ | 58.6 ± 0.7 | 78.3 ± 0.9 | 0.845 ± 0.016 | 0.766 ± 0.009 |
| OCS (4–3) | 8.4 ± 0.4 | 7.0 ± 0.4 | 0.002 ± 0.026 | -0.281 ± 0.023 |
| CS (1–0) | 128.5 ± 1.7 | 173.1 ± 1.2 | 1.186 ± 0.016 | 1.110 ± 0.008 |

s$^{-1}$ cloud. Our results from regions 1 and 2, show that while CS (1–0) and C$^{18}$O (1–0) are brighter in the 50 km s$^{-1}$ cloud and in agreement with the findings of Amo-Baladrón, Martín-Pintado, & Martín (2011); SiO (2–1) is brighter in the 50 km s$^{-1}$ cloud, HNCO 5(0,5)–4(0,4) is brighter in the 20 km s$^{-1}$ cloud and H$^{13}$CO$^+$ (1–0) is consistent between both clouds. Our differences can be explained, by considering that we integrated over a significantly larger region as opposed to their 30 × 30 arcsec boxes from their cloud positions. To further analyse the differences and similarities between regions within and encompassing the 20 and 50 km s$^{-1}$ clouds, relative line abundance ratios were calculated.

### 3.3 Line Ratio Analysis

There are 21 pairs of molecular transition ratios in the 3-mm dataset and 11 pairs of transition ratios in the 7-mm dataset, for both the 20 and 50 km s$^{-1}$ clouds, that satisfy the conditions described in § 2.4. We have also shown that 3-mm C$^{18}$O (1–0) and $^{13}$CS (2–1) are optically thin and assumed this to be the case for 7-mm $^{13}$CS (1–0) as well (see Table 5). A comparison of results between both the 20 and 50 km s$^{-1}$ clouds; as well as regions within those two clouds is established in Tables 8–16 and discussed in this section.

#### 3.3.1 3-mm $\log_{10}$ $^{13}$CS (2–1) and $\log_{10}$ C$^{18}$O (1–0) line ratios

In Fig. 3(a), the $\log_{10}$ $^{13}$CS (2–1) plot, 7 of the molecular transitions are brighter in region 1 compared to region 2; 9 molecular transitions are brighter in region 2 vs. region 1; and 4 transitions show relative abundances in both regions 1 and 2 that are consistent within uncertainties.

Fig. 3(b), the $\log_{10}$ C$^{18}$O (1–0) plot, has yielded effectively the same structure to that found in Fig. 3(a), which implies a constant $^{13}$CS/C$^{18}$O ratio. 8 molecular transitions are brighter in region 1 (where 7 of them are the same as those from the $\log_{10}$ $^{13}$CS plot). Only 4 molecular transitions are brighter in region 2 vs. region 1 (where those exact molecular transitions are also brighter in the $\log_{10}$ $^{13}$CS plot). In contrast, we find that there are 8 molecular transitions which are consistent between both regions, only overlapping with 3 of the molecular transitions from Fig. 3(a). The list of transitions exhibiting those brightness relationships is shown in Table 8.

We note that the HNCO molecular transitions are considerably brighter within region 1 in comparison to 2, for both 3-mm line ratio plots. HNCO is a well known tracer of shocked gas in the Galactic Centre e.g. Martín et al. (2008). Zinchenko, Henkel, & Mao (2000) find evidence to show that there is a correlation between HNCO and SiO, in which they infer that HNCO is potentially a shock tracer and further verification for this is found in Rodríguez-Fernández et al. (2010). Previous observations of HNCO in the Galactic Centre (in Sgr B2), also predict this (Minh & Irvine 2006; Martín et al. 2008). It is therefore unusual to report that in the $\log_{10}$ $^{13}$CS plot, the SiO molecule is brighter in region 1 (the difference between them is small); whereas in the $\log_{10}$ C$^{18}$O plot, they are consistent with one another. Further investigation into the relationship between SiO and HNCO is required.

In addition, molecules that trace high density gas appear to be more prominent in both regions 1 and 2, whereby we notice both differences and similarities between each region. HNCO (Armstrong & Barrett 1985), $CH_3CCH$ (Bergin et al. 1994) and $N_2H^+$(Daniel, Cernicharo, & Dubernet 2006 and references therein) are all high density tracers and much brighter in region 1. However, $HC_3N$ (Li et al. 2012), is also a high density tracer, but it is brighter in region 2 compared to 1 in the $\log_{10}$ $^{13}$CS and $\log_{10}$ C$^{18}$O line ratio plots for $HC_3N$ (10–9) transition; whereas we find that $HC_3N$ (12–11) is brighter in region 2 in the $\log_{10}$ $^{13}$CS line ratio plot, but consistent within errors in the $\log_{10}$ C$^{18}$O line ratio plot. Similarly, $CH_3CN$ is also a dense gas tracer (Bally et al. 1987), where the 6(1)–5(1) transition is significantly brighter in region 2, in both line ratio plots. In contrast, the 5–4 transition is only slightly brighter in region 2, in the $\log_{10}$ $^{13}$CS line ratio plot and consistent within uncertainties between both regions in the $\log_{10}$ C$^{18}$O line ratio plot. These differences between the results of the $^{13}$CS and C$^{18}$O line ratio plots, can be explained by considering the fact that $^{13}$CS traces denser gas than C$^{18}$O, due to their differing critical densities; which is $3 \times 10^5$ cm$^{-3}$ for $^{13}$CS J = 2–1 (e.g. Sanhueza et al. 2012) and $\approx 2 \times 10^3$ cm$^{-3}$ for C$^{18}$O (1–0) (e.g. Yoshida et al. 2010).

#### 3.3.2 7-mm $\log_{10}$ $^{13}$CS (1–0) ratios

In Fig. 4, almost all of the transitions (9 out of 11) are brighter in region 1 vs. 2, 1 is consistent between both regions and 1 transition is brighter in region 2 as opposed to region 1. The molecular transition which is brighter in region 2 versus region 1 is $CH_3OH$ 7(0,7)–6(1,6) A++ at 44.07 GHz, which we suspect is due to methanol maser emission, as this is a Class I maser transition, which has been observed in Sagittarius A; e.g. Pihlström, Sjouwerman, & Fish (2011). Due to the maser nature of this emission, the line ratios will be due to strongly non-thermal processes and thus not comparable to other line ratios. The relationship between each region for the transitions are also presented in Table 8.

**Table 8.** Comparison of brightness results from the 3 and 7-mm $\log_{10}$ line ratio plots, for molecular transitions between region 1 (20 kms $^{-1}$ cloud) and region 2 (50 km s$^{-1}$ cloud).

| $\log_{10}$ line ratio plots | Region 1 > Region 2 | Region 1 ≈ Region 2 | Region 2 > Region 1 |
|---|---|---|---|
| 3-mm $^{13}$CS (2–1) | $CH_3CCH$ (5–4) gp<br>HNCO 4(0,4)–3(0,3)<br>HNC (1–0) gp<br>$N_2H^+$ (1–0) gp<br>HNCO 5(0,5)–4(0,4) gp<br>$^{13}$CO v = 0, (1–0)<br>CO v = 0, (1–0) | c-$C_3H_2$ 2(1,2)–1(0,1)<br>$H^{13}CO^+$ (1–0)<br>$HCO^+$ (1–0)<br>CN v = 0, 1–0, 3/2–1/2 | $H^{13}CN$ (1–0) gp<br>SiO v = 0, (2–1)<br>$C_2H$ (1–0, 3/2–1/2) gp<br>HCN (1–0) gp<br>$HC_3N$ (10–9)<br>$CH_3CN$ (5–4) gp<br>$CH_3OH$ $v_t$ = 0, 0(0,0)–1(-1,1)<br>$HC_3N$ v= 0, (12–11) gp<br>$CH_3CN$ v = 0, 6(1)–5(1) gp |
| 3-mm $C^{18}O$ (1–0) | $CH_3CCH$ (5–4) gp<br>HNCO 4(0,4)–3(0,3)<br>HNC (1–0) gp<br>$N_2H^+$ (1–0) gp<br>HNCO 5(0,5)–4(0,4) gp<br>c-$C_3H_2$ 2(1,2)–1(0,1)<br>$^{13}$CO v = 0, (1–0)<br>CO v = 0, (1–0) | $H^{13}CN$ (1–0) gp<br>$H^{13}CO^+$ (1–0)<br>SiO v = 0, (2–1)<br>HCN (1–0) gp<br>$HCO^+$ (1–0)<br>$CH_3CN$ (5–4) gp<br>$HC_3N$ v= 0, (12–11) gp<br>CN v = 0, 1–0, 3/2–1/2 | $C_2H$ (1–0, 3/2–1/2) gp<br>$HC_3N$ (10–9)<br>$CH_3OH$ $v_t$ = 0, 0(0,0)–1(-1,1)<br>$CH_3CN$ v = 0, 6(1)–5(1) gp |
| 7-mm $^{13}$CS (1–0) | $NH_2CHO$ 2(0,2)–1(0,1) gp<br>SiO v = 0, (1–0)<br>HNCO 2(0,2)–1(0,1) gp<br>CCS 4,3–3,2<br>$HC_3N$ (5–4) gp<br>$C^{34}S$ (1–0)<br>$CH_3OH$ 1(0,1)–0(0,0) A++<br>OCS (4–3)<br>CS (1–0) | $H^{13}CCCN$ (5–4) | $CH_3OH$ 7(0,7)–6(1,6) A++ |

The 7-mm results also show that the shock (HNCO) and high density tracers (e.g. $HC_3N$) are much brighter in the 20 km s$^{-1}$ cloud than in the 50 km s$^{-1}$ cloud, as discussed in the 3-mm line ratio results. In addition, we note that SiO 1–0 (a well known tracer of shocked gas; e.g. Martín-Pintado et al. 1997) and CS, another high density gas tracer, are both brighter in the 20 kms$^{-1}$ cloud. We have also noticed that $CH_3OH$ 1(0,1)–(0,0) A++ is also brighter in the 20 km s$^{-1}$ cloud than in the 50 kms$^{-1}$ cloud. We can infer from the OCS detections of Goldsmith & Linke (1981), that this molecule is most likely optically thin and traces the denser part of the cloud (i.e. the cores) and $C^{34}S$ is similarly optically thin and also a high density tracer (e.g. Goicoechea et al. 2006).

#### 3.3.3 *3 and 7-mm $\log_{10}$ $^{13}$CS ratios comparing the peak position of the 20 km s$^{-1}$ cloud to the entire cloud itself*

In Fig. 5(a), the 3-mm $^{13}$CS line ratio plot comparing the peak position of the 20 km s$^{-1}$ cloud to the whole cloud itself, we find that from the 19 molecular transitions, 15 are brighter in the peak position, 3 molecules are brighter in region 1 and only 1 transition is consistent within errors between the peak and region 1 ($CH_3CCH$ (5–4) gp). These results are unsurprising and are those which we expect, considering that we have compared the densest part of the 20 km s$^{-1}$ cloud (as described in Table 3), with respect to the entire cloud. Considering this, we would like to point out that the largest differences between the peak position and the entire cloud, again arise from shock and high density tracers such as SiO, HNCO, $HC_3N$ and $N_2H^+$. In the three transitions where region 1 was greater than the peak position ($^{13}$CO, $C_2H$ and CO), there were only very minor differences in the relative abundances between the peak position and the entire cloud; and we note that $C_2H$ and CO are ubiquitous molecules.

The 7-mm $\log_{10}$ $^{13}$CS line ratio plot comparing the same two areas, Fig. 5(b), illustrates that there are 5 molecular transitions which are brighter in the peak position, 2 molecules brighter in region 1 than the peak position and 4 molecules which are equivalent within errors between both regions. There are only minor differences in the SiO relative abundances and HNCO is consistent within uncertainties between both regions, indicating that the gas is likely to be uniformly distributed throughout the cloud, therefore suggesting that the whole cloud is shocked and has high density gas. While $HC_3N$ is brighter in the peak position, CS (a well known high density tracer) and $C^{34}S$ are brighter across the entire cloud, compared to the peak position. Therefore, we notice a similar trend between both the 3 and 7-mm plots. A full comparison of results between regions ‘a’ and ‘1’ is found in Table 11.

**Table 9.** 3-mm line statistics of regions 'a' and 'b', the peak and off peak positions in the 20 km s$^{-1}$ cloud respectively: Integrated emissions, $^{13}$CS (2–1) molecular line ratios; see equation (1). We have also listed the log$_{10}$ integrated emission values of molecules from regions 'a' and 'b'.

| Molecule | Region 'a' $\int T_A$ dV (K km s$^{-1}$) | Region 'b' $\int T_A$ dV (K km s$^{-1}$) | Region 'a' log$_{10}$ $^{13}$CS line ratio | Region 'a' log$_{10}$ of molecule | Region 'b' log$_{10}$ of molecule |
|---|---|---|---|---|---|
| c-$C_3H_2$ 2(1,2)–1(0,1) | 17.1 ± 0.6 | 11.5 ± 1.1 | 0.268 ± 0.026 | 1.233 ± 0.015 | 1.060 ± 0.043 |
| $CH_3CCH$ (5–4) gp | 8.7 ± 1.1 | ... | -0.028 ± 0.060 | 0.937 ± 0.056 | ... |
| $H^{13}CN$ (1–0) gp | 42.2 ± 0.8 | 20.7 ± 1.0 | 0.660 ± 0.023 | 1.625 ± 0.009 | 1.316 ± 0.021 |
| $H^{13}CO^+$ (1–0) | 13.5 ± 1.8 | 4.5 ± 1.1 | 0.166 ± 0.061 | 1.131 ± 0.057 | 0.651 ± 0.103 |
| SiO v = 0, (2–1) | 41.9 ± 0.8 | 13.7 ± 1.0 | 0.657 ± 0.023 | 1.622 ± 0.008 | 1.138 ± 0.032 |
| $C_2H$ (1–0, 3/2–1/2) gp | 32.1 ± 3.0 | 19.1 ± 3.7 | 0.542 ± 0.046 | 1.507 ± 0.040 | 1.280 ± 0.085 |
| HNCO 4(0,4)–3(0,3) | 80.6 ± 0.8 | 25.8 ± 0.9 | 0.941 ± 0.022 | 1.906 ± 0.004 | 1.411 ± 0.014 |
| HCN (1–0) gp | 221.4 ± 4.6 | 154.7 ± 2.4 | 1.380 ± 0.023 | 2.345 ± 0.009 | 2.189 ± 0.007 |
| $HCO^+$ (1–0) | 133.7 ± 2.4 | 86.9 ± 1.6 | 1.161 ± 0.023 | 2.126 ± 0.008 | 1.939 ± 0.008 |
| HNC (1–0) gp | 136.3 ± 1.3 | 68.4 ± 1.1 | 1.169 ± 0.022 | 2.135 ± 0.004 | 1.835 ± 0.007 |
| $HC_3N$ (10–9) | 59.2 ± 0.7 | 12.8 ± 0.7 | 0.807 ± 0.022 | 1.772 ± 0.005 | 1.108 ± 0.023 |
| $CH_3CN$ (5–4) gp | 31.8 ± 0.9 | 8.7 ± 0.7 | 0.537 ± 0.025 | 1.502 ± 0.013 | 0.942 ± 0.035 |
| $^{13}$CS (2–1) | 9.2 ± 0.5 | 10.4 ± 0.8 | ... | 0.965 ± 0.021 | 1.018 ± 0.031 |
| $N_2H^+$ (1–0) gp | 113.9 ± 0.8 | 38.5 ± 0.6 | 1.091 ± 0.021 | 2.057 ± 0.003 | 1.585 ± 0.007 |
| $CH_3OH$ v$_t$ = 0, 0(0,0)–1(-1,1) | 16.8 ± 0.5 | 8.1 ± 0.7 | 0.261 ± 0.025 | 1.226 ± 0.013 | 0.906 ± 0.036 |
| $HC_3N$ v = 0, (12–11) gp | 47.5 ± 0.6 | 9.1 ± 0.7 | 0.712 ± 0.022 | 1.677 ± 0.005 | 0.960 ± 0.033 |
| $C^{18}O$ (1–0) | 15.5 ± 0.7 | 16.0 ± 1.0 | 0.226 ± 0.030 | 1.192 ± 0.021 | 1.204 ± 0.028 |
| HNCO 5(0,5)–4(0,4) gp | 96.6 ± 0.8 | 26.5 ± 0.9 | 1.020 ± 0.022 | 1.985 ± 0.004 | 1.424 ± 0.014 |
| $^{13}$CO v = 0, (1–0) | 217.4 ± 5.6 | 216.8 ± 4.5 | 1.372 ± 0.024 | 2.337 ± 0.011 | 2.336 ± 0.009 |
| $CH_3CN$ v = 0, 6(1)–5(1) gp | 21.7 ± 0.8 | ... | 0.371 ± 0.026 | 1.336 ± 0.016 | ... |
| CN v = 0, 1–0, 1/2–1/2, F=3/2–1/2 | 34.7 ± 2.5 | ... | 0.575 ± 0.038 | 1.540 ± 0.031 | ... |
| CN v = 0, 1–0, 3/2–1/2, F=5/2–3/2 | ... | ... | ... | ... | ... |
| CO v = 0, (1–0) | 1460.5 ± 32.3 | 1254.9 ± 28.8 | 2.199 ± 0.023 | 3.165 ± 0.010 | 3.099 ± 0.010 |

**Table 10.** 7-mm line statistics of regions 'a' and 'b', the peak and off peak positions in the 20 km s$^{-1}$ cloud respectively: Integrated emissions, $^{13}$CS (1–0) molecular line ratios; see equation (1). We have also listed the log$_{10}$ integrated emission values of molecules from regions 'a' and 'b'.

| Molecule | Region 'a' $\int T_A$ dV (K km s$^{-1}$) | Region 'b' $\int T_A$ dV (K km s$^{-1}$) | Region 'a' log$_{10}$ $^{13}$CS line ratio | Region 'a' log$_{10}$ of molecule | Region 'b' log$_{10}$ of molecule |
|---|---|---|---|---|---|
| $NH_2CHO$ 2(0,2)–1(0,1) gp | 8.6 ± 0.2 | 3.3 ± 0.2 | -0.190 ± 0.026 | 0.936 ± 0.012 | 0.513 ± 0.032 |
| SiO v = 0, (1–0) | 47.1 ± 0.4 | 20.5 ± 0.3 | 0.547 ± 0.024 | 1.673 ± 0.004 | 1.311 ± 0.007 |
| HNCO 2(0,2)–1(0,1) gp | 34.6 ± 0.3 | 18.4 ± 0.4 | 0.414 ± 0.024 | 1.539 ± 0.004 | 1.264 ± 0.008 |
| $CH_3OH$ 7(0,7)–6(1,6) A++ | 17.3 ± 0.4 | 5.4 ± 0.3 | 0.111 ± 0.026 | 1.237 ± 0.011 | 0.733 ± 0.026 |
| $H^{13}CCCN$ (5–4) | 6.6 ± 0.4 | ... | -0.310 ± 0.035 | 0.816 ± 0.026 | ... |
| CCS 4,3–3,2 | 7.4 ± 0.4 | 2.3 ± 0.3 | -0.258 ± 0.034 | 0.868 ± 0.025 | 0.362 ± 0.057 |
| $HC_3N$ 5–4 gp | 79.9 ± 0.4 | 36.3 ± 0.3 | 0.776 ± 0.024 | 1.902 ± 0.002 | 1.559 ± 0.004 |
| $^{13}$CS(1–0) | 13.4 ± 0.7 | 6.3 ± 0.4 | ... | 1.126 ± 0.024 | 0.800 ± 0.029 |
| $C^{34}S$ (1–0) | 19.8 ± 1.1 | 11.7 ± 0.7 | 0.171 ± 0.034 | 1.297 ± 0.025 | 1.068 ± 0.027 |
| $CH_3OH$ 1(0,1)–0(0,0) A++ | 98.2 ± 1.8 | 52.9 ± 0.9 | 0.866 ± 0.025 | 1.992 ± 0.008 | 1.723 ± 0.007 |
| OCS (4–3) | 11.8 ± 1.0 | ... | -0.055 ± 0.045 | 1.071 ± 0.038 | ... |
| CS (1–0) | 150.0 ± 2.6 | 117.6 ± 1.4 | 1.050 ± 0.025 | 2.176 ± 0.008 | 2.071 ± 0.005 |

### *3.3.4 3 and 7-mm log$_{10}$ $^{13}$CS ratios comparing the peak position of the 50 km s$^{-1}$ cloud to the entire cloud itself*

In Fig. 6(a), a comparison of the peak position in the 50 km s$^{-1}$ cloud to the entirety of that cloud, we found that 8 transitions are brighter in the peak position, 9 transitions are brighter in region 2; and 3 transitions are consistent between both regions. From the 8 transitions which are brighter in the peak position of the 50 km s$^{-1}$ cloud, 7 of them are the same as those found to be brighter in the peak position of the 20 km s$^{-1}$ cloud; this includes (but is not limited to) SiO, HNCO , $HC_3N$ and $CH_3CN$ v = 0, 6(1)–5(1).

$^{13}$CO, CO and $C_2H$ are greater across the 50 km s$^{-1}$ cloud in comparison to the peak emission, as is also the case in the 3-mm line ratio comparison between the peak and 20 km s $^{-1}$ cloud. However, there were 5 other molecular transitions that were brighter across the 50 km s$^{-1}$ cloud that did not follow this trend in the 20 km s$^{-1}$ cloud, such as HCN (1–0) and $HCO^+$ (1–0). Interestingly, we note that $N_2H^+$ and $CH_3CN$ (5–4) are approximately equivalent between both the peak and entire 50 km s$^{-1}$ cloud, which would imply that these two molecules are uniformly distributed amongst the 50 km s$^{-1}$ cloud. Furthermore, in the 3-mm comparison between the peak

**Table 11.** Comparison of brightness results from the 3 and 7-mm $\log_{10}$ line ratio plots, for molecular transitions between region 'a' (peak position in the 20 km $s^{-1}$ cloud) and region 1 (the 20 km $s^{-1}$ cloud).

| $\log_{10}$ line ratio plots | Region 'a' > Region 1 | Region 'a' ≈ Region 1 | Region 1 > Region 'a' |
|---|---|---|---|
| 3-mm $^{13}CS$ (2–1) | $c\text{-}C_3H_2$ 2(1,2)–1(0,1)<br>$H^{13}CN$ (1–0) gp<br>$H^{13}CO^+$ (1–0)<br>SiO v = 0, (2–1)<br>HNCO 4(0,4)–3(0,3)<br>HCN (1–0) gp<br>$HCO^+$ (1–0)<br>HNC (1–0) gp<br>$HC_3N$ (10–9)<br>$CH_3CN$ (5–4) gp<br>$N_2H^+$ (1–0) gp<br>$CH_3OH$ $v_t$ = 0, 0(0,0)–1(-1,1)<br>$HC_3N$ v= 0, (12–11) gp<br>HNCO 5(0,5)–4(0,4) gp<br>$CH_3CN$ v = 0, 6(1)–5(1) gp | $CH_3CCH$ (5–4) gp | $^{13}CO$ v = 0, (1–0)<br>$C_2H$ (1–0, 3/2–1/2) gp<br>CO v = 0, (1–0) |
| 7-mm $^{13}CS$ (1–0) | $NH_2CHO$ 2(0,2)–1(0,1) gp<br>$CH_3OH$ 7(0,7)–6(1,6) A++<br>$H^{13}CCCN$ (5–4)<br>$HC_3N$ (5–4) gp<br>SiO v = 0, (1–0) | CCS 4,3–3,2<br>$CH_3OH$ 1(0,1)–0(0,0) A++<br>HNCO 2(0,2)–1(0,1) gp<br>OCS (4–3) | $C^{34}S$ (1–0)<br>CS (1–0) |

**Table 12.** 3-mm line statistics of regions 'c' and 'd', the peak and off peak positions in the 50 km $s^{-1}$ cloud respectively: Integrated emissions, $^{13}CS$ (2–1) molecular line ratios; see equation (1). We have also listed the $\log_{10}$ values of molecules from regions 'c' and 'd'.

| Molecule | Region 'c' $\int T_A$ dV (K km $s^{-1}$) | Region 'd' $\int T_A$ dV (K km $s^{-1}$) | Region 'c' $\log_{10}$ $^{13}CS$ line ratio | Region 'c' $\log_{10}$ of molecule | Region 'd' $\log_{10}$ of molecule |
|---|---|---|---|---|---|
| $c\text{-}C_3H_2$ 2(1,2)–1(0,1) | 21.1 ± 0.8 | 15.8 ± 0.8 | 0.135 ± 0.020 | 1.325 ± 0.016 | 1.198 ± 0.021 |
| $CH_3CCH$ (5–4) gp | 6.9 ± 0.7 | ... | -0.349 ± 0.048 | 0.841 ± 0.046 | ... |
| $H^{13}CN$ (1–0) gp | 60.7 ± 1.0 | 39.0 ± 0.7 | 0.593 ± 0.015 | 1.784 ± 0.007 | 1.591 ± 0.008 |
| $H^{13}CO^+$ (1–0) | 14.5 ± 3.8 | 9.1 ± 1.7 | -0.028 ± 0.114 | 1.162 ± 0.113 | 0.959 ± 0.082 |
| SiO v = 0, (2–1) | 57.6 ± 0.6 | 24.4 ± 0.6 | 0.570 ± 0.014 | 1.760 ± 0.005 | 1.387 ± 0.011 |
| $C_2H$ (1–0, 3/2–1/2) gp | 80.4 ± 5.0 | 66.9 ± 4.6 | 0.715 ± 0.030 | 1.905 ± 0.027 | 1.825 ± 0.030 |
| HNCO 4(0,4)–3(0,3) | 64.6 ± 0.6 | 35.3 ± 0.5 | 0.620 ± 0.014 | 1.811 ± 0.004 | 1.548 ± 0.007 |
| HCN (1–0) gp | 309.2 ± 7.0 | 238.5 ± 3.7 | 1.300 ± 0.016 | 2.490 ± 0.010 | 2.377 ± 0.007 |
| $HCO^+$ (1–0) | 135.3 ± 2.9 | 113.8 ± 1.5 | 0.941 ± 0.016 | 2.131 ± 0.009 | 2.056 ± 0.006 |
| HNC (1–0) gp | 107.2 ± 1.5 | 81.7 ± 0.8 | 0.840 ± 0.014 | 2.030 ± 0.006 | 1.912 ± 0.004 |
| $HC_3N$ (10–9) | 81.7 ± 0.6 | 37.0 ± 0.6 | 0.722 ± 0.013 | 1.912 ± 0.003 | 1.568 ± 0.007 |
| $CH_3CN$ (5–4) gp | 38.9 ± 1.0 | 23.2 ± 0.8 | 0.399 ± 0.017 | 1.590 ± 0.011 | 1.366 ± 0.015 |
| $^{13}CS$ (2–1) | 15.5 ± 0.5 | 10.3 ± 0.5 | ... | 1.190 ± 0.013 | 1.011 ± 0.022 |
| $N_2H^+$ (1–0) gp | 91.1 ± 0.8 | 49.6 ± 0.6 | 0.769 ± 0.014 | 1.959 ± 0.004 | 1.696 ± 0.005 |
| $CH_3OH$ $v_t$ = 0, 0(0,0)–1(-1,1) | 31.7 ± 0.9 | 15.8 ± 0.9 | 0.310 ± 0.018 | 1.500 ± 0.012 | 1.199 ± 0.024 |
| $HC_3N$ v = 0, (12–11) gp | 62.3 ± 1.2 | 20.3 ± 0.9 | 0.604 ± 0.015 | 1.795 ± 0.008 | 1.308 ± 0.020 |
| $C^{18}O$ (1–0) | 34.5 ± 4.3 | 30.9 ± 2.4 | 0.348 ± 0.056 | 1.538 ± 0.055 | 1.490 ± 0.034 |
| HNCO 5(0,5)–4(0,4) gp | 70.8 ± 1.3 | 31.2 ± 1.2 | 0.660 ± 0.015 | 1.850 ± 0.008 | 1.495 ± 0.017 |
| $^{13}CO$ v = 0, (1–0) | 333.8 ± 4.0 | 303.2 ± 2.9 | 1.333 ± 0.014 | 2.524 ± 0.005 | 2.482 ± 0.004 |
| $CH_3CN$ v = 0, 6(1)–5(1) gp | 39.2 ± 1.8 | 9.4 ± 1.1 | 0.404 ± 0.024 | 1.594 ± 0.020 | 0.975 ± 0.049 |
| CN v = 0, 1–0, 1/2–1/2, F=3/2–1/2 | ... | ... | ... | ... | ... |
| CN v = 0, 1–0, 3/2–1/2, F=5/2–3/2 | 111.1 ± 5.6 | 90.7 ± 4.8 | 0.855 ± 0.026 | 2.046 ± 0.022 | 1.957 ± 0.023 |
| CO v = 0, (1–0) | 1826.2 ± 20.3 | 1769.1 ± 16.9 | 2.071 ± 0.014 | 3.262 ± 0.005 | 3.248 ± 0.004 |

**Table 13.** 7-mm line statistics of regions 'c' and 'd', the peak and off peak positions in the 50 km s$^{-1}$ cloud respectively: Integrated emissions, $^{13}$CS (1–0) molecular line ratios; see equation (1). We have also listed the log$_{10}$ values of molecules from regions 'c' and 'd'.

| Molecule | Region 'c' $\int T_A$ dV (K km s$^{-1}$) | Region 'd' $\int T_A$ dV (K km s$^{-1}$) | Region 'c' log$_{10}$ $^{13}$CS line ratio | Region 'c' log$_{10}$ of molecule | Region 'd' log$_{10}$ of molecule |
|---|---|---|---|---|---|
| $NH_2CHO$ 2(0,2)–1(0,1) gp | 6.5 ± 0.3 | 3.5 ± 0.3 | -0.407 ± 0.022 | 0.812 ± 0.021 | 0.549 ± 0.032 |
| SiO v = 0, (1–0) | 49.4 ± 0.3 | 30.3 ± 0.3 | 0.475 ± 0.009 | 1.694 ± 0.003 | 1.482 ± 0.005 |
| HNCO 2(0,2)–1(0,1) gp | 23.1 ± 0.3 | 20.0 ± 0.3 | 0.144 ± 0.010 | 1.363 ± 0.005 | 1.301 ± 0.007 |
| $CH_3OH$ 7(0,7)–6(1,6) A++ | 64.5 ± 0.5 | 23.8 ± 0.4 | 0.591 ± 0.009 | 1.810 ± 0.003 | 1.376 ± 0.007 |
| $H^{13}CCCN$ (5–4) | 3.1 ± 0.5 | 3.1 ± 0.4 | -0.732 ± 0.069 | 0.487 ± 0.068 | 0.491 ± 0.055 |
| CCS 4,3–3,2 | 7.0 ± 0.3 | 6.5 ± 0.3 | -0.372 ± 0.018 | 0.847 ± 0.016 | 0.814 ± 0.021 |
| $HC_3N$ 5–4 gp | 72.0 ± 0.3 | 50.3 ± 0.4 | 0.638 ± 0.008 | 1.858 ± 0.002 | 1.701 ± 0.003 |
| $^{13}$CS(1–0) | 16.6 ± 0.3 | 15.0 ± 0.4 | . . . | 1.219 ± 0.008 | 1.175 ± 0.013 |
| $C^{34}S$ (1–0) | 22.5 ± 0.6 | 19.2 ± 0.6 | 0.134 ± 0.014 | 1.353 ± 0.011 | 1.283 ± 0.014 |
| $CH_3OH$ 1(0,1)–0(0,0) A++ | 106.7 ± 1.4 | 77.4 ± 1.0 | 0.809 ± 0.010 | 2.028 ± 0.006 | 1.888 ± 0.006 |
| OCS (4–3) | 8.3 ± 0.5 | 7.4 ± 0.8 | -0.301 ± 0.027 | 0.918 ± 0.026 | 0.869 ± 0.048 |
| CS (1–0) | 203.7 ± 1.7 | 176.9 ± 1.2 | 1.090 ± 0.009 | 2.309 ± 0.004 | 2.248 ± 0.003 |

**Table 14.** Comparison of brightness results from the 3 and 7-mm log$_{10}$ line ratio plots, for molecular transitions between region 'c' (peak position in the 50 km s$^{-1}$ cloud) and region 2 (the 50 km s$^{-1}$ cloud).

| log$_{10}$ line ratio plots | Region 'c' > Region 2 | Region 'c' ≈ Region 2 | Region 2 > Region 'c' |
|---|---|---|---|
| 3-mm $^{13}$CS (2–1) | $CH_3CCH$ (5–4) gp | $H^{13}CO^+$ (1–0) | c-$C_3H_2$ 2(1,2)–1(0,1) |
| | SiO v = 0, (2–1) | $CH_3CN$ (5–4) gp | HCN (1–0) gp |
| | $HC_3N$ (10–9) | $N_2H^+$ (1–0) gp | $C_2H$ (1–0, 3/2–1/2) gp |
| | $CH_3OH$ v$_t$ = 0, 0(0,0)–1(-1,1) | | $HCO^+$ (1–0) |
| | $HC_3N$ v= 0, (12–11) gp | | HNC (1–0) gp |
| | HNCO 5(0,5)–4(0,4) gp | | $^{13}$CO v = 0, (1–0) |
| | $CH_3CN$ v = 0, 6(1)–5(1) gp | | CN v = 0, 1–0, 3/2–1/2 |
| | HNCO 4(0,4)–3(0,3) | | CO v = 0, (1–0) |
| | | | $H^{13}CN$ (1–0) gp |
| 7-mm $^{13}$CS (1–0) | $NH_2CHO$ 2(0,2)–1(0,1) gp | HNCO 2(0,2)–1(0,1) gp | CS (1–0) |
| | SiO v = 0, (1–0) | $H^{13}CCCN$ (5–4) | |
| | $CH_3OH$ 7(0,7)–6(1,6) A++ | CCS 4,3–3,2 | |
| | $HC_3N$ (5–4) gp | $C^{34}S$ (1–0) | |
| | $CH_3OH$ 1(0,1)–0(0,0) A++ | OCS (4–3) | |

and 20 km s$^{-1}$ cloud, $CH_3CCH$ 5–4 was consistent between both regions.

The 7-mm $^{13}$CS results of the same two regions, Fig. 6(b) reveal that from the 11 molecular transitions, 5 of them are brighter in the peak position, 1 is brighter in region 2; and 5 are equivalent within errors between both regions. From the 5 transitions that are brighter in the peak position, 4 of them are the same as those which are brighter in the peak position of the 20 km s$^{-1}$ cloud in the 7-mm molecular lines; SiO and $HC_3N$ are included in those molecules. The only example where the 50 km s$^{-1}$ cloud is greater than the peak position, is in the CS molecule. There are 5 transitions which are approximately equal between both the peak and the 50 km s$^{-1}$ cloud, of which three overlap with those found in the 7-mm comparison for the 20 km s$^{-1}$ (HNCO 2(0,2)–1(0,1), OCS (4–3) and CCS 4,3–3,2). A full breakdown of results between regions 'c' and '2' is found in Table 14.

The results from the 3 and 7-mm line ratios, when comparing each cloud and their respective peak positions, have shown us that while there are localised enhancements of both shock and high density tracers within both clouds (and even across each cloud, as appears to be the case in the 20 km s$^{-1}$ cloud); when comparing each cloud respectively, we in fact observe similar trends between both clouds. Consequently, it is possible to infer that there is only a small variation in the chemistry between the 20 and 50 km s$^{-1}$ clouds, further indicating that the gas within the both clouds is likely to be well mixed and even suggesting that there is a possible uniform distribution of the gas between both clouds.

#### 3.3.5 *3 and 7-mm log$_{10}$ integrated emission comparing the peak and off peak positions within the 20 and 50 km s$^{-1}$ clouds*

Comparing the 19 molecular transitions in the 3-mm log$_{10}$ plot of integrated emission between the peak vs. off peak position in the 20 km s$^{-1}$ cloud, Fig. 7(a), 16 are much brighter in the peak position and the remaining 3 molecular transitions ($^{13}$CS (2–1), $C^{18}O$ (1–0) and $^{13}$CO) are all consistent within uncertainties. These results are expected as we would assume the peak position of the cloud to be substantially larger than the off peak position. The 3 molecular transitions that are equal between the peak and off peak positions are widespread molecules. In the 7-mm log$_{10}$ plot of the same two

**Table 15.** Comparison of brightness results from the 3 and 7-mm $\log_{10}$ values of the molecular lines, between regions 'a' and 'b', the peak and off peak positions in the 20 km $s^{-1}$ cloud respectively.

| wavelength | Region 'a' > Region 'b' | Region 'a' ≈ Region 'b' | Region 'b' > Region 'a' |
|---|---|---|---|
| 3-mm | c-$C_3H_2$ 2(1,2)–1(0,1) | $^{13}$CS (2–1) | ... |
| | $H^{13}CN$ (1–0) gp | $C^{18}O$ (1–0) | |
| | $H^{13}CO^+$ (1–0) | $^{13}$CO v = 0, (1–0) | |
| | SiO v = 0, (2–1) | | |
| | $C_2H$ (1–0, 3/2–1/2) gp | | |
| | HNCO 4(0,4)–3(0,3) | | |
| | HCN (1–0) gp | | |
| | $HCO^+$ (1–0) | | |
| | HNC (1–0) gp | | |
| | $HC_3N$ (10–9) | | |
| | $CH_3CN$ (5–4) gp | | |
| | $N_2H^+$ (1–0) gp | | |
| | $CH_3OH$ $v_t$ = 0, 0(0,0)–1(-1,1) | | |
| | $HC_3N$ v= 0, (12–11) gp | | |
| | HNCO 5(0,5)–4(0,4) gp | | |
| | CO v = 0, (1–0) | | |
| 7-mm | $NH_2CHO$ 2(0,2)–1(0,1) gp | ... | ... |
| | SiO v = 0, (1–0) | | |
| | HNCO 2(0,2)–1(0,1) gp | | |
| | $CH_3OH$ 7(0,7)–6(1,6) A++ | | |
| | CCS 4,3–3,2 | | |
| | $HC_3N$ (5–4) gp | | |
| | $^{13}$CS (1–0) | | |
| | $C^{34}S$ (1–0) | | |
| | $CH_3OH$ 1(0,1)–0(0,0) A++ | | |
| | CS (1–0) | | |

regions, Fig. 7(b), all the molecular transitions are considerably brighter in the peak position.

In Fig. 8(a), the 3-mm $\log_{10}$ integrated emission plot comparing the peak and off peak position of the 50 km $s^{-1}$ cloud, we observe the same trend as found in Fig. 7(a), where 19 of the 21 molecular transitions are brighter in the peak position and the remaining 2 transitions, $C^{18}O$ (ubiquitous) and $H^{13}CO^+$ are the equivalent between both regions. In the 7-mm $\log_{10}$ integrated emission comparison of the same two positions, Fig. 8(b), 9 molecular transitions are brighter in the peak postion and the remaining 3 transitions ($H^{13}CCCN$, CCS and OCS) are consistent between each other in both peak and off peak positions.

The full set of results from a $\log_{10}$ comparison of integrated emission between regions 'a' and 'b'; as well as 'c' and 'd', can be found in Tables 15 and 16 respectively.

#### 3.3.6 *Interpretation for enhancement in shock and density tracers in the 20 km $s^{-1}$ cloud*

The 20 km $s^{-1}$ cloud is enhanced in shock (SiO, HNCO) and high density tracers (e.g. $N_2H^+$), which are approximately evenly distributed throughout the cloud. Star formation is unlikely to be causing this. Previous molecular line observations on a Milky Way giant molecular cloud, G333, found that SiO (2–1) emission associated with a dense core that will eventually undergo high-mass star formation, was isolated in one part of the cloud (Lo et al. 2007). Our 3-mm SiO (2–1) v = 0 image shows that this transition is brighter in one part of the 20 km $s^{-1}$ cloud, but it is still distributed throughout the cloud. The 7-mm SiO (1–0) transition is also distributed across the whole cloud. Moreover, the molecular line distribution of $N_2H^+$ in G333 is not widespread as found in the 20 km $s^{-1}$ cloud, instead, it is closely packed isolated clumps (Lo et al. 2009), unlike what we see in the 20 km $s^{-1}$ cloud. The most likely reason for this occurring, is due to the tidal forces generated by Sgr A*, as will be discussed in § 3.7.

### 3.4 Peak emission vs. moment 0 image comparison

In Fig. 9, we have provided a side by side comparison of peak emission and moment 0 (integrated emission) images, for 5 molecules, from the 3 and 7-mm wavebands. In 7-mm SiO, we see a good consistency between the 50 kms$^{-1}$ cloud (region 1) between both images, however, in the 20 km $s^{-1}$ cloud (region 2), we note that while the emission is widespread between both images, the moment 0 image indicates the presence of two components within the cloud; evidence of this has been observed previously (Minh & Irvine 2006; Minh et al. 2013). 3-mm HNCO 4(0,4)–3(0,3), HNCO 5(0,5)–4(0,4) and $N_2H^+$ (1–0) show similar widespread structure between their peak and moment 0 images respectively. This same trend is also present in the 7-mm CS 1–0 peak and moment 0 images. The consistency between SiO, HNCO and $N_2H^+$ between the peak and moment 0 images demonstrates that the entire line is most likely enhanced; and that the peak emission images are an appropriate representation of the molecular line emission.

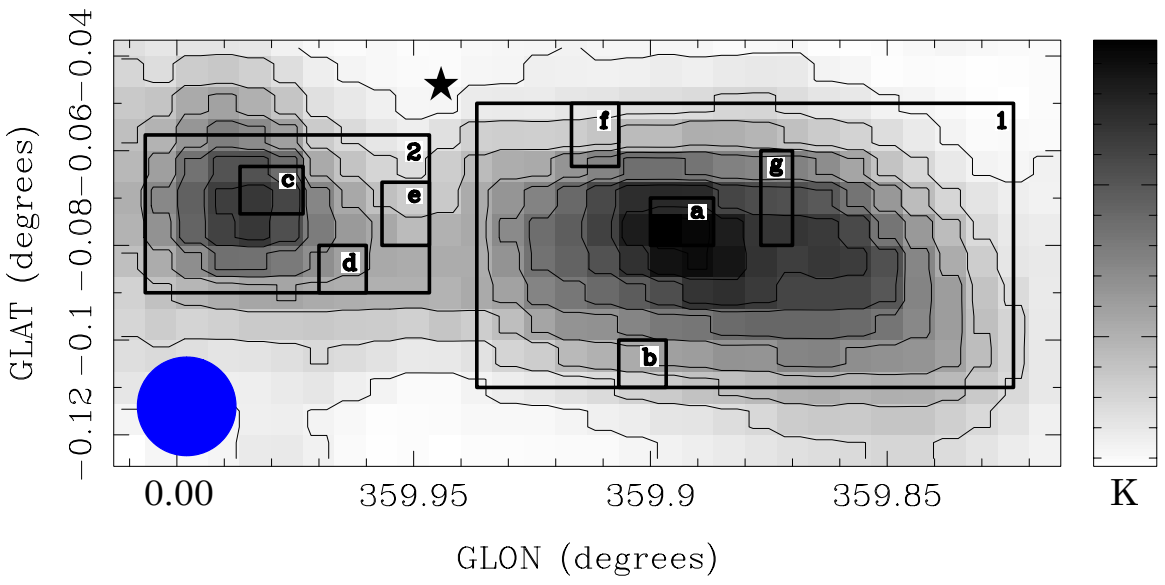

(a) 7-mm SiO 1–0 v = 0 peak emission image

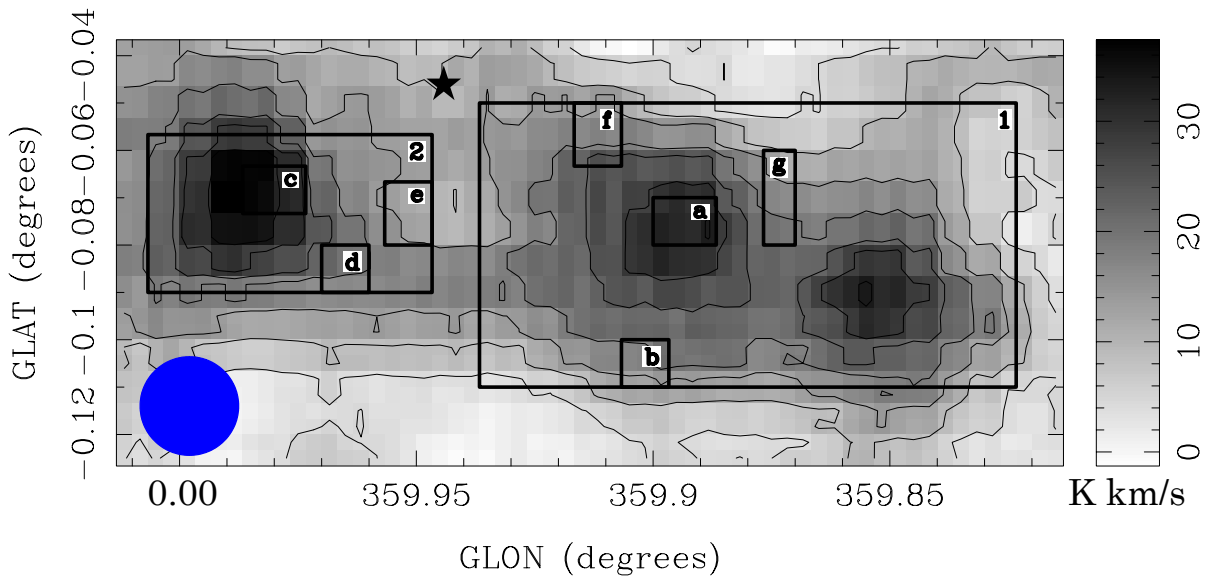

(b) 7-mm SiO 1–0 v = 0 moment 0 image

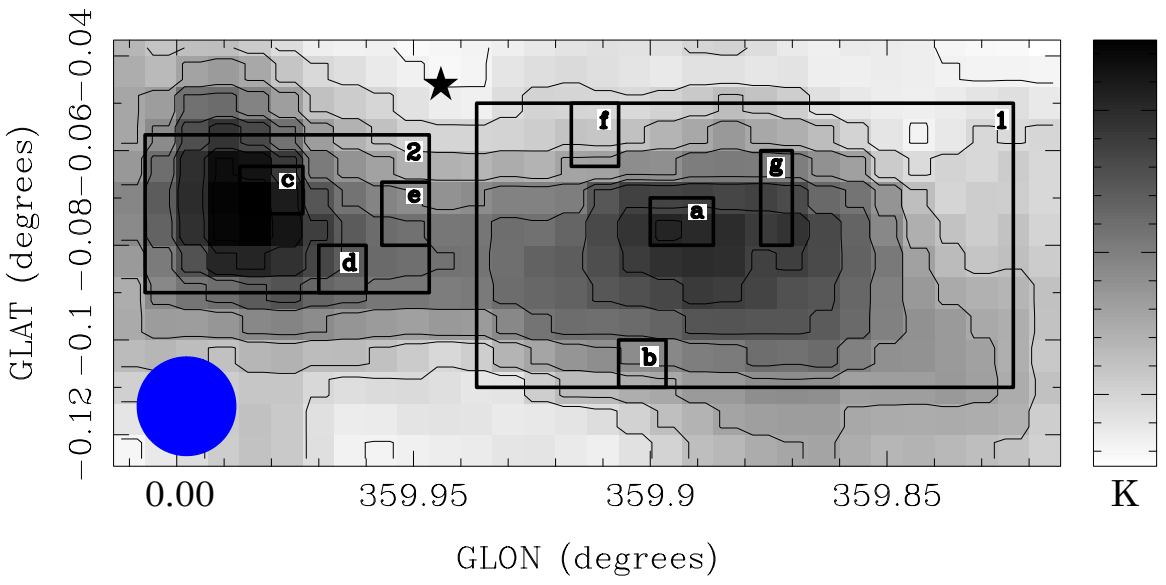

(c) 7-mm CS 1–0 peak emission image

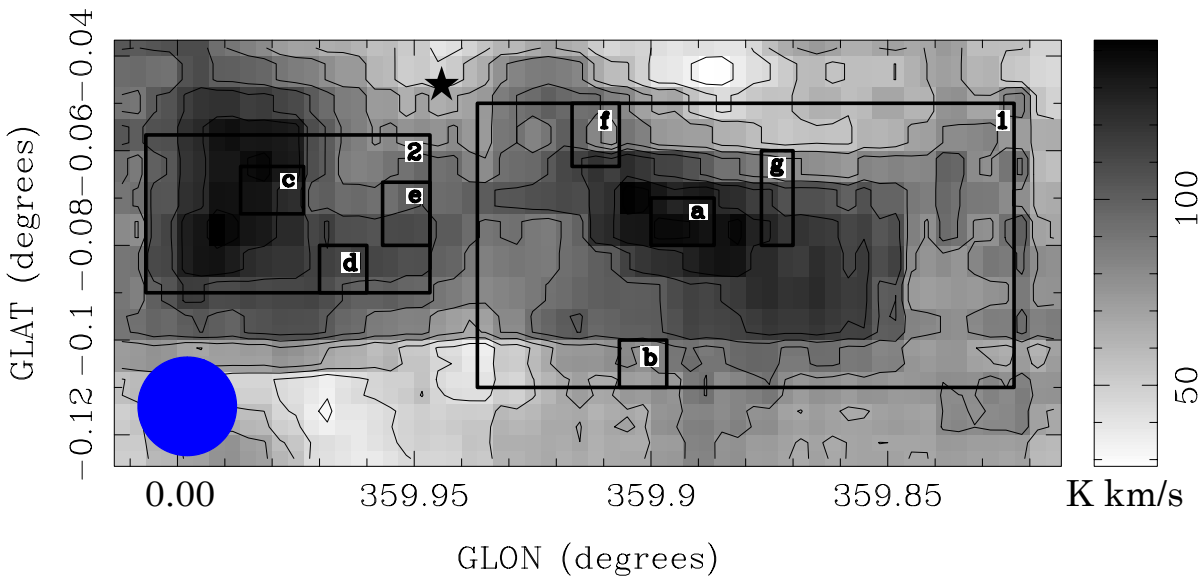

(d) 7-mm CS 1–0 moment 0 image

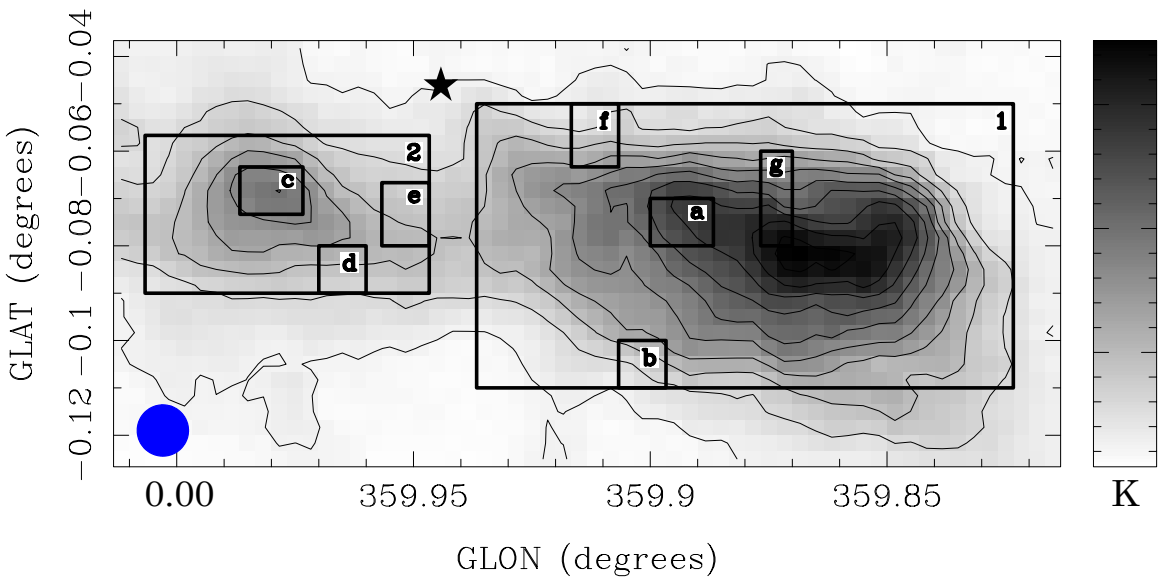

(e) 3-mm HNCO 4(0,4)–3(0,3) peak emission image

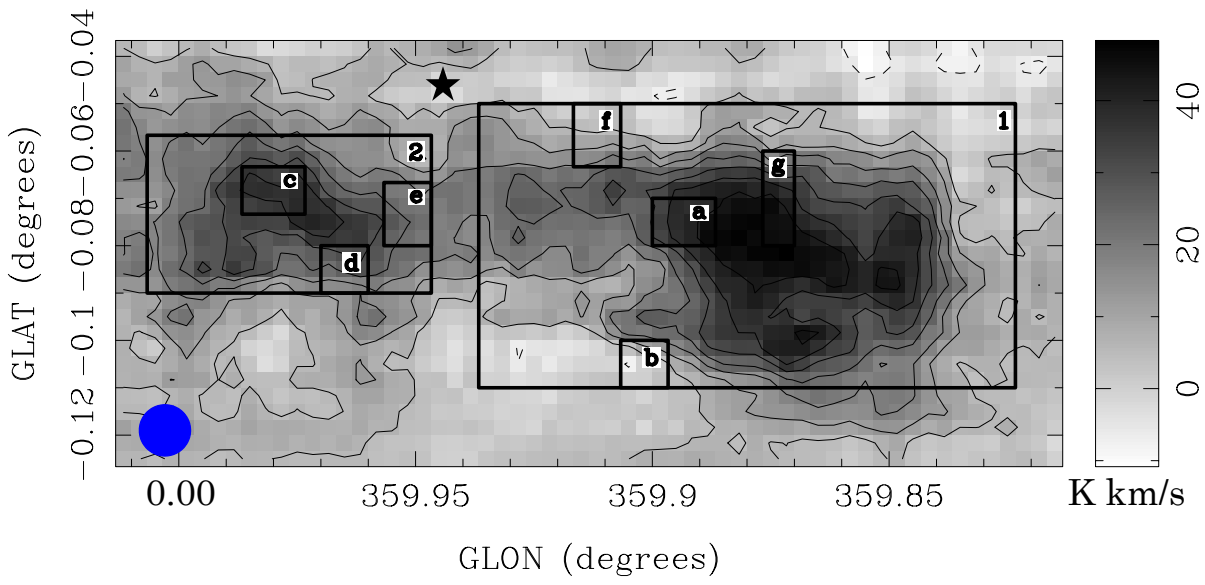

(f) 3-mm HNCO 4(0,4)–3(0,3) moment 0 image

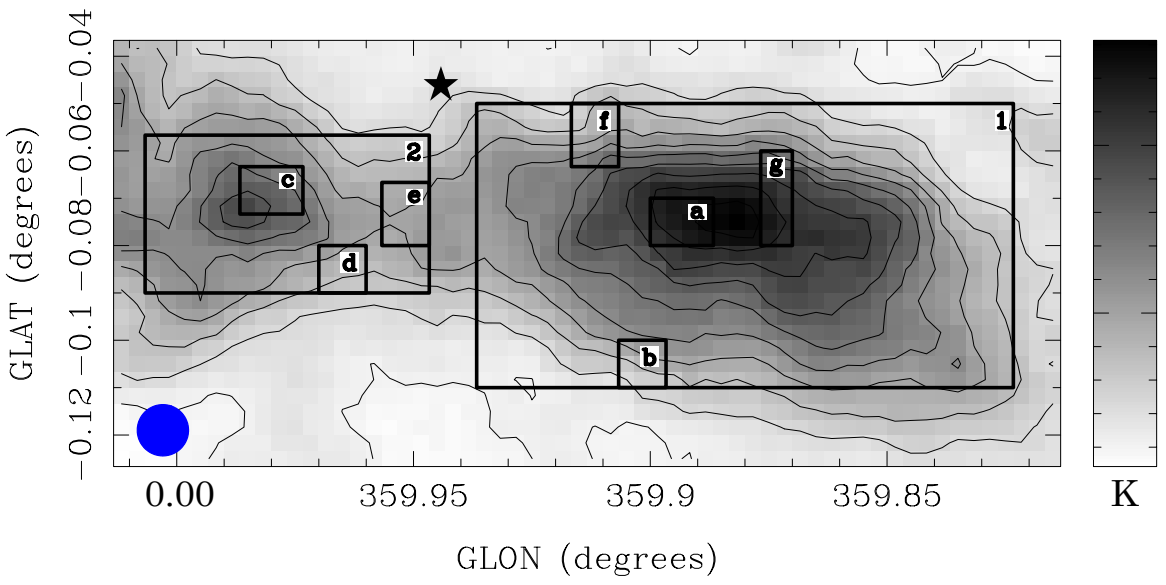

(g) 3-mm $N_2H^+$ (1–0) peak emission image

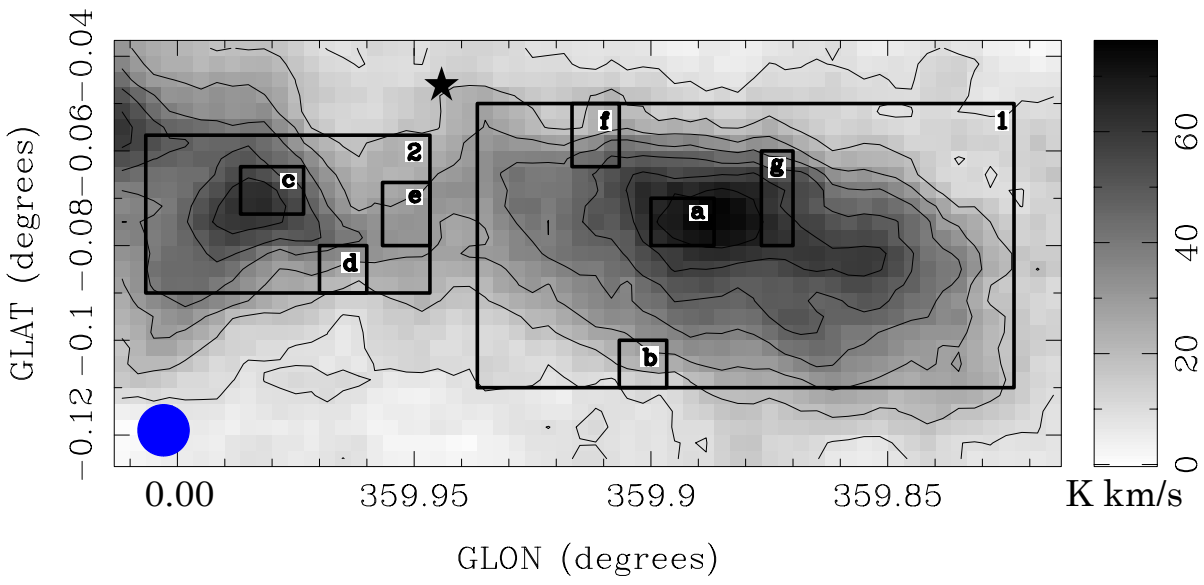

(h) 3-mm $N_2H^+$ (1–0) moment 0 image

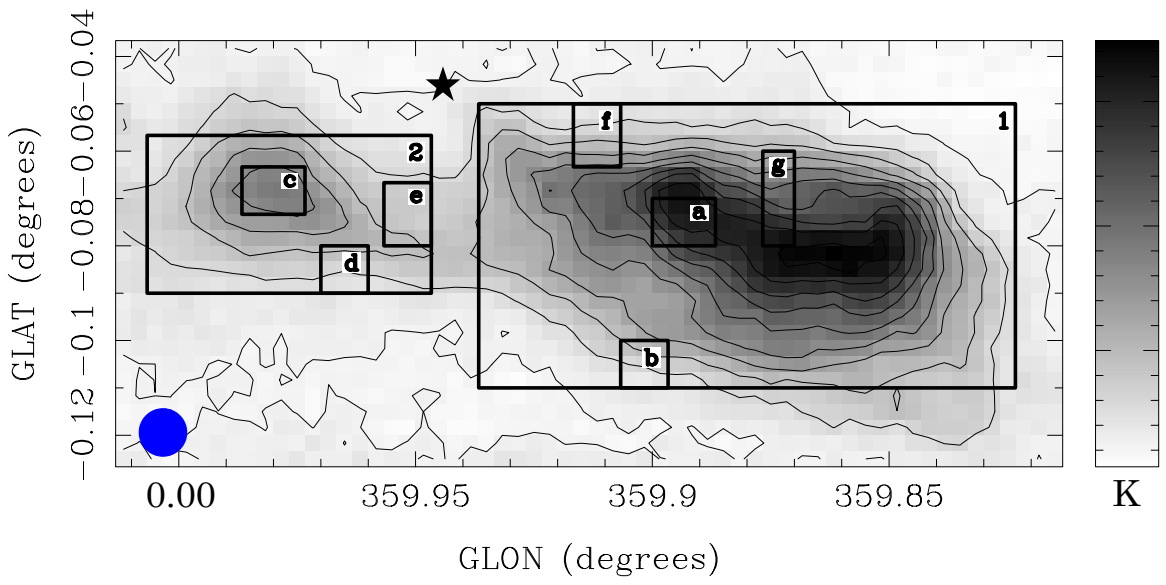

(i) 3-mm HNCO 5(0,5)–4(0,4) peak emission image

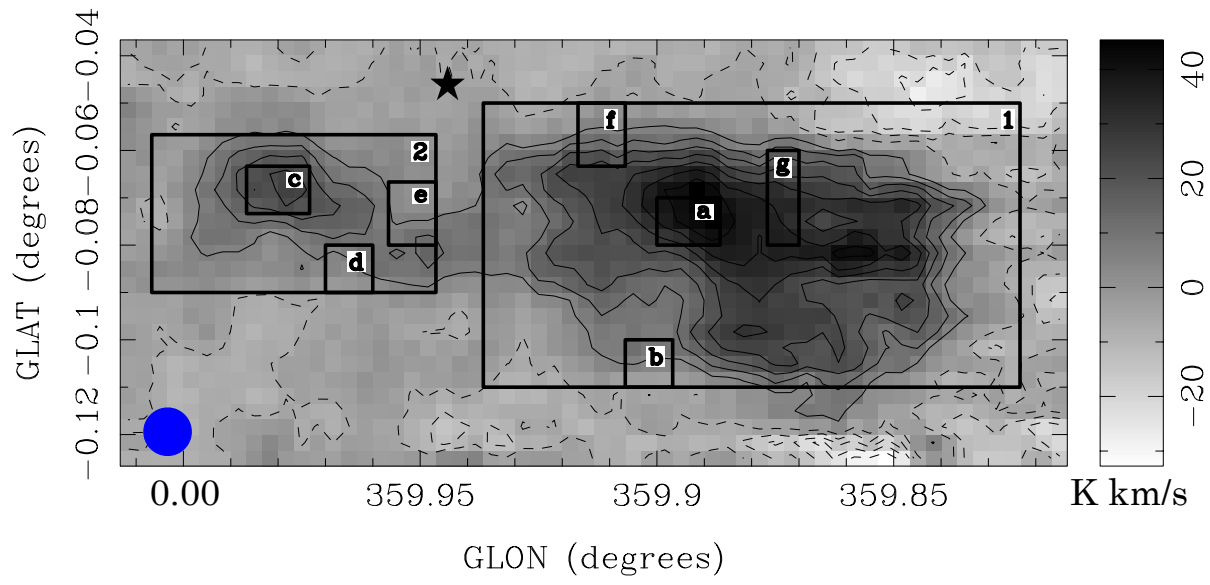

(j) 3-mm HNCO 5(0,5)–4(0,4) moment 0 image

**Figure 9.** A comparison of peak emission and moment 0 images, between a selection of molecules that have show enhancement in the line ratio analysis. A good consistency is found between both types of images in the molecules presented here.

**Table 16.** Comparison of brightness results from the 3 and 7-mm $\log_{10}$ values of the molecular lines, between regions 'c' and 'd', the peak and off peak positions in the 50 km s$^{-1}$ cloud respectively.

| wavelength | Region 'c' > Region 'd' | Region 'c' ≈ Region 'd' | Region 'd' > Region 'c' |
|---|---|---|---|
| 3-mm | c-$C_3H_2$ 2(1,2)–1(0,1) | $H^{13}CO^+$ (1–0) | . . . |
| | $H^{13}CN$ (1–0) gp | $C^{18}O$ (1–0) | |
| | SiO v = 0, (2–1) | | |
| | $C_2H$ (1–0, 3/2–1/2) gp | | |
| | HNCO 4(0,4)–3(0,3) | | |
| | HCN (1–0) gp | | |
| | $HCO^+$ (1–0) | | |
| | HNC (1–0) gp | | |
| | $HC_3N$ (10–9) | | |
| | $CH_3CN$ (5–4) gp | | |
| | $^{13}CS$ (2–1) | | |
| | $N_2H^+$ (1–0) gp | | |
| | $CH_3OH$ $v_t$ = 0, 0(0,0)–1(-1,1) | | |
| | $HC_3N$ v= 0, (12–11) gp | | |
| | HNCO 5(0,5)–4(0,4) gp | | |
| | $^{13}CO$ v = 0, (1–0) | | |
| | $CH_3CN$ v = 0, 6(1)–5(1) gp | | |
| | CN v = 0, 1–0, 3/2–1/2 | | |
| | CO v = 0, (1–0) | | |
| 7-mm | $NH_2CHO$ 2(0,2)–1(0,1) gp | $H^{13}CCCN$ (5–4) | . . . |
| | SiO v = 0, (1–0) | CCS 4,3–3,2 | |
| | HNCO 2(0,2)–1(0,1) gp | OCS (4–3) | |
| | $CH_3OH$ 7(0,7)–6(1,6) A++ | | |
| | $HC_3N$ (5–4) gp | | |
| | $^{13}CS$ (1–0) | | |
| | $C^{34}S$ (1–0) | | |
| | $CH_3OH$ 1(0,1)–0(0,0) A++ | | |
| | CS (1–0) | | |

### 3.5 Line ratio map analysis of enhanced molecules

In Fig. 10, we display five line ratio maps from the 3 and 7-mm datasets, where we have divided the peak emission molecular line maps that have shown vastly differing abundances between the 20 and 50 km s$^{-1}$ clouds, by the respective $^{13}CS$ peak emission molecular line map. Prior to doing this, we clipped each of the line maps used here by masking out low level noise emission, which varied from 0.15–0.35 K depending on the molecule. These maps allow us to examine the distribution of the enhanced molecules within the two clouds.

In the 7-mm SiO 1–0 / 7-mm $^{13}CS$ map, the 20 km s$^{-1}$ cloud is significantly brighter than in the 50 km s$^{-1}$ cloud. The enhanced SiO emission is located within the middle of the 20 km s$^{-1}$ cloud, stretching from the left hand side of region 1 and extending (but not reaching) the right hand side of the aforementioned region. The SiO emission is not uniformly distributed, with multiple areas of bright emission within the 20 kms $^{-1}$ cloud. In contrast, the 50 km s$^{-1}$ cloud only shows a small region of enhancement along the left hand edge of region 2, with the rest of the cloud exhibiting weaker SiO enhancement by a factor of ≈ 2, in comparison to the 20 km s$^{-1}$ cloud. Identical results are present for both clouds in the 7-mm HNCO 2(0,2)–1(0,1) / 7-mm $^{13}CS$ map, where the only differences are that the positions of bright HNCO emission within the 20 km s$^{-1}$ cloud are similar, but not spatially distributed in the same manner as found in the 7-mm SiO / 7-mm $^{13}CS$ map; and the emission within the 50 km s$^{-1}$ cloud is not as bright as it was in the 7-mm SiO 1–0 / 7-mm $^{13}CS$ map.

In the 3-mm HNCO 4(0,4)–3(0,3) / 3-mm $^{13}CS$ line ratio map, it is clear that the emission is brighter and more widespread in a large portion of the 20 km s$^{-1}$ cloud than in the 50 km s$^{-1}$ cloud. The emission in the 20 km s$^{-1}$ cloud is brightest around the middle right hand side of the cloud and it extends outwards in a non uniform manner, such that it covers a large portion of the region 1 box. On the other hand, the overall distribution of HNCO 4(0,4)–3(0,3) in the 50 km s$^{-1}$ cloud is approximately uniformly weak in comparison to emission in the 20 km s$^{-1}$ cloud. The 3-mm $N_2H^+$ 1–0 / 3-mm $^{13}CS$ map exhibits a similar trend as the HNCO 4(0,4)–3(0,3) line ratio map, with bright $N_2H^+$ emission from the 20 km s$^{-1}$ cloud spatially distributed in a similar manner, but with differing bright emission locations that cluster in area at the centre of the cloud emission. The 3-mm HNCO 5(0,5)–4(0,4) / 3-mm $^{13}CS$ line ratio map is identical to the 3-mm HNCO 4(0,4)–3(0,3) line ratio map.

These 5 line ratio maps provide convincing evidence that the 20 km s$^{-1}$ cloud is enhanced in shock and high density tracers in comparison to the 50 km s$^{-1}$ cloud. The enhancement covers a wide area that is roughly the same size across the 20 km s$^{-1}$ cloud in each of the line ratio maps. We have also shown while that the 50 km s$^{-1}$ cloud exhibits enhancement as well, it is lower than that found in the 20 km s$^{-1}$ cloud.

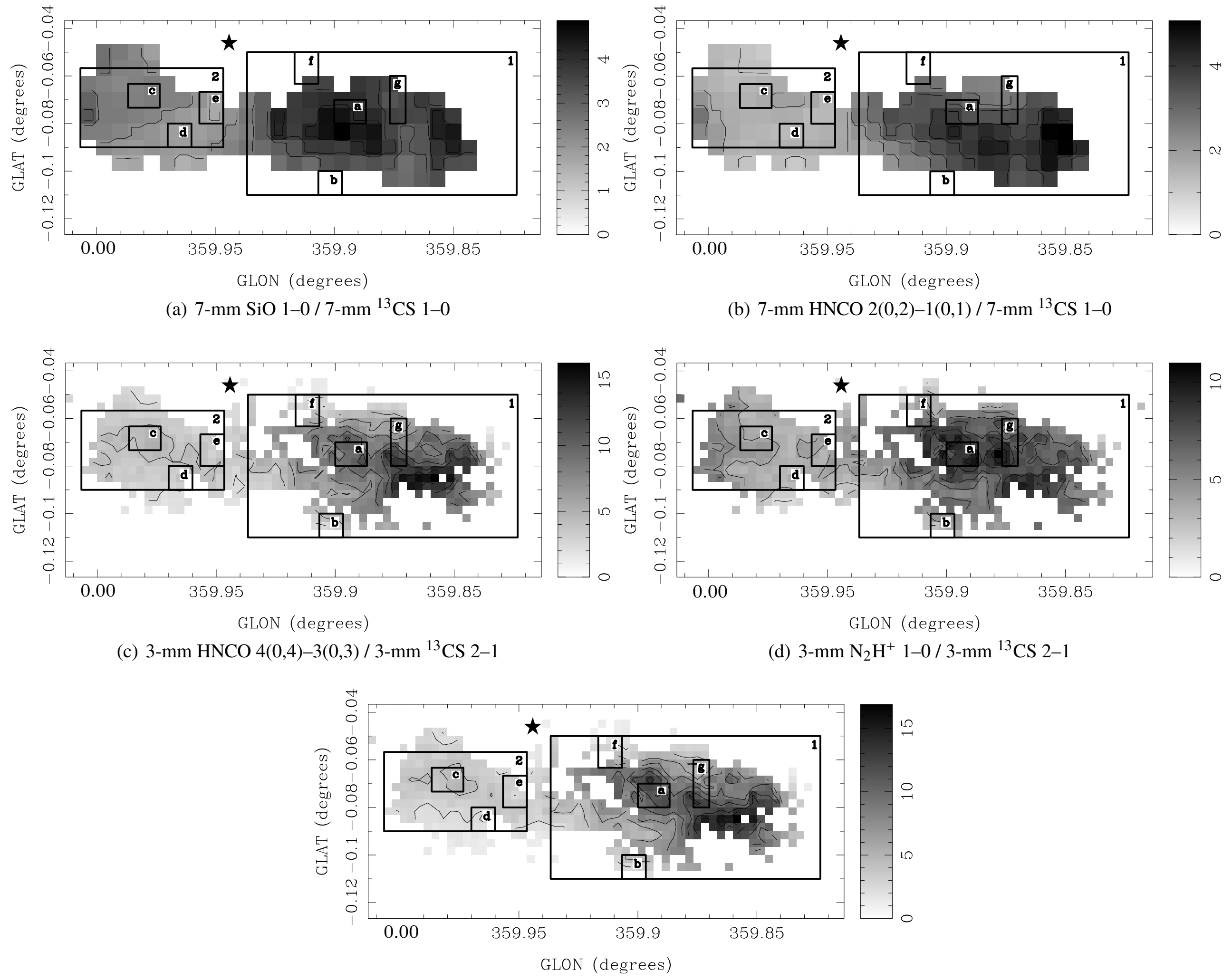

(a) 7-mm SiO 1–0 / 7-mm $^{13}$CS 1–0

(b) 7-mm HNCO 2(0,2)–1(0,1) / 7-mm $^{13}$CS 1–0

(c) 3-mm HNCO 4(0,4)–3(0,3) / 3-mm $^{13}$CS 2–1

(d) 3-mm $N_2H^+$ 1–0 / 3-mm $^{13}$CS 2–1

(e) 3-mm HNCO 5(0,5)–4(0,4) / 3-mm $^{13}$CS 2–1

**Figure 10.** Five peak emission line ratio maps of a selection of molecules which show enhancement in the 20 km s$^{-1}$ cloud (region 1). Each of the 5 peak emission molecular line maps, 7-mm SiO (1–0), 7-mm HNCO 2(0,2)–1(0,1), 3-mm HNCO 4(0,4)–3(0,3), 3-mm $N_2H^+$ (1–0) and 3-mm HNCO 5(0,5)–4(0,4), were divided by the respective $^{13}$CS peak emission map. Each image (including the ratio molecules) were clipped to avoid low level noise, which would affect the final result. As these are line ratio maps, the units are dimensionless.

### 3.6 Kinematics of molecular clouds near Sagittarius A*

In Fig. 11, we present nine examples of PV diagrams (3 molecules, CS 1–0, HNCO 2(0,2)–1(0,1) and SiO 1–0, with 3 slices each), from the 7-mm dataset. These PV diagrams were made using PVEXTRACTOR[9], following the 3 paths (A, B and C), from left to right, as shown in Fig. 2(a). In the 20 km s$^{-1}$ cloud, which is located at an offset of $\geqslant 0.05$ degrees, we find that this cloud exhibits an extended velocity gradient across all nine PV diagrams. In the three 'A' slices, we find that the 20 km s$^{-1}$ cloud is located between 0 to 20 km s$^{-1}$. In slices 'B' and 'C', the velocity gradient is more pronounced, with the gradient located from $\approx -10$ to + 40 km s$^{-1}$. In all three molecules and their respective slices, the elongated velocity structure extends $\approx$ 0.13 degrees. In contrast, the 50 km s$^{-1}$ cloud, which has a velocity gradient centralised between $\approx$ 40 to 60 km s$^{-1}$ for the bulk of the emission (and extending as far as 80 km s$^{-1}$ e.g. in Fig. 11(a), which is slice 'A' in 7-mm CS), is compact and only located in a narrow range from $\approx$ 0 to 0.05 degrees, in all the PV diagrams.

Similarly to Fig. 11, in Fig. 12, we present another nine examples of PV diagrams; this time from the 3-mm dataset, which includes H$^{13}$CN (1–0), $N_2H^+$ (1–0) and CO (1–0). In all three molecules, similar trends are observed as those discussed in the 7-mm PV diagrams.

A point of similarity between the 3 and 7-mm PV diagrams is the presence of two components in the 20 km s$^{-1}$ cloud. In the 7-mm HNCO slice 'B' PV diagram, Fig. 11(e), one is centred on $\approx$ 8 km s$^{-1}$ at an offset of 0.14 degrees; and the other at $\approx$ 18 km s$^{-1}$ with an offset position of 0.105 degrees. In the 3-mm $N_2H^+$ slice 'C' PV diagram, Fig. 12(h), the two components are located at the approximately the same positions, but each of them have a velocity which is between 4 to 6 km s$^{-1}$ larger. We note that despite the integrated emission image of 7-mm SiO 1–0 v = 0, Fig. 9(b), clearly

[9] https://github.com/radio-astro-tools/pvextractor

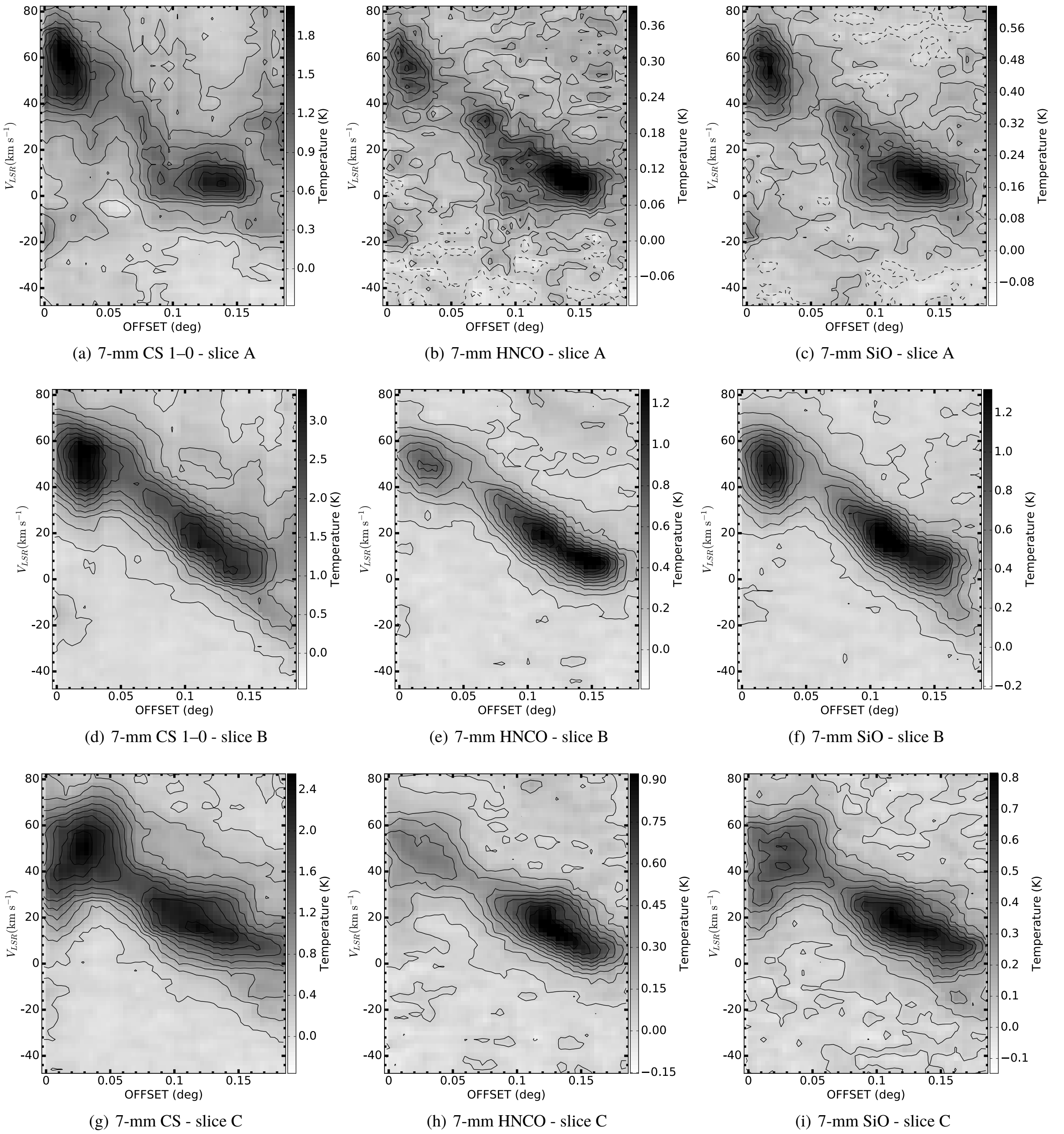

(a) 7-mm CS 1–0 - slice A (b) 7-mm HNCO - slice A (c) 7-mm SiO - slice A

(d) 7-mm CS 1–0 - slice B (e) 7-mm HNCO - slice B (f) 7-mm SiO - slice B

(g) 7-mm CS - slice C (h) 7-mm HNCO - slice C (i) 7-mm SiO - slice C

**Figure 11.** PV diagrams from a selection of bright 7-mm molecules. The 3 molecules presented here are: CS (1–0), HNCO 2(0,2)–1(0,1) and SiO 1–0. Each slice has been labelled as A, B or C; and refers to the lines drawn on Fig. 2(a), which were used to extract each PV diagram.

illustrating two components, it is difficult to discern this feature in the SiO PV diagrams.

There is good consistency between the 18 PV diagrams of the 6 molecules from the 3 and 7-mm datasets. We can infer from this result that both clouds exist within well defined velocity limits and it appears that the gas within the clouds is well mixed. Evidence for this arises from the similarity between the PV diagrams of slices B and C in all 6 molecules. While the PV diagrams from slices A appear to deviate from this trend slightly, as there isn't a significant amount of variation between each of those PV diagrams and as slice A, cuts through the 'top edge' of both clouds; this trend still holds.

The molecular ridge is located between the two aforementioned clouds. The PV diagrams also reveal a small and narrow component present between the velocities of 0 to –20 km s$^{-1}$, which we interpret to be part of the Northern Ridge streamer (NR), as displayed in Fig 1. Furthermore, self absorption is also present at $\approx$ 0 km s$^{-1}$, which is illustrated in the PV diagrams of 3-mm CO v = 0, 1–0. The absorption feature at 0 km s$^{-1}$ in Sgr A has been previously observed in CO (Liszt et al. 1977) and $HCO^+$ (Linke, Stark, & Frerking 1981). Liszt et al. (1977) found that this feature belongs to gas that lies between the sun and the Galactic Centre.

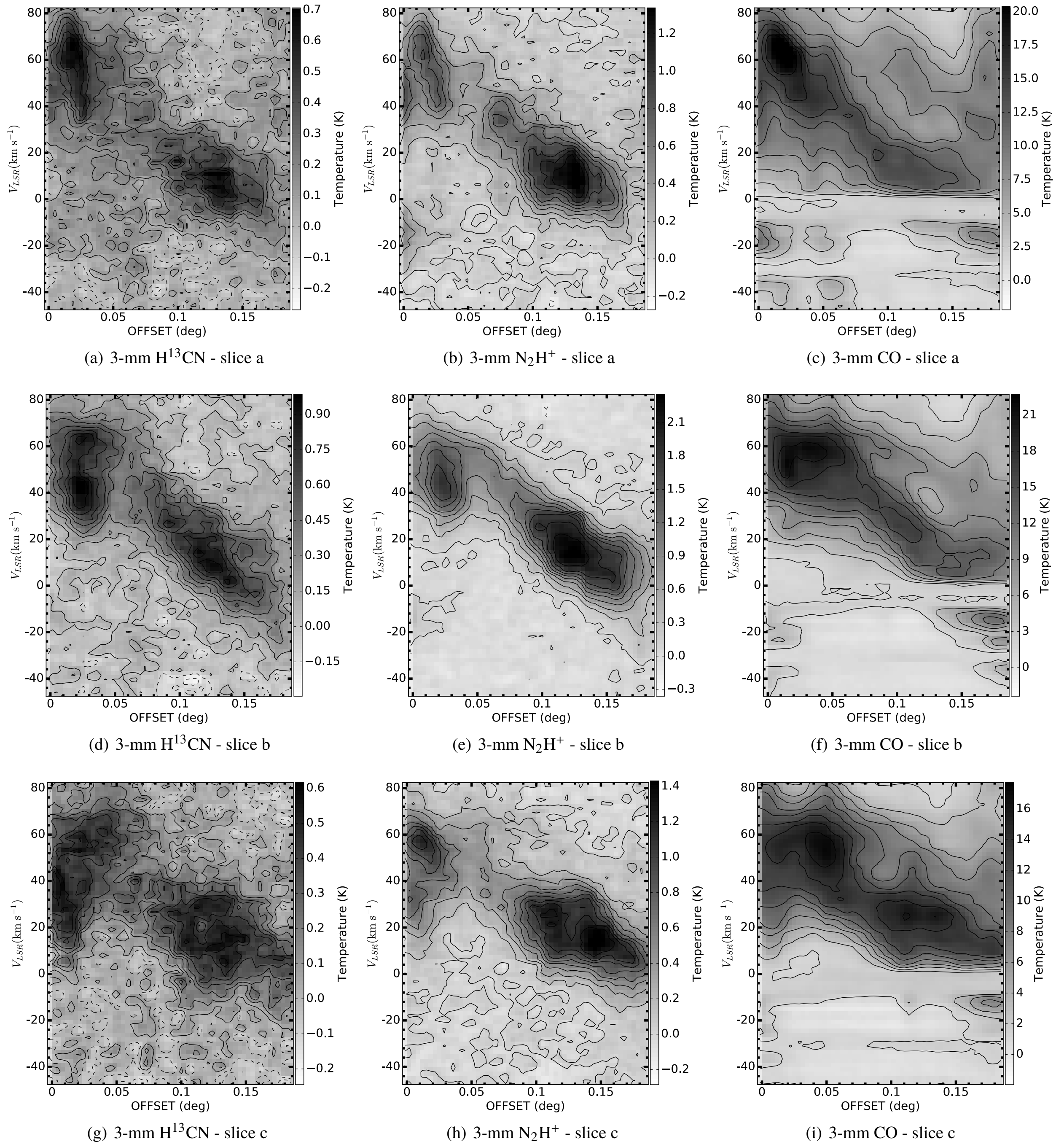

(a) 3-mm $H^{13}CN$ - slice a (b) 3-mm $N_2H^+$ - slice a (c) 3-mm CO - slice a

(d) 3-mm $H^{13}CN$ - slice b (e) 3-mm $N_2H^+$ - slice b (f) 3-mm CO - slice b

(g) 3-mm $H^{13}CN$ - slice c (h) 3-mm $N_2H^+$ - slice c (i) 3-mm CO - slice c

**Figure 12.** PV diagrams from a selection of bright 3-mm molecules. The 3 molecules shown here are: $H^{13}CN$ 1–0, $N_2H^+$ 1–0 and CO 1–0. Each slice is labelled as A, B or C; and refers to the lines drawn on Fig. 2(a), which were used to extract each PV diagram.

There is considerable uncertainty in the nature of the molecular ridge in the Galactic Centre and its relationship between the 20 and 50 km s$^{-1}$ clouds (Ferrière 2012 and references therein) as well as its interaction with a supernova remnant[10]. In Fig. 1, Ferrière 2012 deduced that the molecular ridge appears to join both of the aforementioned clouds. Coil & Ho (2000) resolved the molecular ridge with a synthesized beam of ≈ 14″× 9″. Whilst the beam sizes of our datasets are larger than that, we can still see evidence of the ridge across our range of molecules in the peak brightness emission images and PV diagrams. This has allowed us to constrain its nature, through a series of visual diagnostics.

[10] A thorough discussion of the interaction between a supernova remnant located at 359.92°, –0.09° ($l$, $b$), the molecular ridge and 20 km s$^{-1}$ cloud can be found in Coil & Ho (2000).

In the peak brightness emission images, visual inspection of the structure between the 20 and 50 km $s^{-1}$ clouds, where the molecular ridge is found, show that in many molecules, all three molecular components are effectively connected with one another, e.g. 7-mm CS (1–0) and $CH_3OH$ 1(0,1)–0(0,0) A++. This 'connectivity' is also present in 3-mm $H^{13}CN$ (1–0), $HC_3N$ (10–9), $CH_3CN$ (5–4), $N_2H^+$ (1–0) and 7-mm $HC_3N$ (5–4). Additionally, it can be difficult to distinguish between the 20 km $s^{-1}$ cloud and the molecular ridge as two distinct features. This can be seen in 3-mm $^{13}CO$ (1–0), $^{13}CS$ (2–1), CO (1–0) and 0.65-mm CO (4–3) (see online supplementary material). Furthermore in dense tracers such as HCN (1–0), this ridge is also blended between both clouds.

In the PV diagrams, the molecular ridge in 7-mm CS (1–0) and 3-mm CO (1–0) appear combined with the 20 and 50 km $s^{-1}$ clouds across all PV slices. Other molecules such as $H^{13}CN$ (1–0) and $N_2H^+$ (1–0), which trace different properties of the cloud, also demonstrated this trend, to varying degrees. We have found that there is good agreement between the peak brightness emission images and their corresponding position velocity diagrams.

Longmore et al. (2013) presented a large scale schematic of molecular clouds in the Galactic Centre, in fig. 2 of their paper. Their 'top down view' model, shows the spatial positions of these clouds with respect to the 'infinity' shaped cloud structure, as found in Molinari et al. (2011). Their diagram shows the position of that structure as well as the molecular clouds we have analysed. As the clouds pass by the location of Sgr A*, we would expect it to have tidal effects on the molecular clouds (Kruijssen, Dale, & Longmore 2015 and references therein). The peak brightness emission images and the clear velocity gradient in our large sample of molecules indicates so.

A comparison of our datasets to models of the interstellar gas close to Sgr A* has yielded the following: our results are in agreement with the positions of the 20 and 50 km $s^{-1}$ clouds, as well as the molecular ridge from the model presented in Fig. 1. Whilst we also see evidence of the molecular ridge in many molecules, it is difficult to refer to this feature as independent from both clouds, as displayed in our peak brightness emission images and PV diagrams.

Therefore, we are proposing there is evidence to suggest that the molecular ridge and 20 km $s^{-1}$ cloud are part of one long extended cloud, which is possibly connected to the 50 km $s^{-1}$ cloud as well.

### 3.7 Tidal Shearing Stability Against Disruption

An equation to quantify the minimum density required of a cloud, such that it will withstand the tidal shearing force from Sgr A*, has been formulated by Tsuboi et al. (2011) and references therein. This was given by:

$$n_{\rm min}({\rm cm^{-3}}) \approx 2.5 \times 10^7 \left(\frac{1}{r({\rm pc})^3} + \frac{1}{r({\rm pc})^{1.8}}\right) \quad (7)$$

Note that the equation above already has the mass of Sgr A* factored into it; and the only missing variable is $r$, which is the distance between the cloud and the centre of our Galaxy. This equation is a slight variation from the original paper and is to be used as an order of magnitude test only (Tsuboi, M., private communication). Tsuboi et al. (2011) found $n_{\rm min}$ to range between $\approx 1.2 \times 10^5$ – 2.5 $\times 10^6$ cm$^{-3}$, when substituting a projected distance of r $\approx$ 4 – 20 pc for the 20 km $s^{-1}$ cloud into equation (7), inferring that cloud may be disrupted by tidal shear. However, the work of Kruijssen, Dale, & Longmore (2015), have determined the distance of the 20 km $s^{-1}$ cloud to the Galactic Centre, to be $67^{+67}_{-20}$ pc and the 50 km $s^{-1}$ cloud as $62^{+67}_{-20}$ pc. With these new values, the range of $n_{\rm min}$ for the 20 km $s^{-1}$ cloud is approximately $2.5 \times 10^4$ cm$^{-3}$ (at 47 pc) to 3.7 $\times 10^3$ cm$^{-3}$ (at 134 pc). Therefore, we propose it is reasonable to suggest that if $n(H_2) < n_{\rm min}$ for the 20 km $s^{-1}$ cloud, as defined by region 1, then we can expect that this cloud is being tidally sheared by the gravitational potential, arising from the presence of Sgr A*.

To demonstrate this, we must first evaluate the average density for each molecule, which is:

$$n({\rm molecule}) = \frac{\int N\, dA}{{\rm volume(cm^3)}} \quad (8)$$

The total number of molecules in our regions ($\int N\, dA$) was calculated by multiplying our value of $N$ (cm$^{-2}$) by the area of each cloud. The two areas, corresponding to regions 1 (20 km $s^{-1}$ cloud) and 2 (50 km $s^{-1}$ cloud), are $\approx 1.23 \times 10^{39}$ cm$^2$ and $\approx 3.63 \times 10^{38}$ cm$^2$ respectively. The volume of the cloud was deduced, by first establishing the deconvolved full-width half maxima major (x) and minor (y) axes. These were computed by fitting a 2D Gaussian, with the MIRIAD task `imfit` to a moment 0 (integrated emission) image. The 7-mm $HC_3N$ (5–4) molecular transition provided the best fit to the 20 km $s^{-1}$ cloud, in region 1. The radius of the z axis is equal to:

$$0.5 \times \sqrt{x \times y} \quad (9)$$

This (equation 9) effectively allows us to estimate the *physical depth* of the cloud. The values of the deconvolved axes in x, y and z are $\approx$ 358, 137 and 222 (arcseconds) respectively. These axes can be converted to centimetres using the Reid et al. (2009) approximation of 1′= 2.3 pc; and 1 pc $\approx 3.09 \times 10^{18}$ cm. The best volume formula to use, for the 20 km $s^{-1}$ cloud, is that of an ellipsoid (the structure of which can be easily recognized in Fig. 1). This is given by: $\frac{4}{3}\pi abc$. Where a, b and c are the radii of the deconvolved x, y and z axes. This yields a volume of $9.47 \times 10^{57}$ cm$^3$.

The last step, was converting $n$(molecule) to $n(H_2)$. Assuming $\int N\, dA$, $N(H_2)$ and the volume are constant, for each region, then the average $H_2$ density, for all the molecules is:

$$n(H_2) = \frac{n({\rm molecule})}{X_R} \quad (10)$$

We found that $n(H_2) \approx 7.0 \times 10^3$ cm$^{-3}$. This is the average $H_2$ density across the entire cloud, as defined by the size of our boxed region (1) in Fig. 2(a). This value does not rule out the possibility that within this cloud, smaller 'substructures' e.g. dense cores, are likely to exist which can withstand the tidal shearing condition i.e. $n(H_2) > n_{\rm min}$. We can infer this from our peak temperature maps, as we can see that high density tracers such as CS (1–0) and HCN (1–0) are brighter in differing extents of the cloud, which are not enhanced in other molecules (see online supplementary material). To spatially resolve and quantify those dense cores, which are likely to be smaller than our beams, we need to investigate them with higher resolution data e.g. from the Atacama Large Millimeter/submillimeter Array. Indeed, Lu et al. 2015 detected five clumps with 17 associated dense cores (1.3-mm continuum emission) and 18 $H_2O$ masers within the 20 km $s^{-1}$ cloud, using high-resolution interferometric observations from the Submillimeter Array and Karl G. Jansky Very Large Array, with clean beams of 4.9″ × 2.8″ and 3.5″ × 2.4″ respectively. However, by calculating the mean density, we are treating the cloud as one body and by integrating over a wide area, we incorporate all the likely 'substructures', which consequently does not affect our result for $n(H_2)$. This value is lower than $n_{min}$, which at the closest distance of 47 pc has a value of $\approx 2.5 \times 10^4$ cm$^{-3}$. We have deduced that the closest distance between the 20 km s$^{-1}$ cloud and Galactic Centre, is the most likely scenario. The presence of an extended velocity gradient in this cloud, as shown in the PV diagrams is additional verification of this. Similarly, the peak brightness emission images of the 20 km s$^{-1}$ cloud have illustrated elongated emission; therefore demonstrating a good correlation to the PV diagrams. Furthermore, shock and high density tracers are enhanced in the 20 km s$^{-1}$ cloud, in comparison to the 50 km s$^{-1}$ cloud, as shown in the analysis of the relative abundances between both clouds in § 3.3 and the discussion of the line ratio maps of enhanced molecules in § 3.5; which can be explained due to the passage of this cloud close to Sgr A* (e.g. Kruijssen, Dale, & Longmore 2015). It is therefore reasonable to expect that the 20 km s$^{-1}$ cloud is not stable against tidal shear from Sgr A*.

In Fig. 1, a molecular feature known as the southern streamer, connects part of the 20 km s$^{-1}$ cloud to the circumnuclear ring. Strong evidence of this was found in $NH_3$ (1,1) and (2,2) transition maps of Coil & Ho (1999). The fine structure of the streamer cannot be easily resolved in our data, as Coil & Ho (1999) used a beam size of ≈ 14″ × 9″, which is smaller than the beams we used (see Table 1). This is complicated further when the close proximity of the streamer to Sgr A* is also taken into account. The mean density of the 20 km s$^{-1}$ cloud, as measured in $H_2$, renders the structure of the southern streamer irrelevant in conclusion regarding the tidal shearing.

## 4 CONCLUSIONS

The 20 and 50 km s$^{-1}$ molecular clouds, as well as a connecting feature known as the molecular ridge, located in our Galactic Centre close to Sgr A*, were systematically examined. This was achieved using molecular line data from the Mopra 22-metre and NANTEN2 4-metre radio telescopes. A statistical and visual analysis showed the following new observational results when comparing the physical and chemical properties of each cloud and their relationship between each other and the molecular ridge:

Both 3-mm $\log_{10}$ $^{13}$CS (2–1) and $\log_{10}$ C$^{18}$O (1–0) line ratio plots have shown that shock and high density tracers such as HNCO and $N_2H^+$ are brighter within the 20 km s$^{-1}$ cloud (region 1). The 7-mm $\log_{10}$ $^{13}$CS (1–0) line ratio plot also indicates this, with many molecules being brighter in region 1. However, we also note that there are a large number of 3-mm molecules which have line ratios that are consistent between both regions and greater in the 50 km s$^{-1}$ cloud (region 1).

A comparison between the peak positions (regions 'a' and 'c') with respect to both the 20 and 50 km s$^{-1}$ (regions 1 and 2) in the $\log_{10}$ $^{13}$CS line ratios for the 3 and 7-mm molecules, found that while the majority of molecules are preferentially larger in the peak positions of the 20 and 50 km s$^{-1}$ clouds, we have also shown that a few molecules such as SiO and HNCO, which probe the shock and density conditions in the clouds, have in fact very similar relative abundances between the entire cloud and peak emission in the 20 km s$^{-1}$ cloud and that the CS molecule is brighter across region 1 than in the peak postion. A similar trend is observed in the 50 km s$^{-1}$, where we find a few molecules such as $N_2H^+$ and $CH_3CCH$ (5–4) are consistent between the peak position and the entire cloud. Similarly, the CS molecule is brighter across the entirety of the 50 km s$^{-1}$ than in the peak position. Concordantly, we have suggested that there is a good possibility that there are only small variations in the chemistry between both the 20 and 50 km s$^{-1}$ clouds.

The line ratio maps produced when dividing shock and high density molecules by the $^{13}$CS optically thin tracer, has shown there is good evidence that the 20 km s$^{-1}$ cloud is enhanced in those types of molecular tracers, throughout a widespread region in the aforementioned cloud. A comparison between the peak emission and moment 0 images, indicates the same trend and also shows that the entire emission line for those molecules is enhanced in the 20 km s$^{-1}$ cloud. While the 50 km s$^{-1}$ cloud does indeed also show enhancement (in particular, with 7-mm SiO 1–0); the degree of enhancement for molecules in that cloud is far less than that detected within the 20 km s$^{-1}$ cloud.

The PV diagrams have revealed a good consistency across both the 20 and 50 km s$^{-1}$ clouds, in all three slices (A, B and C), for each of the 6 bright molecules presented. As a result, we have inferred that the gas in both clouds appear to be well mixed.

Additionally, we have examined the peak brightness emission temperature maps and corresponding position velocity diagrams across both clouds. We inferred from our observations, that a feature known as the molecular ridge, which is considered to be separate to both the 20 and 50 km s$^{-1}$ cloud, appears to be combined with the 20 km s$^{-1}$ cloud and may be connected to the 50 km s$^{-1}$ cloud as well.

Furthermore, a 'western peak', present within the 20 km s$^{-1}$ cloud in the NANTEN2 CO (4–3) image and only visible in four transitions from the 3 and 7-mm datasets was discussed and concluded to be a cold dense and not very optically thick core.

Finally, we have assessed the possibility that the 20 km s$^{-1}$ cloud is being tidally sheared by Sagittarius A*. The mean hydrogen density of this cloud, $n(H_2)$, was compared and found to be lower than $n_{min}$, the minimum density required to withstand tidal shearing from Sgr A*. In conjunction with the elongated structure and wide velocity gradient of this cloud, as well as the enhancement in both shock and density tracers, the analysis we have carried out indicates it is plausible that this cloud, as one body, is not stable against tidal shear from Sgr A*.

## ACKNOWLEDGEMENTS

This research has made use of NASA's Astrophysics Data System (ADS). In addition, the SIMBAD database, operated at CDS, Strasbourg, France and the NED IPAC database have been utilized as well.

Jonathan P. Marshall is supported by a UNSW Vice-Chancellor's Postdoctoral Research Fellowship. Leonardo Bronfman acknowledges support from CONICYT Project PFB-06 and FONDECYT project 1120195. Nadia Lo's postdoctoral fellowship is supported by a CONICYT/FONDECYT postdoctorado, under project no. 3130540.

The data was obtained using the Mopra radio telescope, a part of the Australia Telescope National Facility which is funded by the Commonwealth of Australia for operation as a National Facility managed by CSIRO. The University of New South Wales (UNSW) digital filter bank (the UNSW-MOPS) used for the observations with Mopra was provided with support from the Australian Research

Council (ARC), UNSW, Sydney and Monash Universities, as well as the CSIRO.

We thank Steven N. Longmore, J. M. Diederik Kruijssen, Timothy A. Davis, Thomas P. Robitaille, Adam Ginsburg and Lisa Harvey-Smith for helpful discussions. Additionally, we are grateful to Katia Ferrière for allowing us to reproduce her diagram. We would also like to thank the referee Paul T.P. Ho, for comments and suggestions that have greatly improved this manuscript.

This paper has been typeset from a TEX/LATEX file prepared by the author.

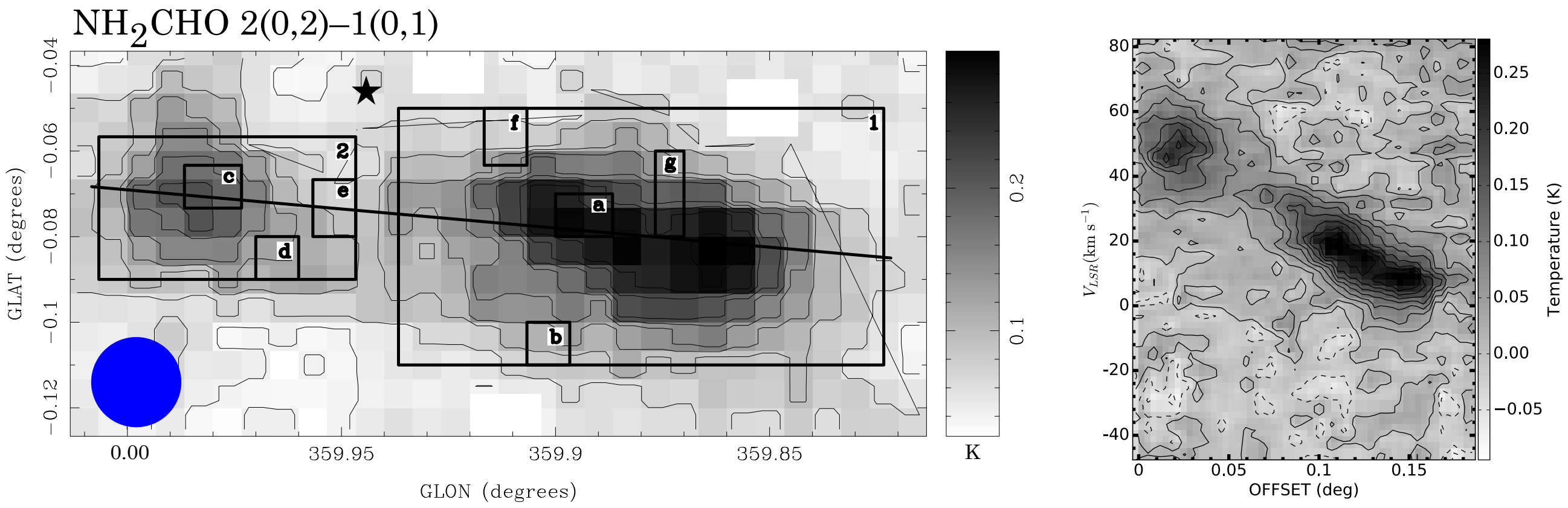

(a) (b)

**Figure 1:** (a) 7-mm $NH_2CHO$ 2(0,2)–1(0,1) peak emission image. Contour levels: 0.04, 0.07, 0.09, 0.12, 0.15, 0.17, 0.20, 0.23, 0.26, 0.28 (K). (b) 7-mm $NH_2CHO$ 2(0,2)–1(0,1) PV diagram.

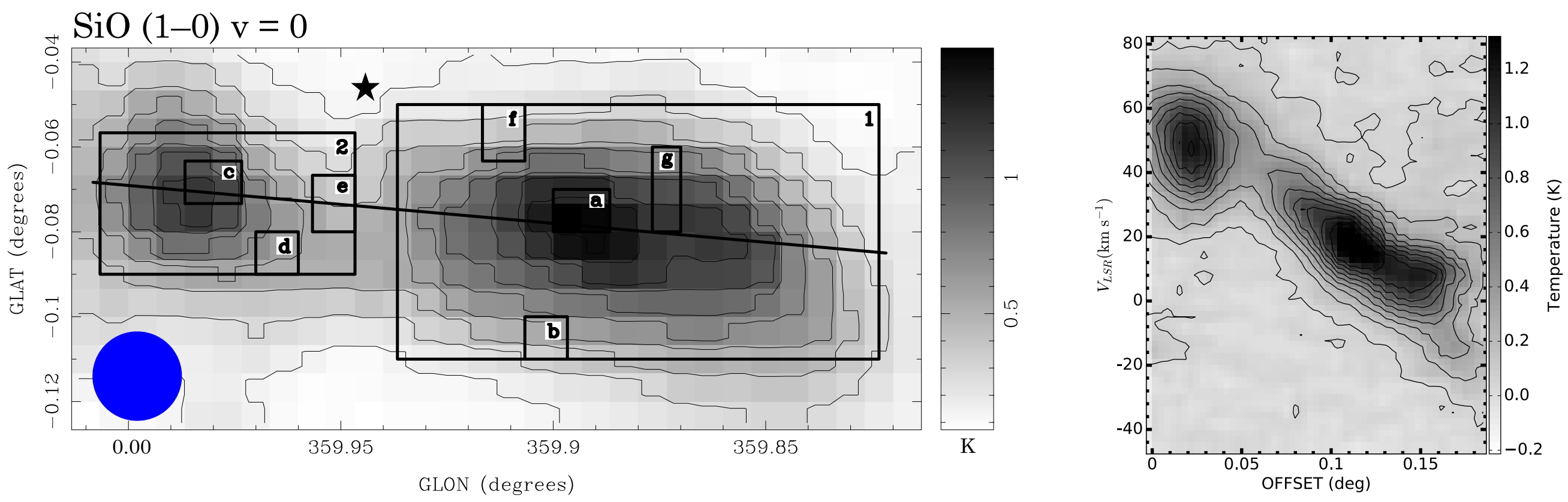

(a) (b)

**Figure 2:** (a) 7-mm SiO (1–0) v = 0 peak emission image. Contour levels: 0.14, 0.28, 0.42, 0.56, 0.70, 0.85, 0.99, 1.13, 1.27, 1.41 (K). (b) 7-mm SiO (1–0) v = 0 PV diagram.

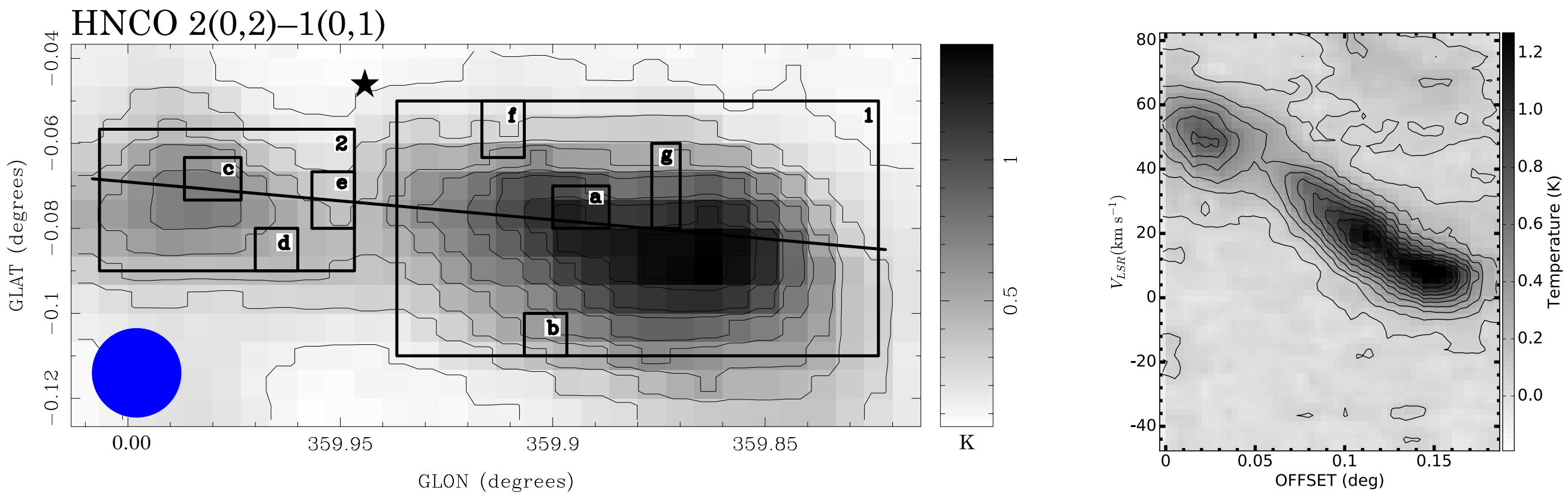

(a) (b)

**Figure 3:** (a) 7-mm HNCO 2(0,2)–1(0,1) peak emission image. Contour levels: 0.12, 0.26, 0.39, 0.53, 0.66, 0.80, 0.94, 1.07, 1.21, 1.34 (K). (b) 7-mm HNCO 2(0,2)–1(0,1) PV diagram.

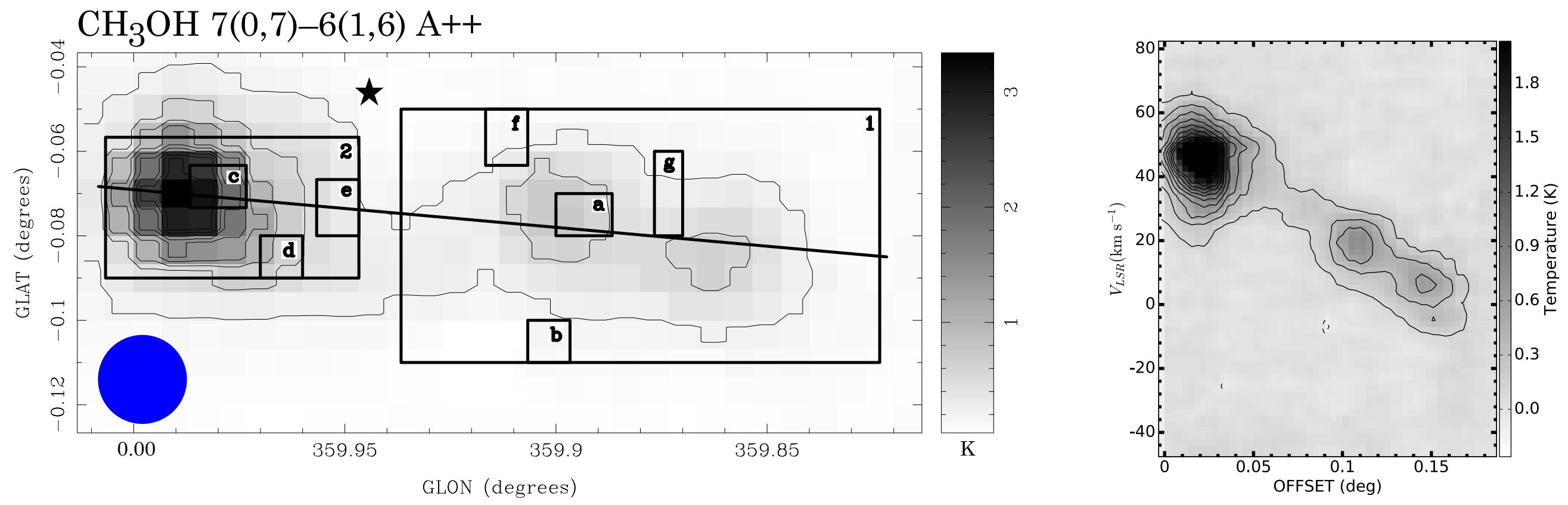

(a) (b)

**Figure 4:** (a) 7-mm $CH_3OH$ 7(0,7)–6(1,6) A++ peak emission image. Contour levels: 0.21, 0.54, 0.87, 1.20, 1.53, 1.86, 2.19, 2.52, 2.85, 3.18 (K). (b) 7-mm $CH_3OH$ 7(0,7)–6(0,6) A++ PV diagram.

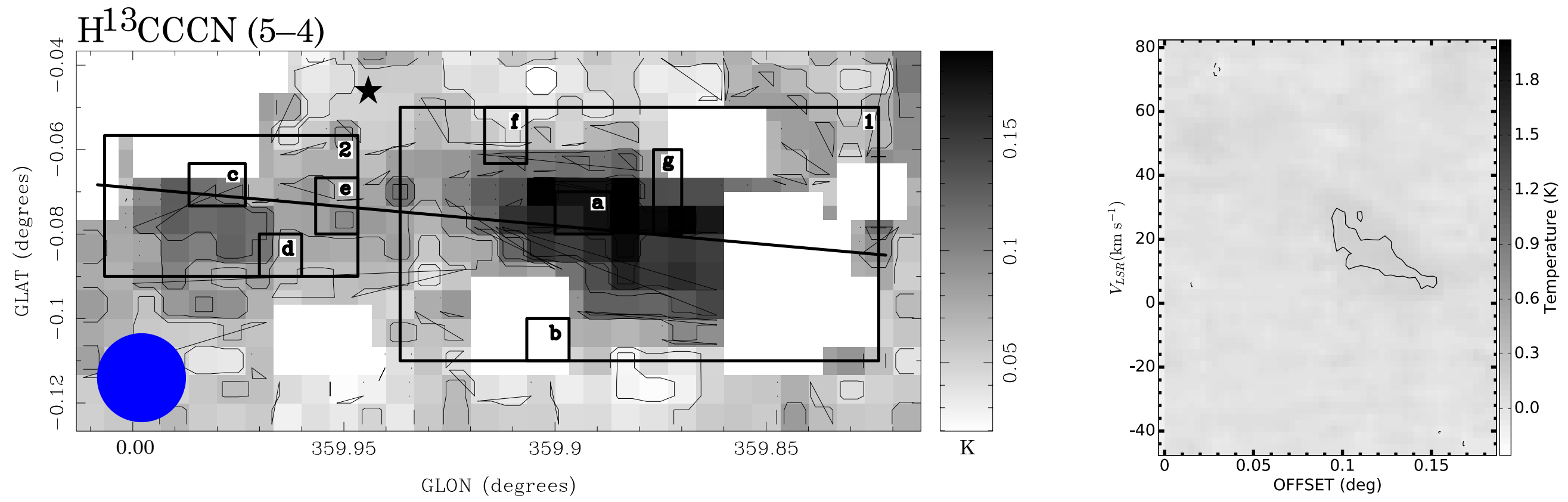

(a) (b)

**Figure 5:** (a) 7-mm $H^{13}CCCN$ (5–4) peak emission image. Contour levels: 0.03, 0.04, 0.06, 0.08, 0.10, 0.11, 0.13, 0.15, 0.16, 0.18 (K). (b) 7-mm $H^{13}CCCN$ (5–4) PV diagram.

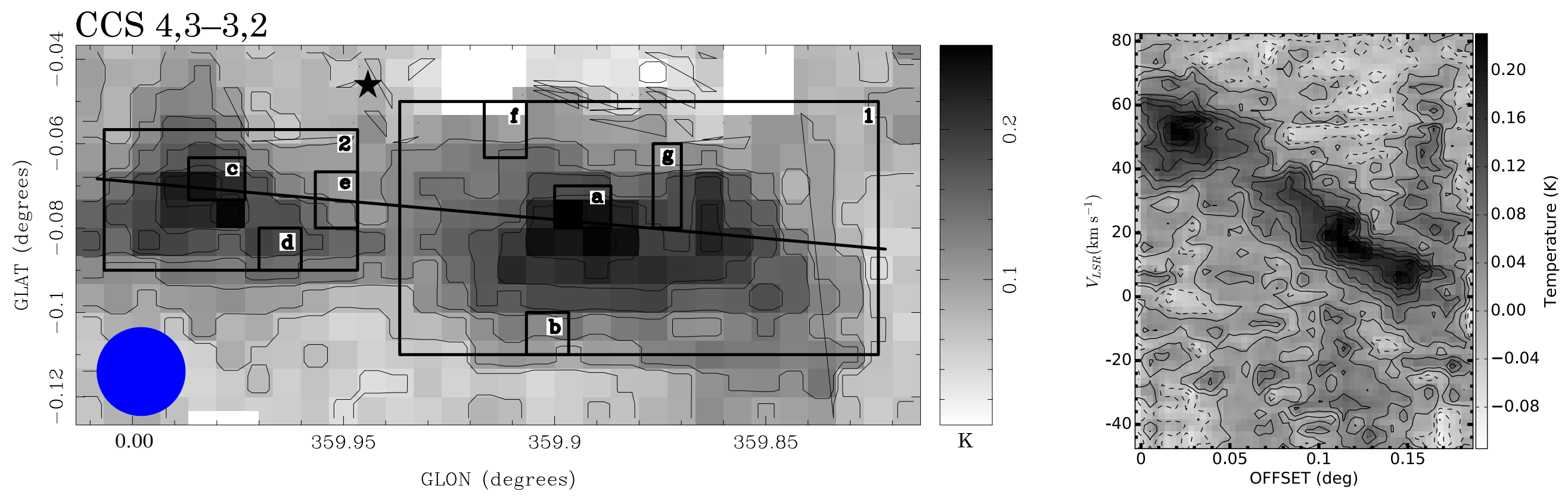

(a) (b)

**Figure 6:** (a) 7-mm CCS 4,3–3,2 peak emission image. Contour levels: 0.02, 0.04, 0.07, 0.09, 0.12, 0.14, 0.17, 0.19, 0.22, 0.24 (K). (b) 7-mm CCS 4,3–3,2 PV diagram.

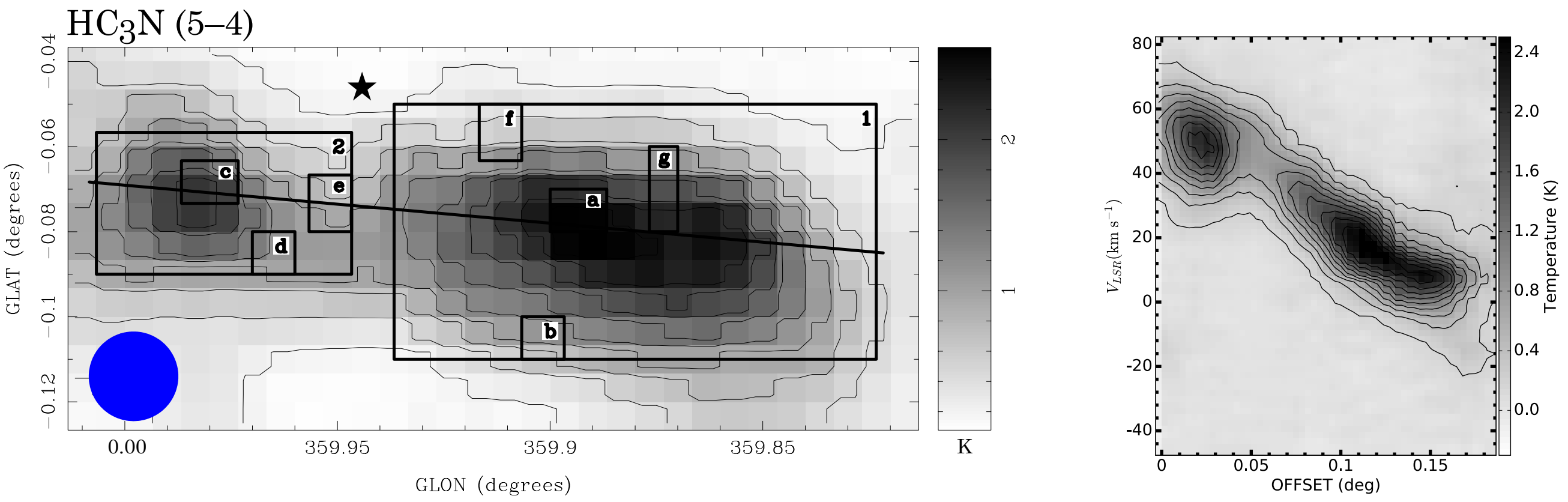

(a) (b)

**Figure 7:** (a) 7-mm $HC_3N$ (5–4) peak emission image. Contour levels: 0.20, 0.46, 0.71, 0.96, 1.22, 1.47, 1.72, 1.98, 2.23, 2.48 (K). (b) 7-mm $HC_3N$ (5–4) PV diagram.

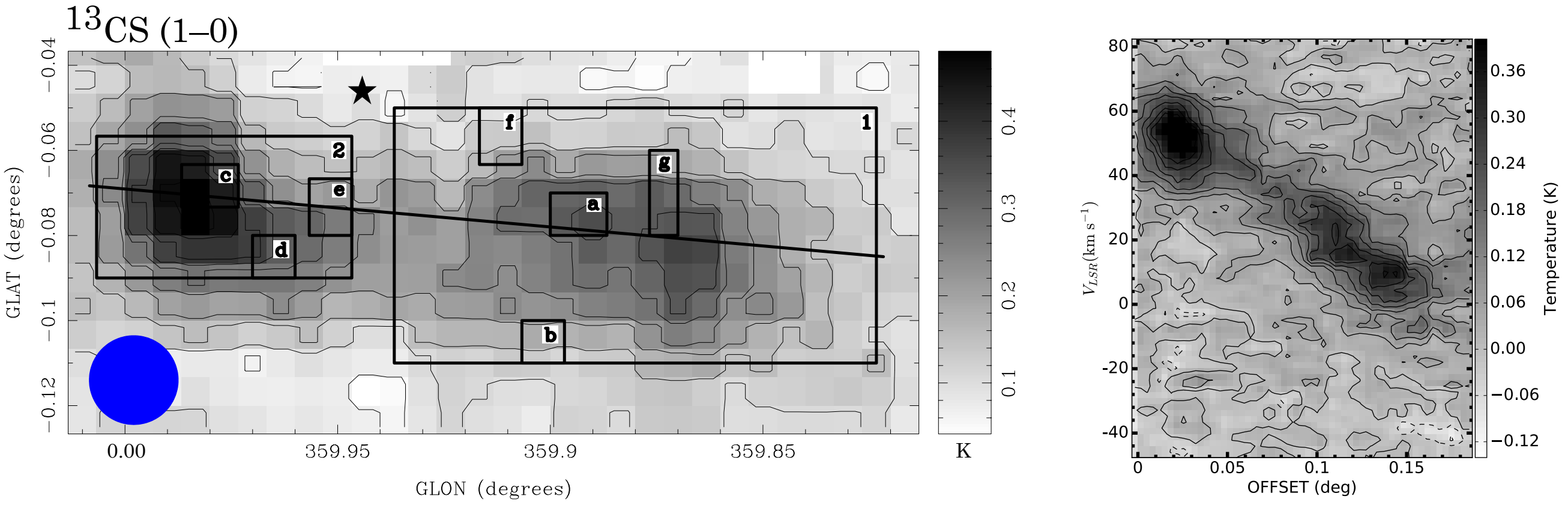

(a) (b)

**Figure 8:** (a) 7-mm $^{13}CS$ (1–0) peak emission image. Contour levels: 0.06, 0.11, 0.15, 0.20, 0.24, 0.28, 0.33, 0.37, 0.41, 0.46 (K). (b) 7-mm $^{13}CS$ (1–0) PV diagram.

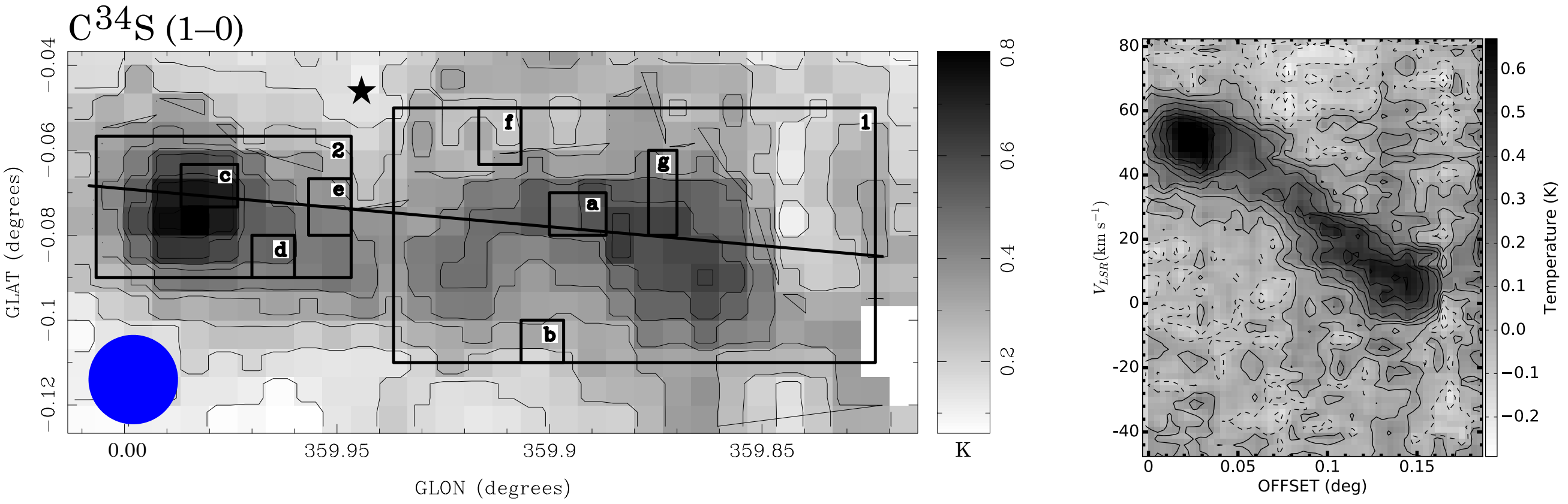

(a) (b)

**Figure 9:** (a) 7-mm $C^{34}S$ (1–0) peak emission image. Contour levels: 0.10, 0.17, 0.25, 0.32, 0.39, 0.47, 0.54, 0.62, 0.69, 0.77 (K). (b) 7-mm $C^{34}S$ (1–0) PV diagram.

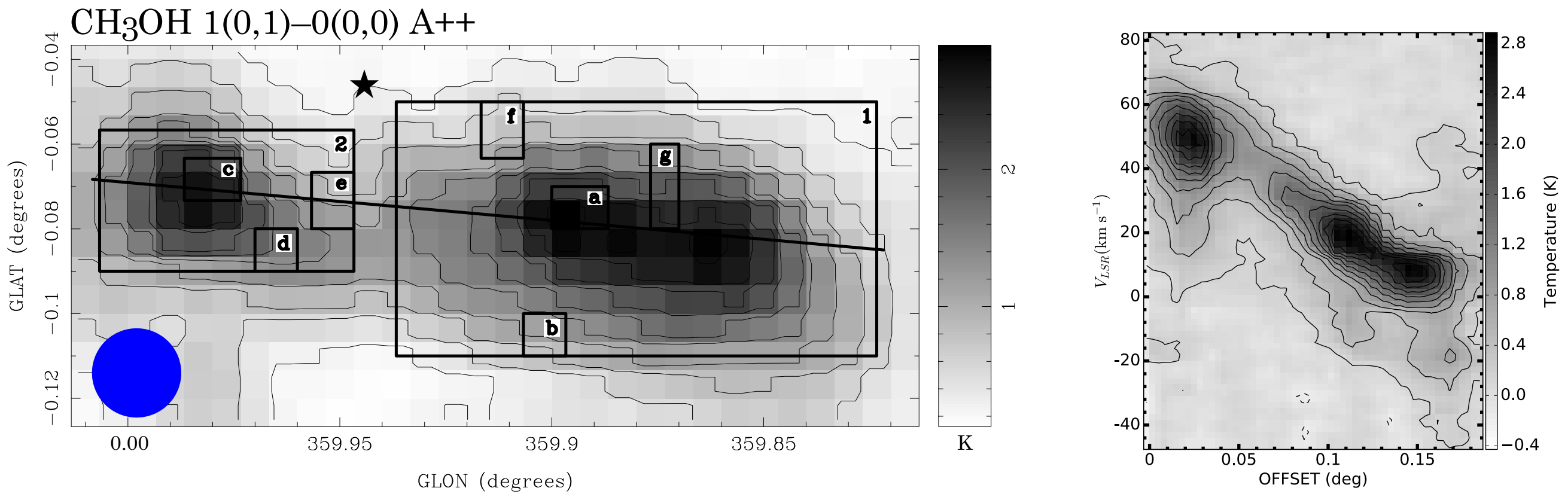

(a) (b)

**Figure 10:** (a) 7-mm $CH_3OH$ 1(0,1)–0(0,0) A++ peak emission image. Contour levels: 0.26, 0.54, 0.82, 1.10, 1.38, 1.65, 1.93, 2.21, 2.49, 2.77 (K). (b) 7-mm $CH_3OH$ 1(0,1)–0(0,0) A++ PV diagram.

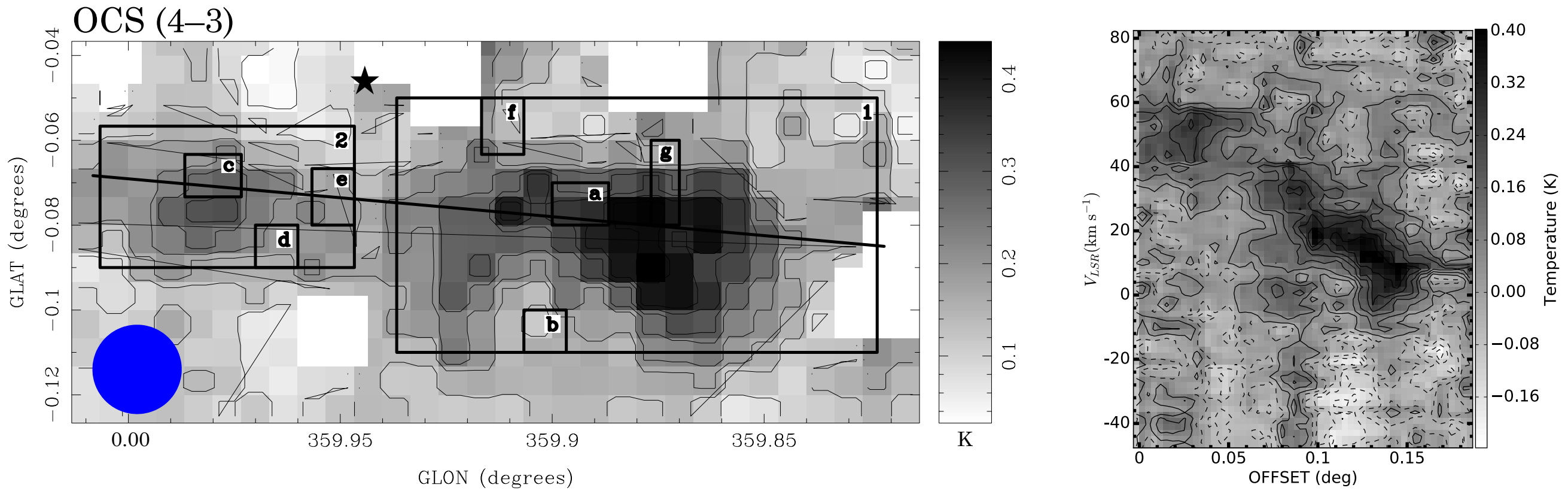

(a) (b)

**Figure 11:** (a) 7-mm OCS (4–3) peak emission image. Contour levels: 0.05, 0.09, 0.13, 0.17, 0.21, 0.25, 0.30, 0.34, 0.38, 0.42 (K). (b) 7-mm OCS (4–3) PV diagram.

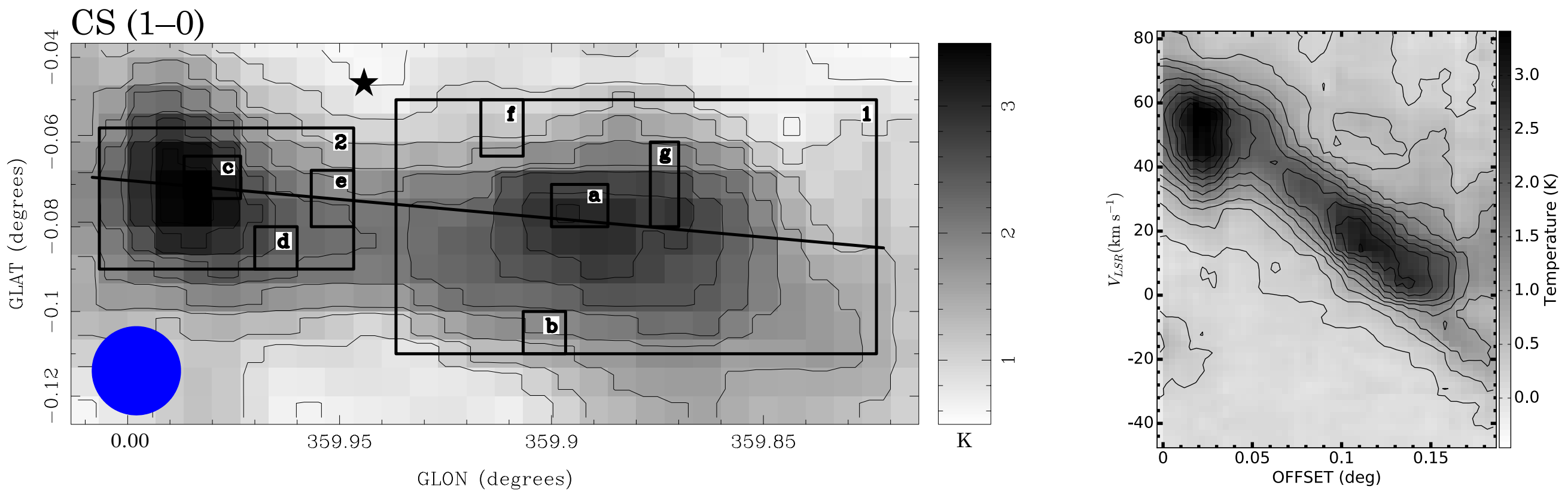

(a) (b)

**Figure 12:** (a) 7-mm CS (1–0) peak emission image. Contour levels: 0.64, 0.94, 1.24, 1.54, 1.84, 2.14, 2.44, 2.74, 3.04, 3.34 (K). (b) 7-mm CS (1–0) PV diagram.

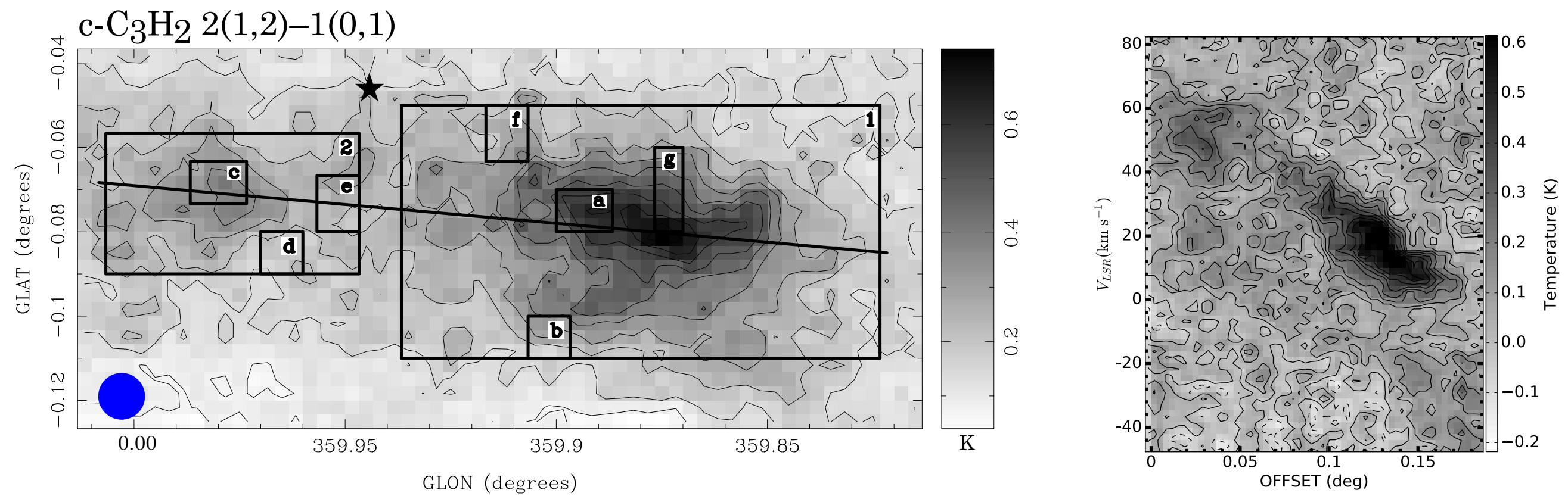

(a) (b)

**Figure 13:** (a) 3-mm c-$C_3H_2$ 2(1,2)–1(0,1) peak emission image. Contour levels: 0.07, 0.14, 0.21, 0.28, 0.35, 0.42, 0.49, 0.56, 0.63, 0.70 (K). (b) 3-mm c-$C_3H_2$ 2(1,2)–1(0,1) PV diagram.

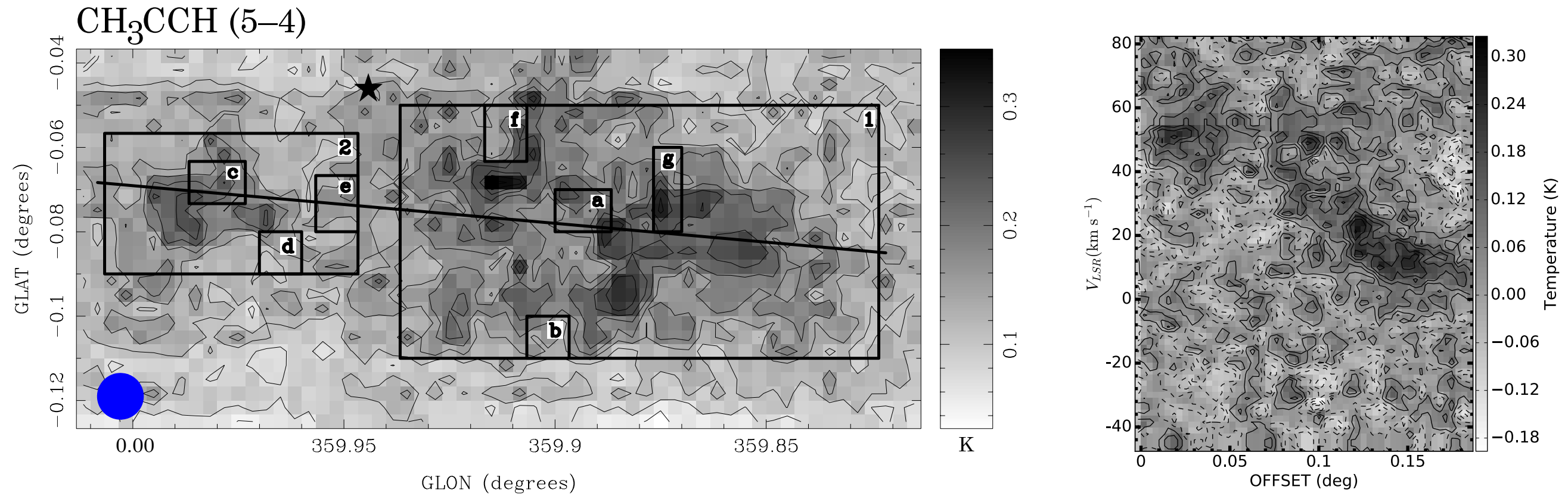

(a) (b)

**Figure 14:** (a) 3-mm $CH_3CCH$ (5–4) peak emission image. Contour levels: 0.05, 0.08, 0.11, 0.14, 0.17, 0.20, 0.24, 0.27, 0.30, 0.33 (K). (b) 3-mm $CH_3CCH$ (5–4) PV diagram.

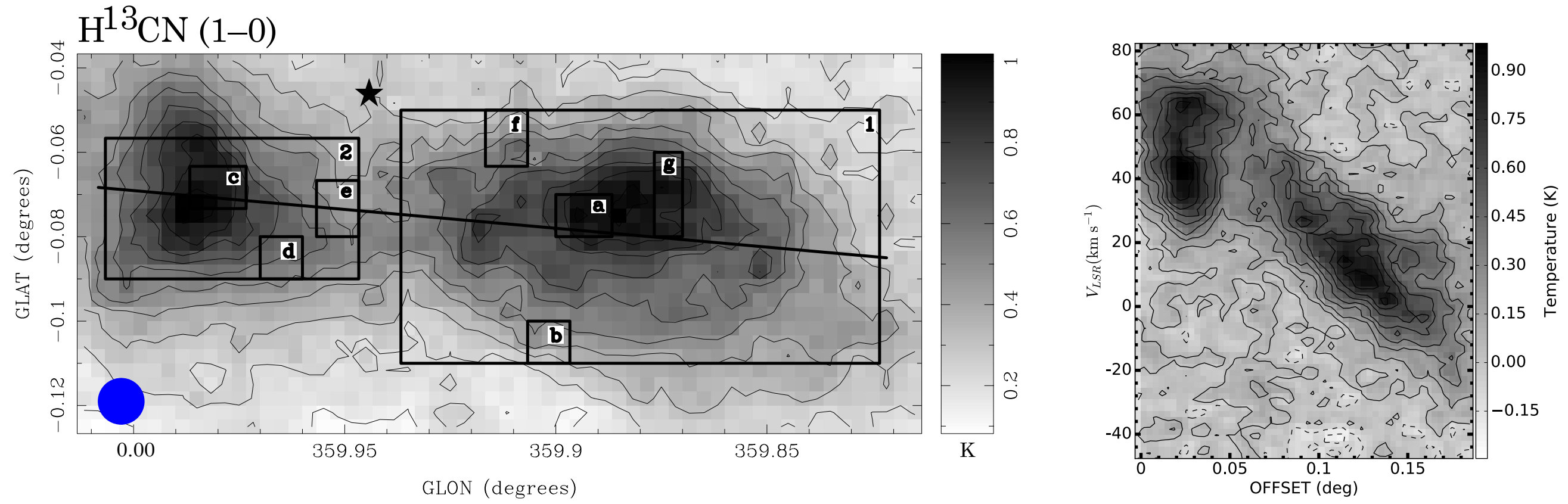

(a) (b)

**Figure 15:** (a) 3-mm $H^{13}CN$ (1–0) peak emission image. Contour levels: 0.13, 0.22, 0.32, 0.41, 0.50, 0.60, 0.69, 0.78, 0.88, 0.97 (K). (b) 3-mm $H^{13}CN$ (1–0) PV diagram.

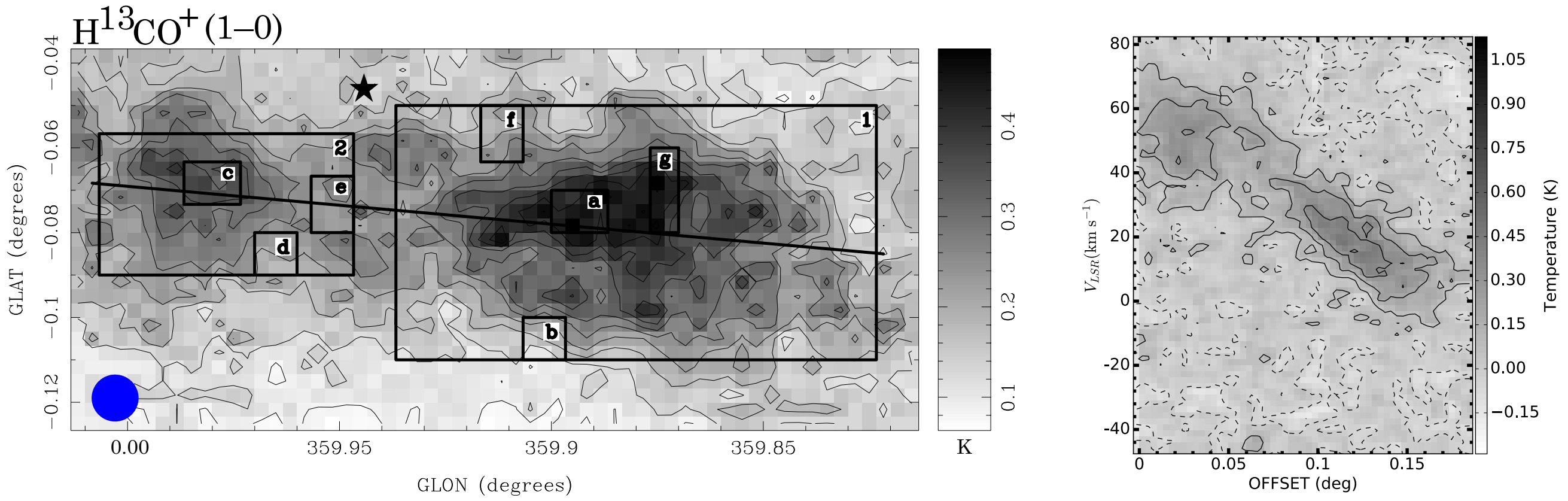

(a) (b)

**Figure 16:** (a) 3-mm $H^{13}CO^+$ (1–0) peak emission image. Contour levels: 0.08, 0.13, 0.17, 0.21, 0.25, 0.30, 0.34, 0.38, 0.42, 0.46 (K). (b) 3-mm $H^{13}CO^+$ (1–0) PV diagram.

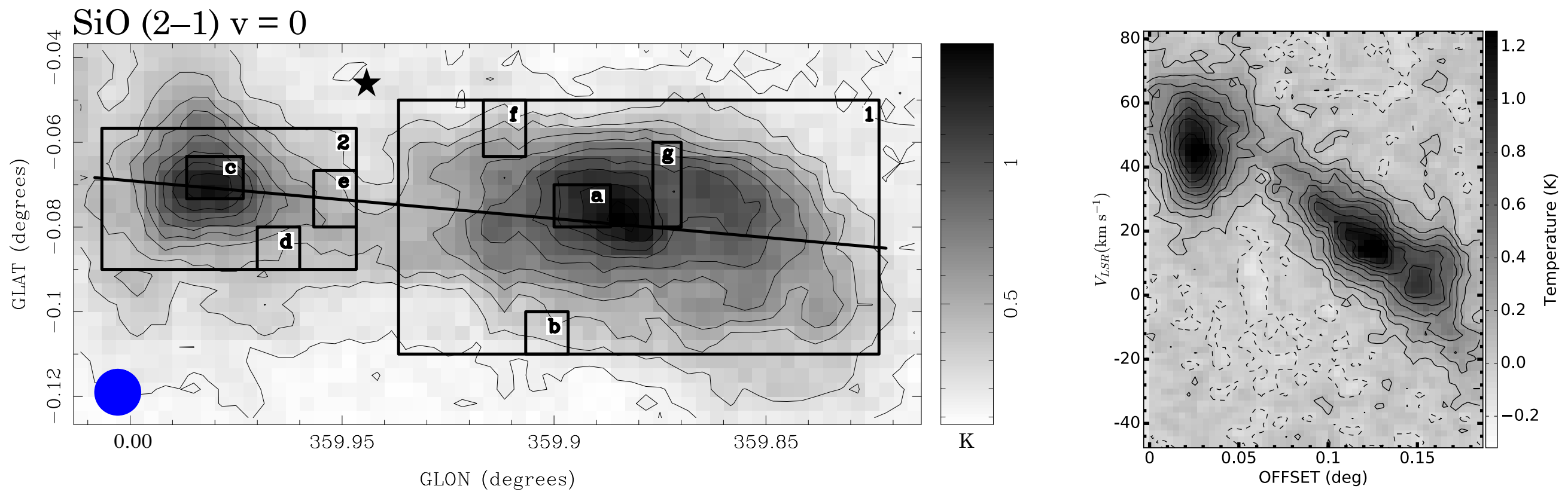

(a) (b)

**Figure 17:** (a) 3-mm SiO (2–1) v = 0 peak emission image. Contour levels: 0.14, 0.28, 0.41, 0.54, 0.68, 0.81, 0.95, 1.08, 1.22, 1.35 (K). (b) 3-mm SiO (2–1) v = 0 PV diagram.

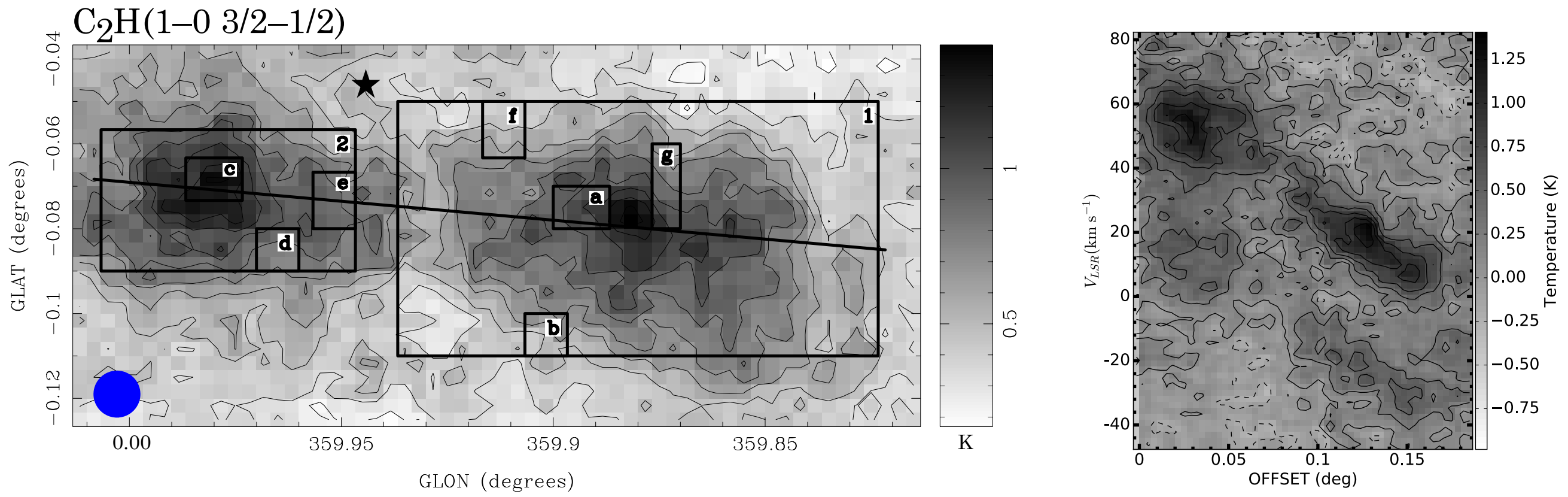

(a) (b)

**Figure 18:** (a) 3-mm $C_2H$ (1–0 3/2–1/2) peak emission image. Contour levels: 0.23, 0.35, 0.48, 0.60, 0.72, 0.85, 0.97, 1.09, 1.21, 1.34 (K). (b) 3-mm $C_2H$ (1–0 3/2–1/2) PV diagram.

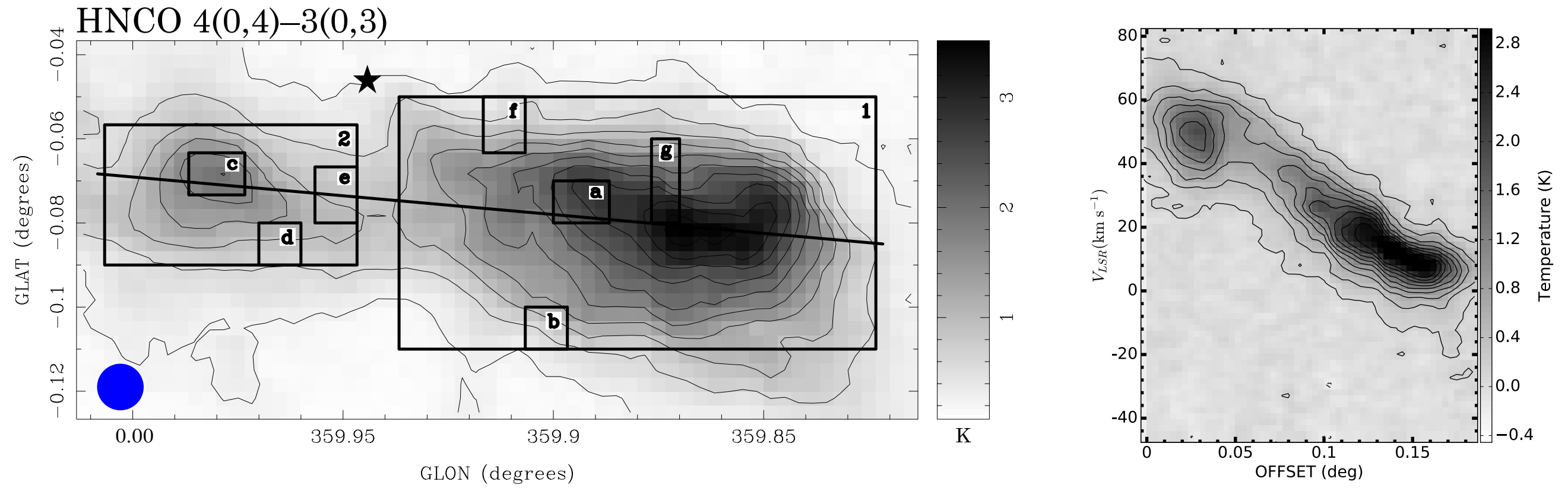

(a) (b)

**Figure 19:** (a) 3-mm HNCO 4(0,4)–3(0,3) peak emission image. Contour levels: 0.25, 0.59, 0.94, 1.28, 1.62, 1.97, 2.31, 2.66, 3.00, 3.34 (K). (b) 3-mm HNCO 4(0,4)–3(0,3) PV diagram.

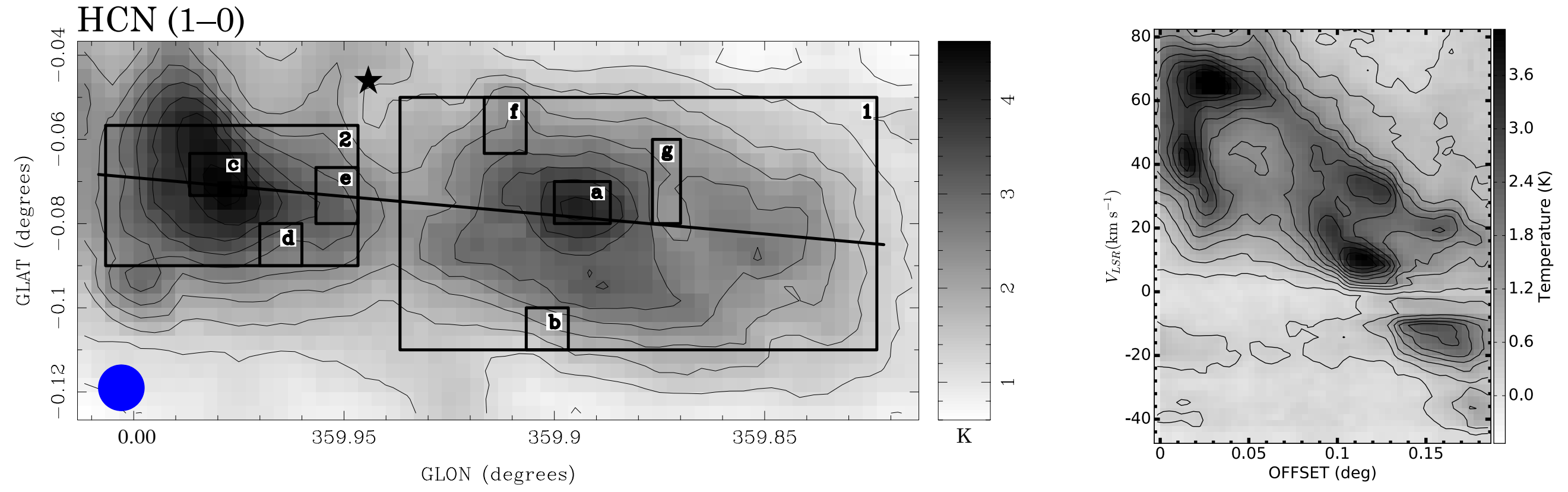

(a) (b)

**Figure 20:** (a) 3-mm HCN (1–0) peak emission image. Contour levels: 0.80, 1.20, 1.61, 2.01, 2.41, 2.81, 3.21, 3.62, 4.02, 4.42 (K). (b) 3-mm HCN (1-0) PV diagram.

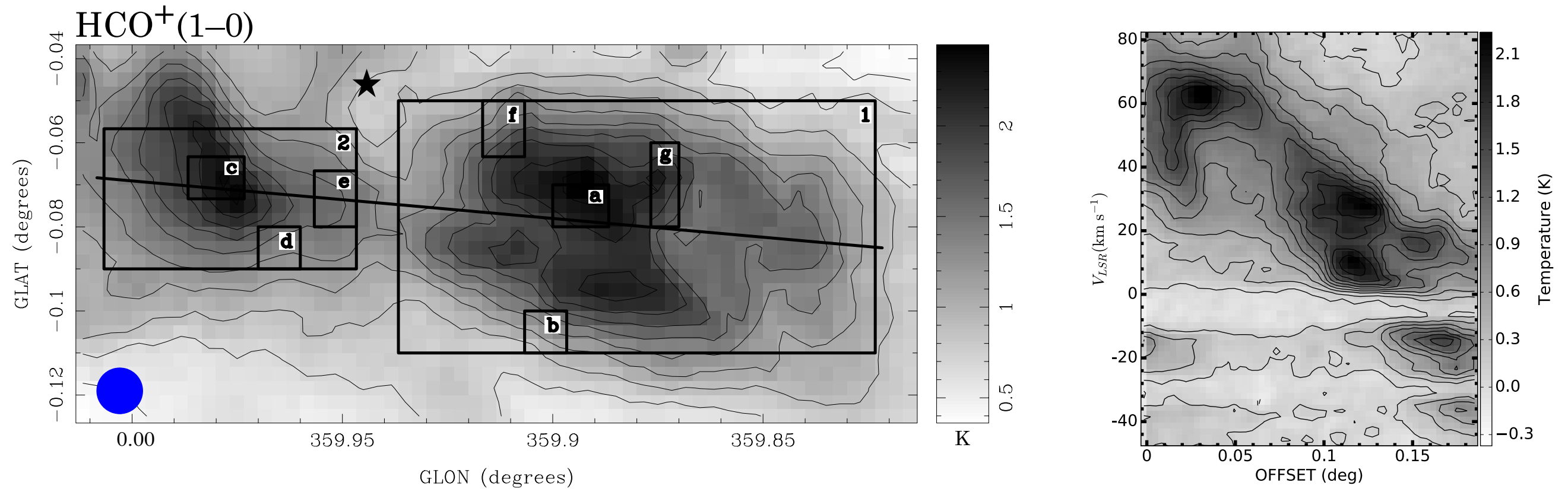

(a) (b)

**Figure 21:** (a) 3-mm $HCO^+$ (1–0) peak emission image. Contour levels: 0.46, 0.67, 0.88, 1.09, 1.30, 1.51, 1.72, 1.93, 2.14, 2.34 (K). (b) 3-mm $HCO^+$ (1–0) PV diagram.

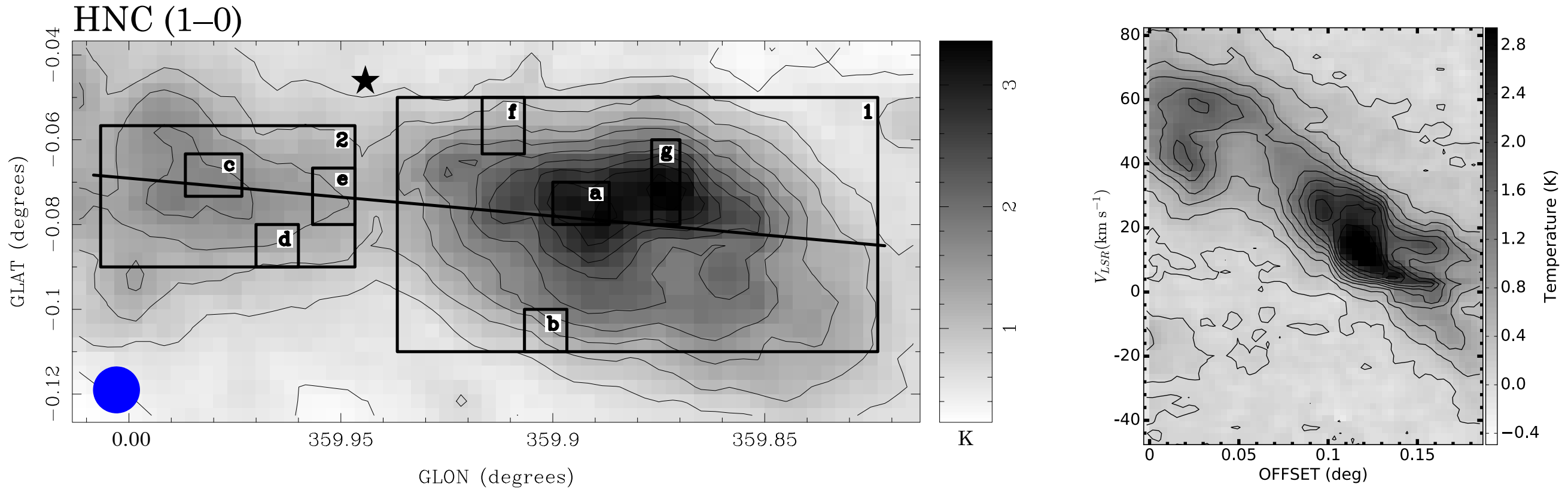

(a)

(b)

**Figure 22:** (a) 3-mm HNC (1–0) peak emission image. Contour levels: 0.39, 0.70, 1.01, 1.33, 1.64, 1.95, 2.27, 2.58, 2.89, 3.21 (K). (b) 3-mm HNC (1–0) PV diagram.

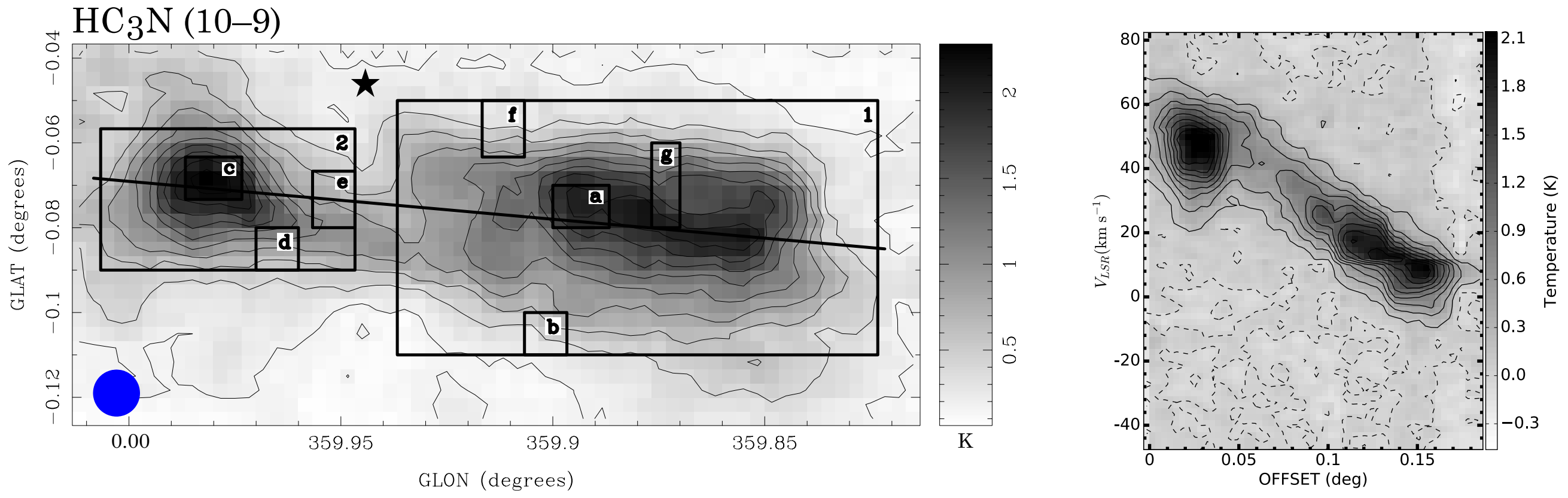

(a)

(b)

**Figure 23:** (a) 3-mm $HC_3N$ (10–9) peak emission image. Contour levels: 0.17, 0.39, 0.62, 0.84, 1.06, 1.28, 1.51, 1.73, 1.95, 2.17 (K). (b) 3-mm $HC_3N$ (10–9) PV diagram.

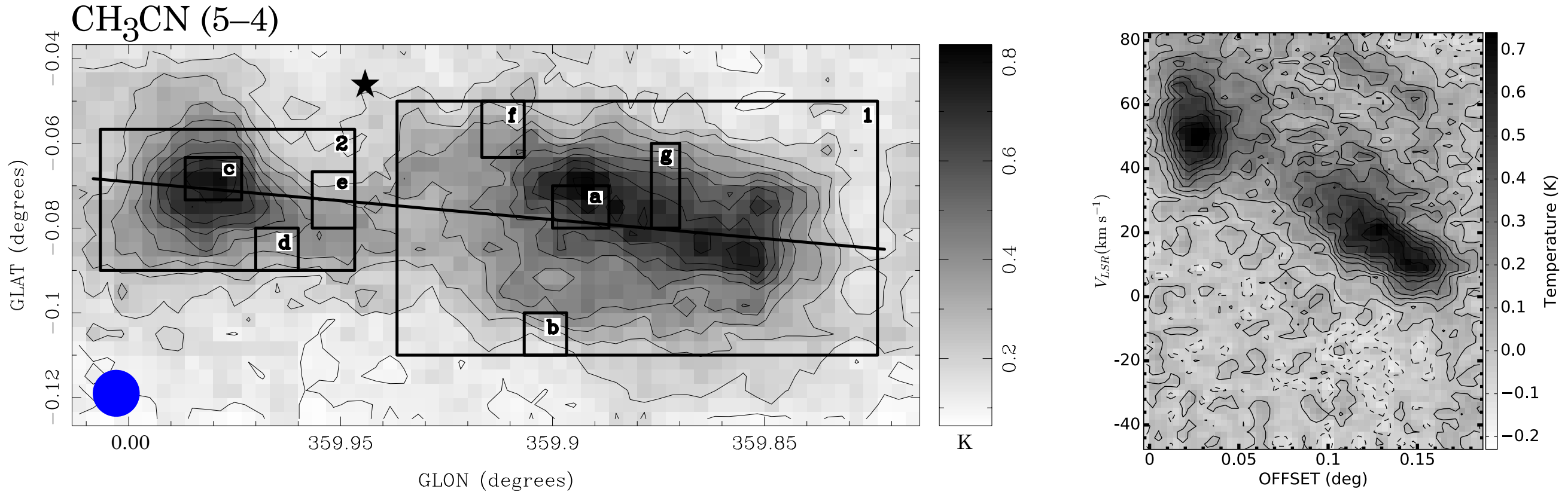

(a)

(b)

**Figure 24:** (a) 3-mm $CH_3CN$ (5–4) peak emission image. Contour levels: 0.10, 0.18, 0.26, 0.33, 0.41, 0.49, 0.57, 0.64, 0.72, 0.80 (K). (b) 3-mm $CH_3CN$ (5–4) PV diagram.

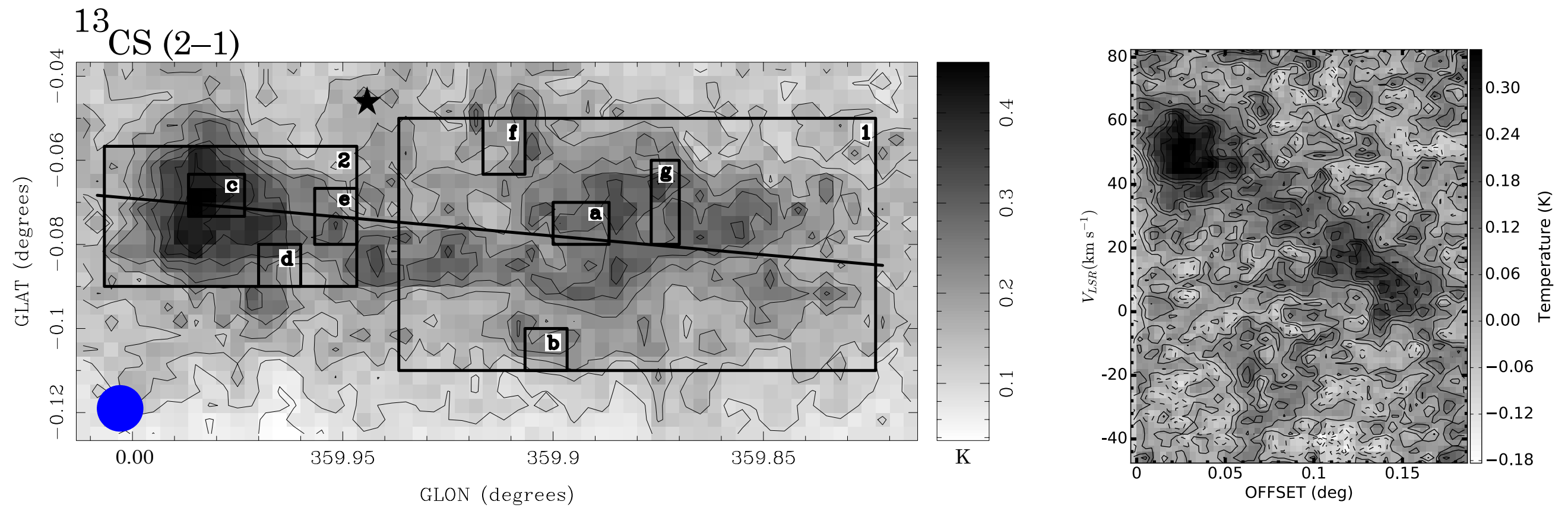

(a) (b)

**Figure 25:** (a) 3-mm $^{13}$CS (2–1) peak emission image. Contour levels: 0.06, 0.10, 0.14, 0.18, 0.23, 0.27, 0.31, 0.35, 0.39, 0.44 (K). (b) 3-mm $^{13}$CS (2–1) PV diagram.

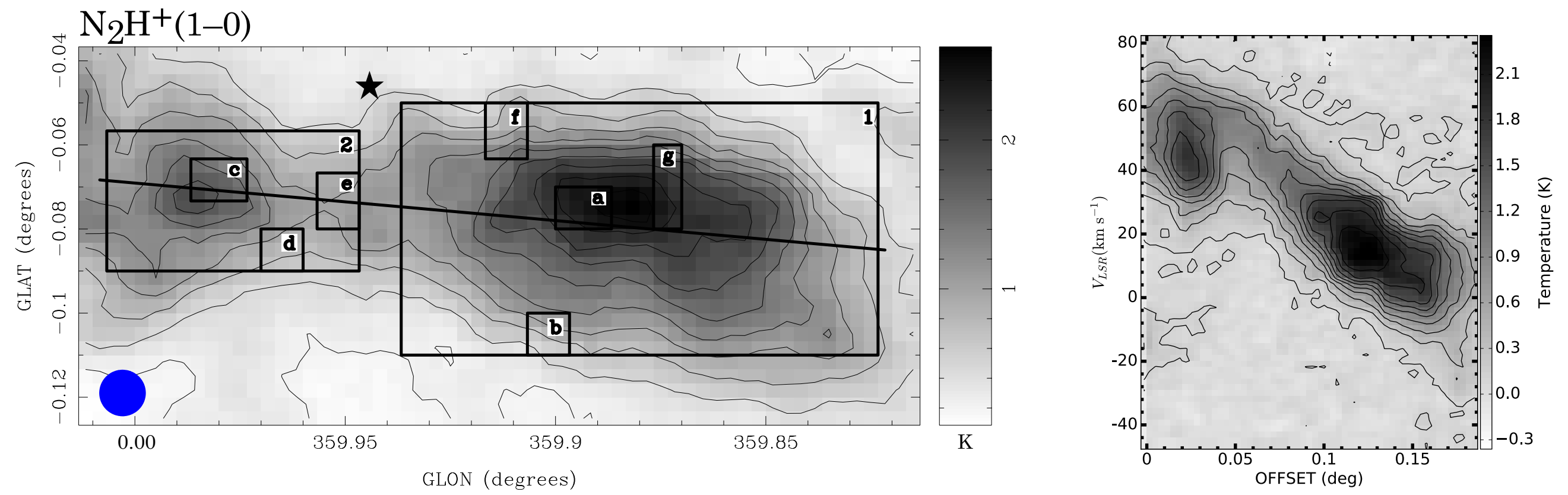

(a) (b)

**Figure 26:** (a) 3-mm $N_2H^+$ (1–0) peak emission image. Contour levels: 0.21, 0.47, 0.72, 0.97, 1.23, 1.48, 1.74, 1.99, 2.25, 2.50 (K). (b) 3-mm $N_2H^+$ (1–0) PV diagram.

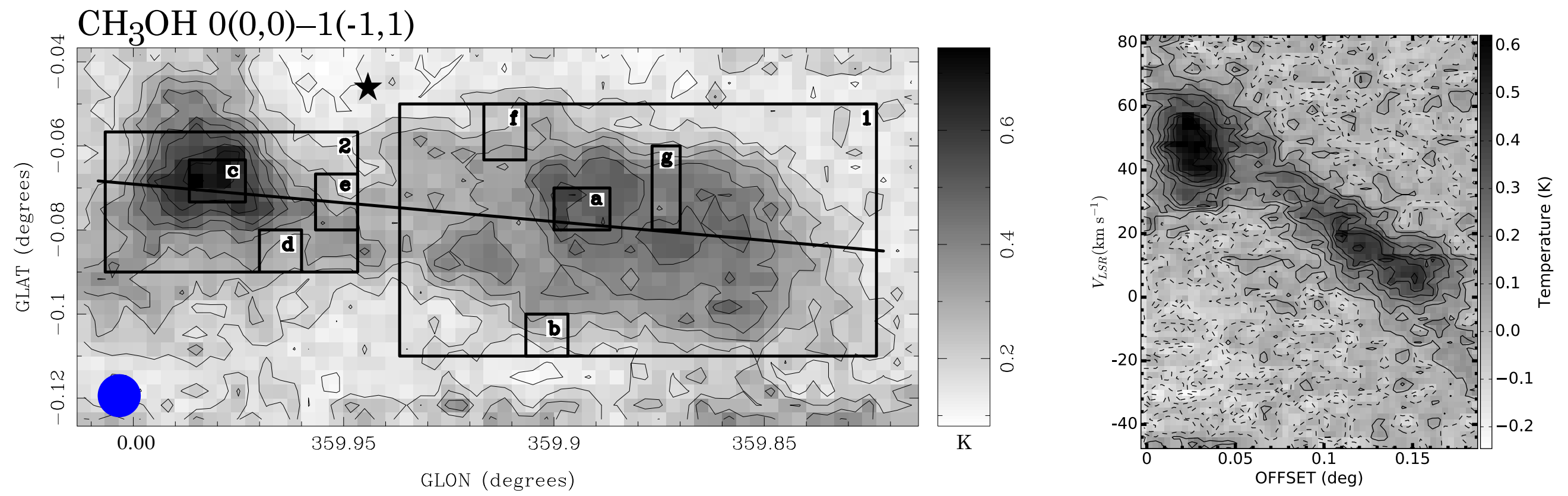

(a) (b)

**Figure 27:** (a) 3-mm $CH_3OH$ 0(0,0)–1(-1,1) peak emission image. Contour levels: 0.11, 0.18, 0.25, 0.31, 0.38, 0.45, 0.51, 0.58, 0.65, 0.71 (K). (b) 3-mm $CH_3OH$ 0(0,0)–1(-1,1) PV diagram.

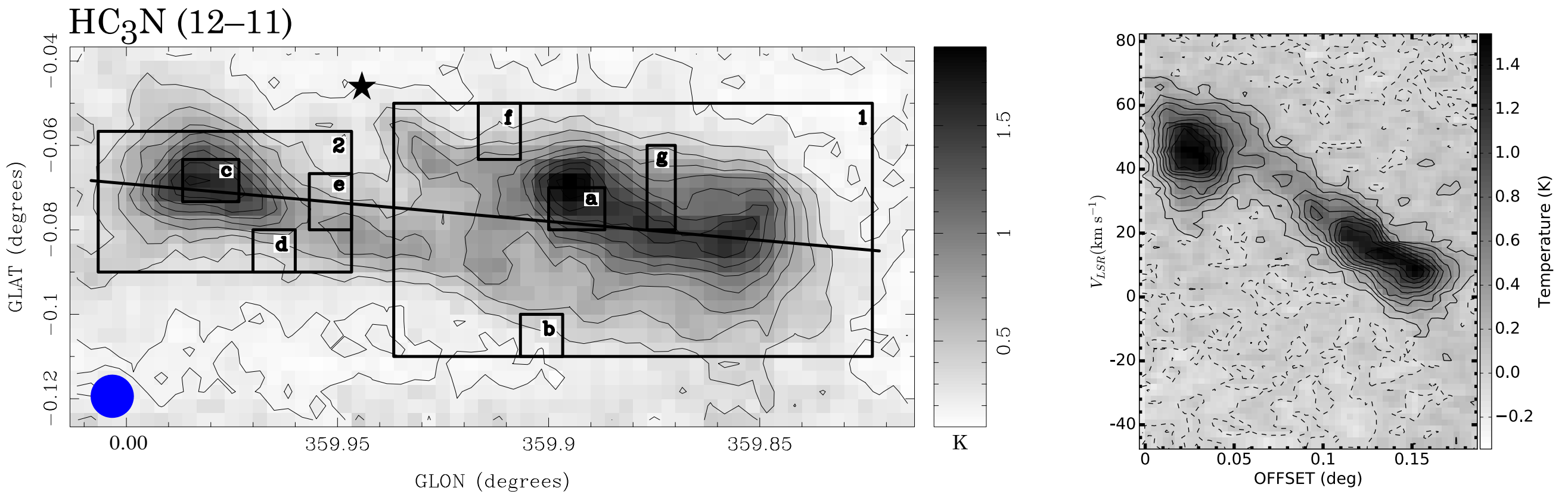

(a) (b)

**Figure 28:** (a) 3-mm $HC_3N$ (12–11) peak emission image. Contour levels: 0.19, 0.37, 0.54, 0.72, 0.90, 1.07, 1.25, 1.42, 1.60, 1.78 (K). (b) 3-mm $HC_3N$ (12–11) PV diagram.

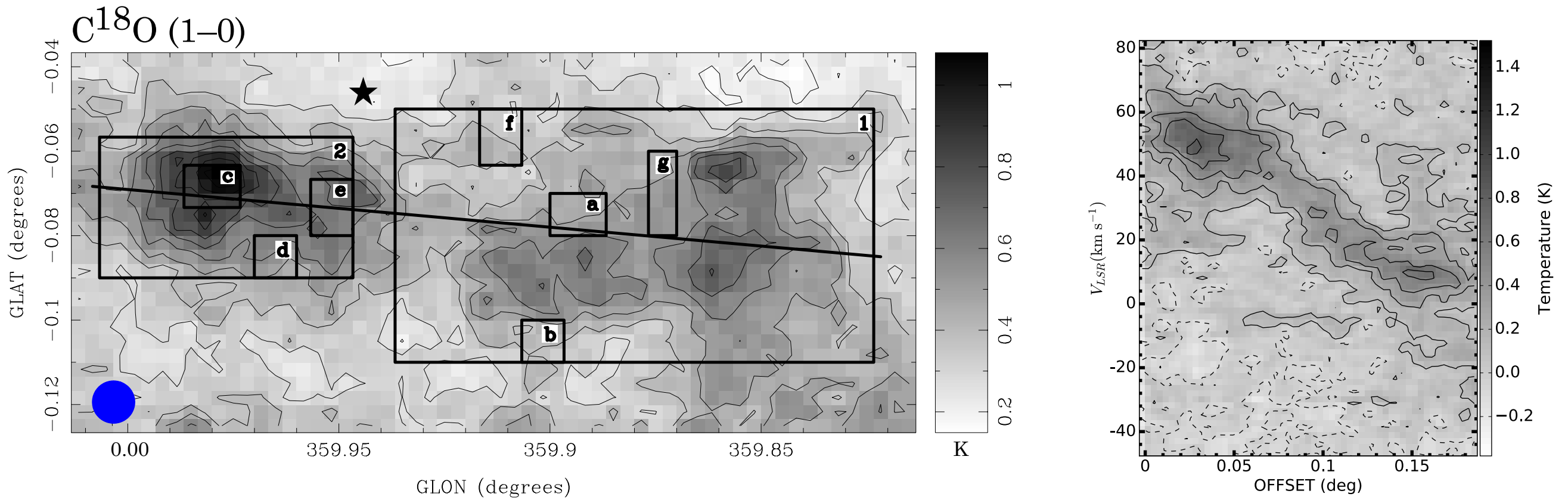

(a) (b)

**Figure 29:** (a) 3-mm $C^{18}O$ (1–0) peak emission image. Contour levels: 0.20, 0.29, 0.38, 0.47, 0.57, 0.66, 0.75, 0.85, 0.94, 1.03 (K). (b) 3-mm $C^{18}O$ (1–0) PV diagram.

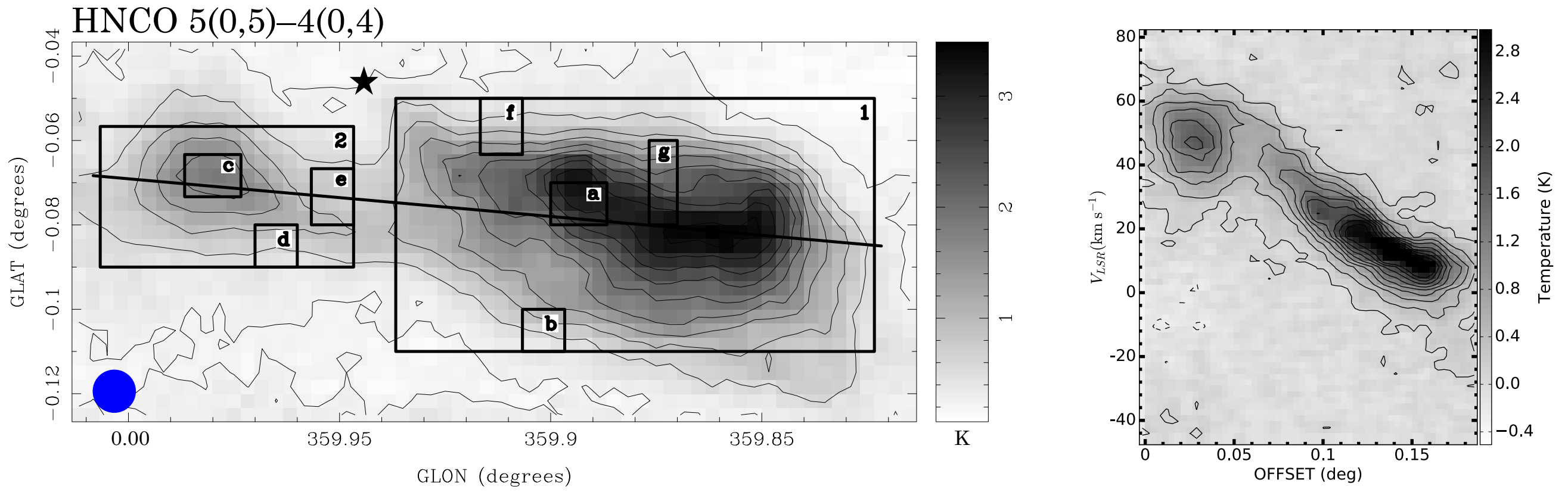

(a) (b)

**Figure 30:** (a) 3-mm HNCO 5(0,5)–4(0,4) peak emission image. Contour levels: 0.24, 0.58, 0.92, 1.26, 1.61, 1.95, 2.29, 2.63, 2.98, 3.32 (K). (b) 3-mm HNCO 5(0,5)–4(0,4) PV diagram.

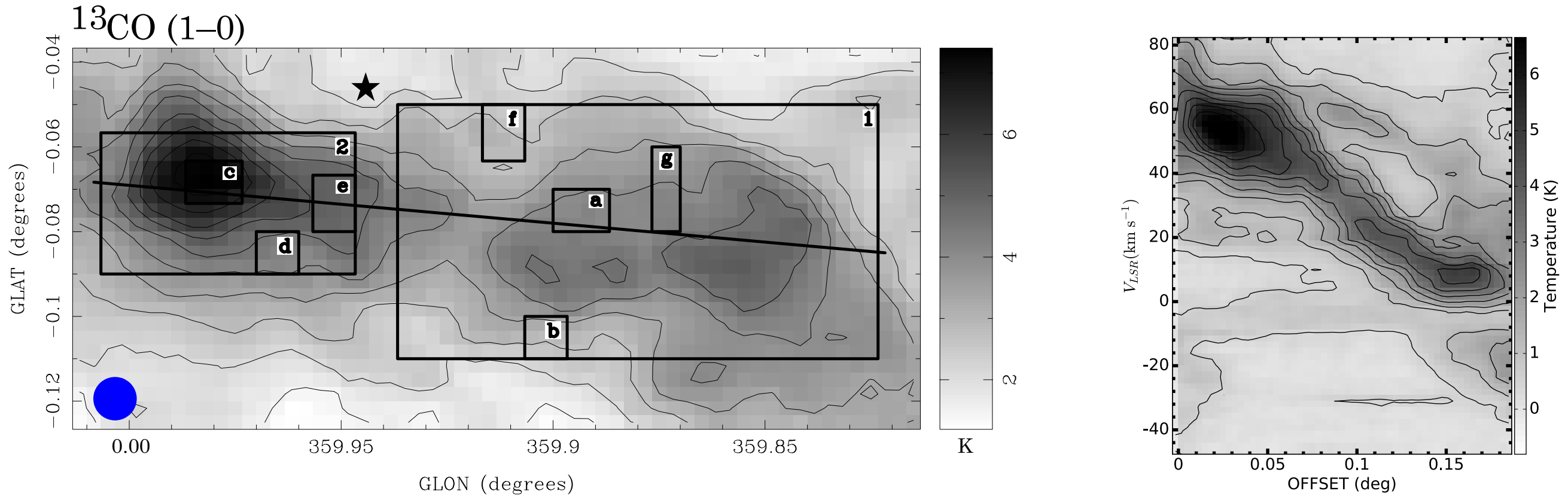

(a) (b)

**Figure 31:** (a) 3-mm $^{13}$CO (1–0) peak emission image. Contour levels: 1.50, 2.12, 2.75, 3.37, 3.99, 4.61, 5.23, 5.85, 6.48, 7.10 (K). (b) 3-mm $^{13}$CO (1–0) PV diagram.

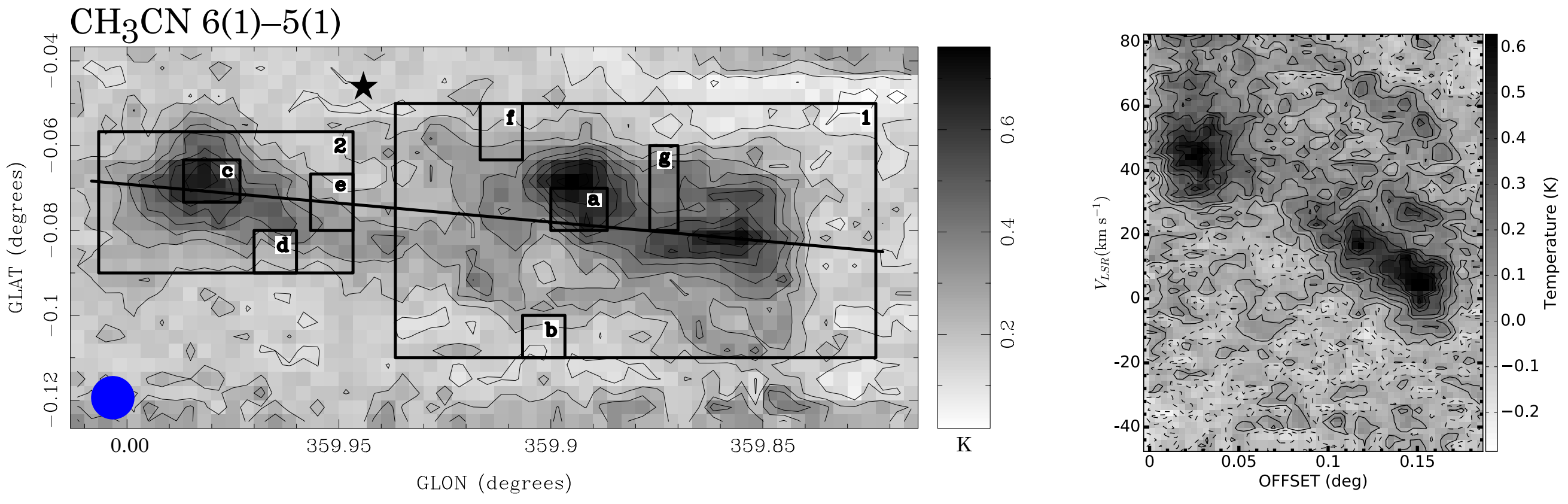

(a) (b)

**Figure 32:** (a) 3-mm $CH_3CN$ 6(1)–5(1) peak emission image. Contour levels: 0.05, 0.13, 0.20, 0.28, 0.35, 0.43, 0.50, 0.58, 0.65, 0.72 (K). (b) 3-mm $CH_3CN$ 6(1)–5(1) PV diagram.

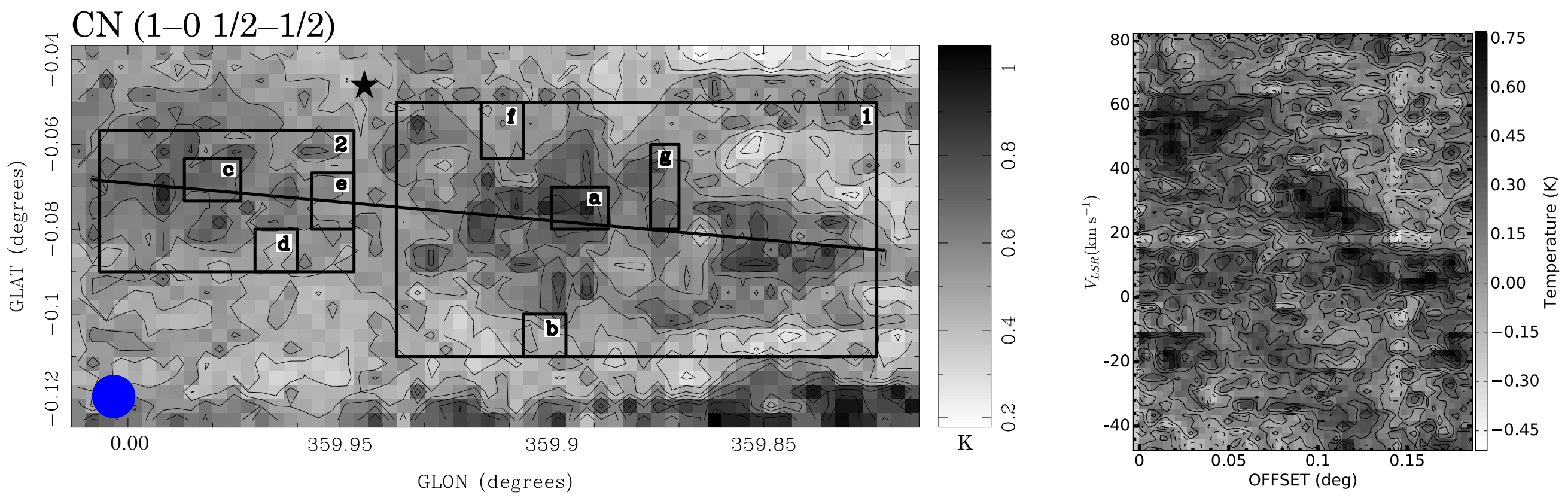

(a) (b)

**Figure 33:** (a) 3-mm CN (1–0 1/2–1/2) peak emission image. Contour levels: 0.22, 0.31, 0.40, 0.48, 0.57, 0.66, 0.75, 0.83, 0.92, 1.01 (K). (b) 3-mm CN (1–0 1/2–1/2) PV diagram.

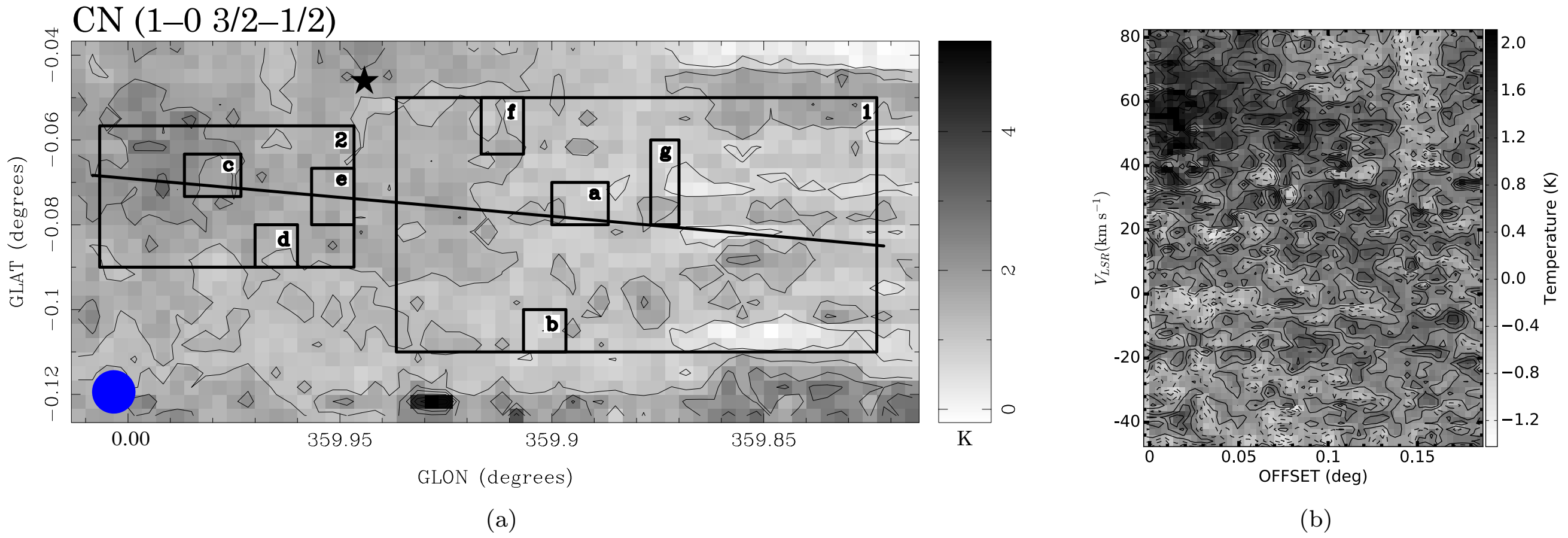

**Figure 34:** (a) 3-mm CN (1–0 3/2–1/2) peak emission image. Contour levels: 0.66, 1.33, 1.99, 2.65, 3.31, 3.98, 4.64, 5.30 (K). (b) 3-mm CN (1–0 3/2–1/2) PV diagram.

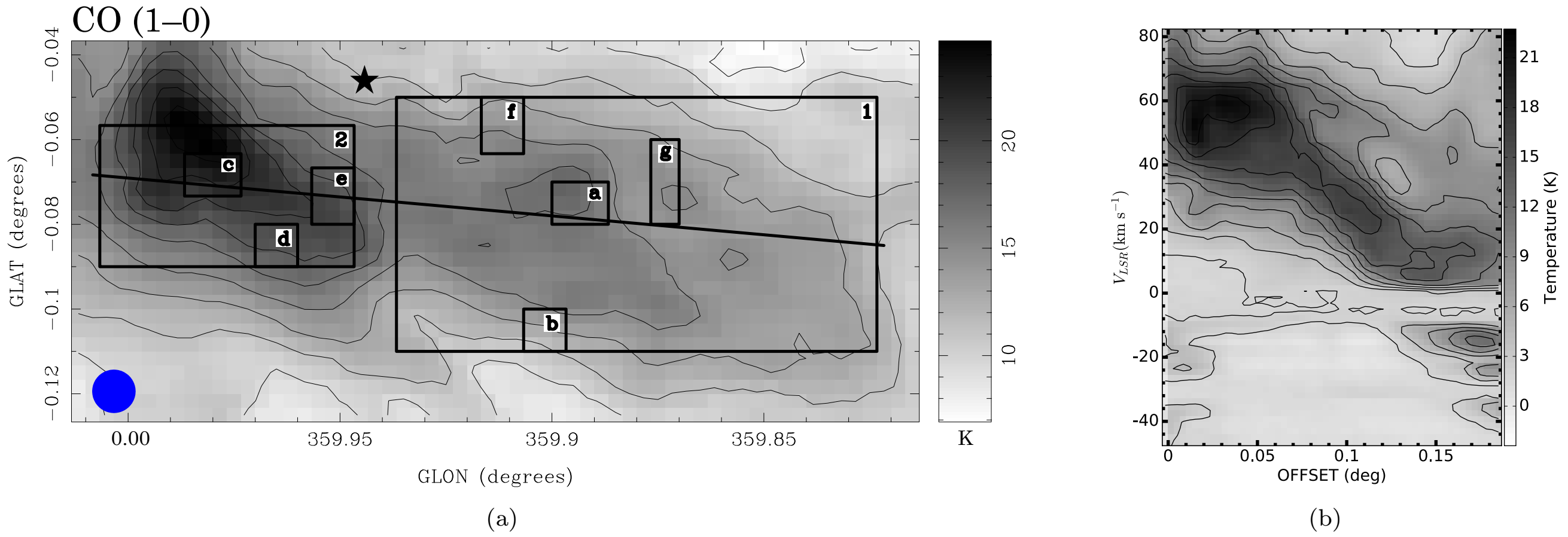

**Figure 35:** (a) 3-mm CO (1–0) peak emission image. Contour levels: 7.79, 9.57, 11.34, 13.12, 14.89, 16.67, 18.44, 20.22, 21.99, 23.77 (K). (b) 3-mm CO (1–0) PV diagram.

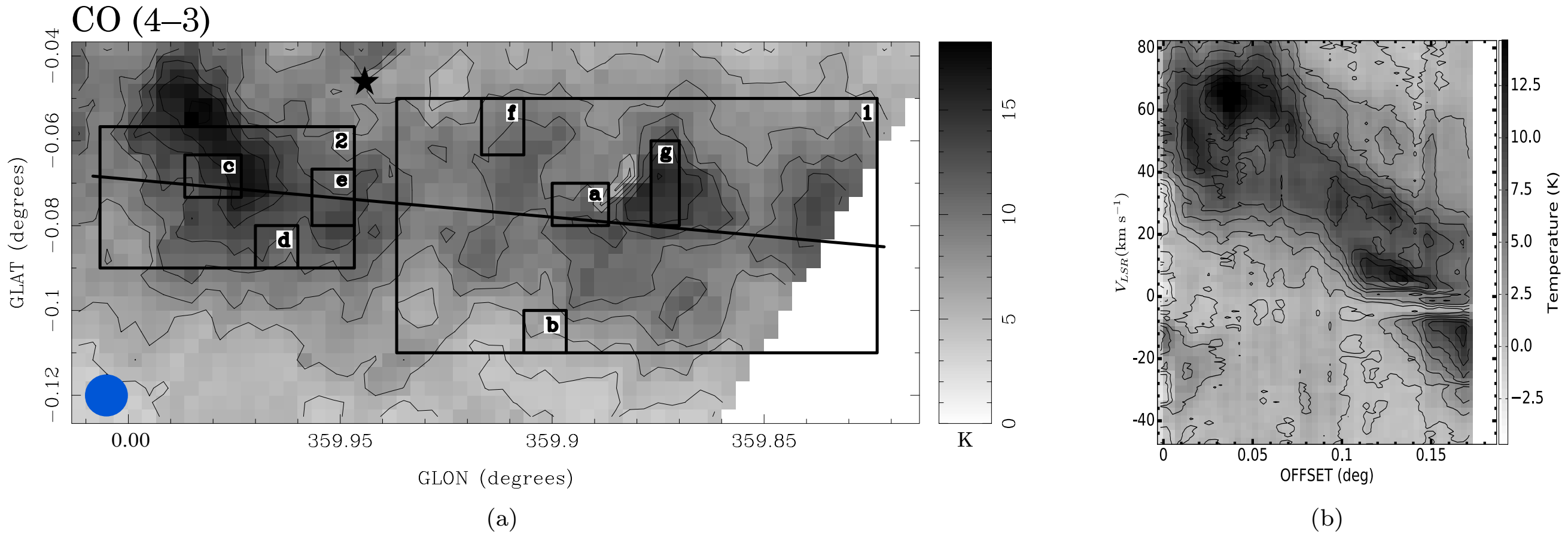

**Figure 36:** (a) 0.65-mm CO (4–3) peak emission image. Contour levels: 0.91, 2.74, 4.57, 6.39, 8.22, 10.05, 11.87, 13.70, 15.53, 17.35 (K). (b) 0.65-mm CO (4–3) PV diagram.